\documentclass[12pt,a4paper,titlepage,oneside]{book}
\usepackage{graphicx}
\usepackage[latin1]{inputenc}
\usepackage{amssymb}
\usepackage[reqno]{amsmath}
\usepackage{a4wide}
\usepackage{fancyhdr}
\usepackage{marvosym}
\usepackage{sectsty}
\usepackage{cancel}
\usepackage{palatino}
\usepackage{mathpazo}
\usepackage{mathptmx}
\usepackage{epsfig}
\usepackage{amssymb}
\usepackage{color}
\definecolor{grey}{rgb}{.85,.85,.85}

\newcommand{\Tr}{\textrm{Tr}}

\newcommand{\csquarebox}[1]{\fbox{\hbox to #1{\vbox to #1{}\hss}}}

\begin{document}

\sectionfont{\fontfamily{phv}\selectfont}
\subsectionfont{\fontfamily{phv}\selectfont}
\subsubsectionfont{\fontfamily{phv}\selectfont}

\pagenumbering{roman}
\begin{titlepage}

\begin{center}
\vspace*{10mm}
{\Huge ENTANGLEMENT - A QUEER IN THE  \\
WORLD OF \\
QUANTUM : SOME PERSPECTIVES\\[3cm]}

{\large Thesis Submitted to the University of Calcutta\\
For the Degree of \\
Doctor of Philosophy (Science)\\
2007\\[3cm]}

{\large By\\ INDRANI  CHATTOPADHYAY \\ Department of Applied Mathematics\\
University College of Science and Technology\\
University of Calcutta\\
92, A.P.C. Road\\
Kolkata- 700009\\INDIA\\[1cm]

}

\end{center}
\cleardoublepage
\end{titlepage}

\pagestyle{fancy}
\newcommand{\helv}{%
\fontfamily{phv}\fontseries{b}\fontsize{12}{14}\selectfont}

\renewcommand{\chaptermark}[1]%
{\markboth{\MakeUppercase{\thechapter.\ #1}}{}}
\renewcommand{\sectionmark}[1]%
{\markright{\MakeUppercase{\thesection.\ #1}}}

\fancyhf{}
\fancyhead[LE,RO]{\rm\thepage}
\fancyhead[LO]{\nouppercase{\rm\rightmark}}
\fancyhead[RE]{\nouppercase{\rm\leftmark}}
\fancyhead[LE,RO]{\helv\thepage}
\fancyhead[LO]{\nouppercase{\helv\rightmark}}
\fancyhead[RE]{\nouppercase{\helv \leftmark}}
\fancypagestyle{plain}{\fancyhead{}
\renewcommand{\headrulewidth}{0pt}}

\renewcommand{\chaptermark}[1]{\markboth{#1}{}} 

\chapter*{Acknowledgements}
\thispagestyle{plain}
\addcontentsline{toc}{chapter}{Acknowledgements}

The author would like to thank her supervisor Dr. Debasis Sarkar,
for proper guidance and help without which it would be truly
impossible to pursue the research work. The author is also grateful to
all the senior scientists of this field or related areas for
various discussions, helpful advices and inspirations. Specially,
the author acknoledges Dr. G. Kar and group with all the good-old
longtime sessions at ISI, Kolkata. The author would sincerely thank
Dr. Arun Pati for invitation at the Institute of Physics. Discussions
about the `No-Go Theorems' at IOP are really open new paths to go on.
Its' a pleasure to be in contact with the renowned research
personalities like, Prof. P. Joag, Dr. S. Ghosh, Dr. S. Bandyopadhyay,
Dr. A. Majumder, Dr. N. Nayek, Dr. P. Agarwal, Dr. P. Parashar.

The author feels proud to be a student and a research scholar of the
Department of Applied Mathematics, University of Calcutta. The
author is thankful to her respected teachers, office stuffs and
library stuffs and other scholars of this department. The author
acknowledges the former and present Heads of the Department of Applied
Mathematics for their kind assistances to pursue the research work.
The author feels honor to acknowledge Prof. S. C. Bose for his
immense help in all respect, with all the academic discussions in
various topics, encouragement and moral support. The author also
acknowledges Prof. S. Sen, Prof. B. Bagchi and Prof. P. R.
Ghosh for their constant encouragements, academic help and positive
attitudes.

The author thanks all of her family members for their love and
encouragements.

The author is grateful to the Honorable Vice-Chancellor, University
of Calcutta for the kind permission for early submission of the
thesis. The author acknowledges CSIR, India for the financial support
by providing Junior Research Fellowship for two years and
subsequently the Senior Research Fellowship for one year during her
research period. Lastly, the author acknowledges all her colleagues at Netaji
Subhash Open University.

\vspace{1cm}

\noindent June, 2007 \hspace{8.5cm} \textbf{(Indrani Chattopadhyay)}

\noindent Kolkata

\cleardoublepage
\pagebreak

\thispagestyle{empty}

\tableofcontents

\mainmatter
\renewcommand{\chaptermark}[1]{\markboth{\thechapter.\ #1}{}}  

\baselineskip 16pt
\chapter{Preface}\label{preface}

\section{General Introduction}
\emph{Quantum Information and Computation} is a recent and exciting
field of interdisciplinary research largely connected with quantum
physics, advances in mathematics and computer science. This is a
rapidly growing field and theoretical development of the subject
becomes very important. This subject is originated and related with
other fields of research like \emph{Quantum
Foundations} \cite{Beltrametti,14b,48}, \emph{Quantum
Optics} \cite{optics}, and \emph{Quantum Dots}. Surprisingly, a
special kind of quantum non-locality termed as
\emph{Entanglement}, plays a key role with its vast applicability on
all those areas \cite{Alber,Benenti,13a,Bruss}. After
Bell's work \cite{BellEPR,BellBook} that addresses EPR paradox \cite{EPR}, the
power of quantum entanglement reflected through many experiments.
Bell's work encourages a lot of people to thought about other proofs
of non-locality to prove or disprove Contextual Hidden Variable
theory. It took really a long time to draw the attention of the
physical community about the features of Entangled systems. In quantum
information and computation, general
interests for using the structure of entangled systems grew after
the very famous work of Bennett group that proposes a secure
cryptographic protocol \cite{crypto}. Some of the major areas of
research about the theory of entanglement are related with the
quantification \cite{plenio1,UniqueMeasure,VedralPlenioRippin,Vedral2},
characterization \cite{Bruss,Vedral1,Vidalpure} and also the uses of
entanglement for performing different computational
tasks \cite{densecoding,teleport,ErrorCorr}. Specifically, the
second and partly the third area are suitable to experience various
counter intuitive phenomena of quantum mechanics. Often the behavior
of entangled system surprises the scientific society in many senses.
Surprise is not much surprising for results of quantum experiments.
Entanglement is gradually becoming a queer in the physical world.

In this thesis our intention is to investigate two peculiar
features of entanglement. First is the notion of
\emph{Incomparability} \cite{majorization} and second one is the
\emph{Activable Bound Entanglement} \cite{bcbe,Smolin}. Existence of
incomparable pair of pure bipartite states is a peculiarity of pure
state entanglement manipulation by Local Operations and Classical
Communications (in short, LOCC), starting from $3\times 3$ systems.
Incomparability shows that in such a simple structure of the state
space, there lies some unexplained phenomena regarding local
evolution of non-local systems. In this simplest possible structure
the incomparability is so strong that all the earlier methods for
resolving incomparability with certainty, became unsuccessful. We
have investigated about such unresolved classes. Incorporating free
entanglement, it became possible to transform deterministically one
pure bipartite entangled state to another, which are incomparable in
nature, by collective LOCC. Obviously not only the presence but the
non-recoverable use of entanglement enhances the process.

Next we use the phenomena of existence of incomparable pair of
states, to show some impossibilities regarding the allowable local
operations defined on quantum systems. Firstly, we consider the exact
spin-flipping operation defined on three arbitrary input qubits.
Then we consider two large classes of operations, one is general
anti-unitary operations and the other one is general
angle-preserving operations and both are detected to be impossible
by the existence of incomparable pair of pure bipartite states. Thus
we propose incomparability as a new detector of impossible
operations in quantum mechanics.

Proceeding further to mixed entanglement states we found a general
class of bound entangled states. Bound entanglement is an
interesting observation in some mixed quantum systems. Though it is
unusable in the sense of distilling out the hidden entanglement by
LOCC, but still capable in performing some quantum computational
tasks. Surprisingly it is seen to be more powerful than free
entanglement in some cases. We found there exists exactly four orthogonal
activable bound entangled states in any even number of qubit systems \cite{bcbe}.
They have some peculiar characteristics including local indistinguishability.
As an application, we have tried to use the general class of activable bound
entangled states to a data hiding scheme to hide two classical bits of
information \cite{hideclass}. However, the scheme has some limitations
regarding security against possible classical or quantum attack. An important
feature of our protocol is, the number of parties concerned can
be increased in pairs as large as possible. A nice Bell correlation
is responsible for that feature. This work is another example of using
multipartite bound entanglement in information processing tasks.

In short our work represents some interesting phenomena in
quantum information and computation theory through the notion of
entangled systems. The work is directed to find new characteristics
of entanglement and we use it to perform some tasks of
information processing. In many perspectives of information theory,
local manipulations of a non-local system is not always very
predictable. The work is largely related with the investigation on
evolution of entangled systems under LOCC. In conclusion,
our work highlights on some special features of entangled
systems in both pure and mixed level together with some new
implementations.

\section{Outline of the Thesis}

In the second chapter, we describe the notations and basic ideas of
quantum information and computation theory. Those prior knowledge
will provided here for independent study of this work for an
uninitiated reader. In the third chapter, we introduce the concept
of entanglement, with elementary characteristics and applicability
as an information theoretic resource of the system. We also describe
the constraints on possible local evolution of entangled system.
Examples are shown to emphasize on the importance of those
constraints to detect impossible local operations defined on single
system, via the evolutions of joint system under those operations.
In the fourth chapter, we describe majorization process and its
applicability on quantum mechanics. Then we describe Nielsen's
criteria for inter-conversion of a pair of pure bipartite states by
LOCC and existence of incomparable pair of states. We provide here
two methods for using free entanglement as a resource, to transform
pairs of incomparable states \cite{catlys}. Next, we investigate
the root of existence of such incomparable pairs and their power to
sense various local operations physical or non-physical. In the fifth
chapter, we consider flipping operations defined on a single system. Here, we show
the idea how incomparability of two pure bipartite states may used
to detect various classes of impossible operations on a single
system. We consider the exact flipping operation defined on a
minimum number of three arbitrary qubit states and show that the
operation is a valid quantum operation only when the three input
states lies on a great circle of a qubit \cite{incomflip}. To
investigate whether the anti-unitary nature of flipping operation is
responsible for this feature, we further consider the general class
of anti-unitary operators defined on only three arbitrary qubits,
which again shows the same result in the sixth chapter \cite{impos}.
We also consider here a general class of angle-preserving operations,
that also give rise to a similar result, i.e., the operation is
physical only on a great circle of the Bloch sphere. Next,
we describe the notion of local indistinguishability of a set of
orthogonal states of composite system in the seventh chapter. The
states are \emph{Activable Bound Entangled} \cite{bcbe}
in nature. We emphasis on the difference between discrimination of a
whole set of states with discrimination of some subsets of the whole
set. Local indistinguishability of quantum systems exhibits various
surprising notions. It reveals other non-local characters of quantum
systems beside of entanglement. Our multi-partite activable bound
entangled states also enriched of the local indistinguishable kind
of non-locality \cite{hideclass}. Thus, we search with this nature of
our system and attempts to apply this feature to build some secure
data hiding protocols. Lastly, we discuss the aspects of data hiding
and limitations regarding our protocol in the eighth chapter.

\chapter{Quantum Information Theory: Preliminaries}

\section{Mathematical Preliminaries}Before of describing the main
work, we first briefly mention some of the basic notions of
mathematical terms widely used in quantum information
theory \cite{fano,NielsenChuang}.

\textbf{Linear vector space:} A linear vector space is a set of
elements, called vectors, which is closed under addition and
multiplication by scalars. Thus, if we denote, $|\psi \rangle$,
$|\phi \rangle$ as vectors belonging to a certain vector space, then
their superposition $a|\psi \rangle~+~b|\phi \rangle$, is also a
vector, where $a,b$ are any scalars. We shall consider scalers in
general complex numbers.

\textbf{Linearly independent vectors:} A set of non-zero vectors
$\{| \varphi_i \rangle; ~i=1,2,\cdots,n\}$ are linearly independent
iff there does not exists a set of scalars $\{a_i;
~i=1,2,\cdots,n\}$ not all zero such that $\sum_{i=1}^n a_i | v_i \rangle~=~0$(zero vector).\\
The maximum number of linearly independent vectors of a linear
vector space is called the dimension of the space and any maximal
set of linearly independent vectors is called a basis of that space.

\textbf{Norm:} Norm is a function $\|\cdot \|  $ that associates to
each vector of a linear vector space $V$ a non-negative value which
satisfies the following conditions:

(i) For any vector $| v \rangle \in V$, $\|| v \rangle \| \geq 0$,
where equality holds if and only if the vector is the zero(null)
vector.

(ii) For any vector $| v \rangle \in V$ and scaler $a$,  $\||a v
\rangle \| = |a| \|| v \rangle \|$.

(iii)(Triangle inequality:) For any two vectors $| v \rangle, |w
\rangle \in V$, $\|| v \rangle + |w \rangle \| \leq \|| v \rangle
\|+ \| |w \rangle \|$.

A linear vector space endowed with a norm is known as a normed
linear space.

\textbf{Linear operator:} A linear operator from a vector space $V$
to another vector space $W$ is defined by a map $A:V\rightarrow~ W$
which is linear in its inputs, i.e., for any set of vectors $\{| v_i
\rangle; ~i=1,2,\cdots,n\}$ and scalars $\{ a_i ; ~i=1,2,\cdots,n\}$
we have,
\begin{equation}
A(\sum_{i=1}^n a_i | v_i \rangle)~=~\sum_{i=1}^n a_i A(| v_i
\rangle)
\end{equation}
All matrices are linear operators acting on some suitable vector
spaces.

\textbf{Inner product:} For any linear vector space $V$ over the
field of scalers $F$(real or complex) inner product is a mapping
from $V \times V \longrightarrow F$ which associates a scalar
denoted by $(\psi,~\phi)$ (or, $\langle \psi | \phi \rangle$), with
every ordered pair of vectors $(|\psi \rangle, | \phi \rangle)$ of
this space. It must also satisfy the following properties stated below.

(I) Positivity: $(\psi,~\psi)\geq 0,$ for any  $|\psi \rangle \in V$
where equality holds if and only if $|\psi \rangle $ is zero vector.

(II) Skew-symmetry: $(\psi,~\phi)~=~(\phi,~\psi)^\ast $, where
$\ast$ denotes complex conjugation.

(III) Linearity: $(\psi,~c_1\phi_1+ c_2\phi_2)~=
~c_1(\psi,~\phi_1)~+~ c_2(\psi~,~\phi_2)$ for every $c_1,c_2 \in F.$

Clearly, inner product as defined above is anti-linear in it's first
argument. A linear vector space endowed with an inner product is
generally known as inner product space. Every inner product space is
naturally a normed linear space. Two vectors $| \psi \rangle, | \phi
\rangle,$ are said to be orthogonal if and only if $(\psi,~\phi)~=~0
$. A basis of an inner product space is said to be an orthonormal
basis if all vectors in the basis are mutually orthogonal and with
unit norm. We now describe some of the linear operators that are
required in our work.

\emph{Adjoint of an operator:} Corresponding to any linear operator
$A$ acting on an inner product space $H$, there exists a unique
linear operator $A^\dagger$ acting on $H$, known as adjoint
operator, so that
\begin{equation}
\langle u |A | v \rangle ~=~\langle (A^\dagger u )| v
\rangle,~~~\forall~~ | u \rangle, | v \rangle \in H.
\end{equation}

\emph{Normal operator:} An operator $A$ is said to be Normal if $A
A^\dagger =A^\dagger A$.

\emph{Hermitian operator:} An operator $A$ is said to be hermitian
or self-adjoint if $A^\dagger =A$.

A normal operator is hermitian if and only if all eigenvalues of
this operator are real. All eigenvalues of a hermitian operator are
real. All hermitian operators are by definition normal.

\emph{Unitary operator:} A normal operator U is said to be unitary
if $U^\dagger U ~=~ I$

If $U$ is an unitary operator, then it is possible to express it as
$U~=~e^{iA}$ where $A$ is a hermitian operator. Unitary operators
have eigenvalues of the form $\exp(i\alpha)$ for some real $\alpha$.

\emph{Positive operator:} An operator $A$ acting on an inner product
space $H$ is said to be a positive operator if $\langle v |A | v
\rangle $ is a real non-negative number for any vector $| v \rangle
\in H.$

A positive operator is necessarily hermitian and eigenvalues are
nonnegative.

\textbf{Commutator:} Commutator of two operators $A$ and $B$ acting
on the same vector space is defined as
\begin{equation}
[A,B] ~=~AB-BA
\end{equation}
Two operators are said to commute each other, if $[A,B] ~=~0$, i.e.,
if $AB=BA$.

\textbf{Anti-commutator:} Anti-Commutator of two operators $A$ and
$B$, defined on the same vector space is
\begin{equation}
\{A,B\} ~=~AB+BA
\end{equation}
Thus two operators are said to anti-commute with one another, if
$\{A,B\} ~=~0$.

\textbf{Pauli matrices:} Pauli matrices are three $2\times2$
matrices defined as

\begin{equation}
\sigma_x ~=~\left(%
\begin{array}{cc}
  0 & 1 \\
  1 & 0 \\
\end{array}%
\right),~ ~\sigma_y ~=~\left(%
\begin{array}{cc}
  0 & -i \\
  i & 0 \\
\end{array}%
\right),~ ~\sigma_z ~=~\left(%
\begin{array}{cc}
  1 & 0 \\
  0 & -1 \\
\end{array}%
\right) \label{pauli}
\end{equation}\\
\emph{Properties of Pauli matrices:}

(i) $\sigma_x^2~=~\sigma_y^2~=~\sigma_z^2~=~I$

(ii) $\sigma_x \sigma_y~=~i\sigma_z~,~\sigma_y
\sigma_z~=~i\sigma_x~,~\sigma_z \sigma_x~=~i\sigma_y$

(iii) $\{\sigma_i,\sigma_j\} ~=~\delta_{ij} ~ 2I ~~\forall~i,j \in
\{x,y,z\}$, i.e., $\{\sigma_i,\sigma_j\} ~=~0$ if $i~\neq~j$.

We now state some important theorems which are required for our
discussions and have many applications in various disciplines
including quantum information theory \cite{Preskill}.

\textbf{Spectral decomposition theorem:} Any normal operator $N$
acting on an inner product space $V$ is diagonal with respect to
some orthonormal basis of $V$. Also, every diagonalizable operator
is normal.

\textbf{Simultaneous diagonalization theorem:} Any two hermitian
operator $A$ and $B$ acting on an inner product space $V$ will
commute with each other (i.e., $[A,B]=0$), if and only if they are
simultaneously diagonalizable (i.e., there exists an orthonormal
basis of $V$ so that both of $A$ and $B$ are diagonal with respect
to this basis).

\textbf{Polar and Singular value decompositions:} These are two
prescription for decomposing a linear operator into simpler
parts \cite{NielsenChuang}. The structure of a general linear
operator is quite complicated to study, by reducing it in terms of
positive operators and unitary operators their action can be
realized in a much more physical way.

\textbf{Polar decomposition theorem:} For every linear operator $A$
there exists an unitary operator, say $U$, and correspondingly two
positive operators $P$ and $Q$, such that $A$ can be uniquely
represented by,
\begin{equation}
A=UP=QU; ~~P\equiv \sqrt{A^{\dagger} A}~ , ~Q\equiv \sqrt{A
A^{\dagger}}
\end{equation}

\textbf{Singular value decomposition theorem:} For every square
matrix $A$ there exists two unitary matrices $U, ~V$ and a
non-negative diagonal matrix $D$ such that $A$ can be decomposed as,
$A~=~UDV$.

The diagonal elements of $D$ are known as singular values.

\textbf{Linear functional:} Corresponding to a normed linear space
$V$ there exists a \emph{dual space of linear functionals} defined
on $V$. A linear functional $\emph{F}$ assigns a scalar value (here,
real or complex) denoted by $F(\phi)$ to every vector $|\phi \rangle
\in V$, such that
\begin{equation}
\begin{array}{lcl}
F(a \psi~+~b \phi )~=~  a F( \psi) ~+~b F( \phi )
\end{array}
\end{equation}
for every pair of vectors $|\psi \rangle,~|\phi \rangle \in V$ and
any pair of scalars $a,b \in F$. The set of all bounded
(equivalently, continuous) linear functionals acting on $V$, forms a
linear space, say, $V'$ which is also a normed linear space
(actually, a Banach Space) where the sum of two functionals is
defined as $(F_1~+~F_2)~(\phi)~=~F_1~(\phi)~+~F_2~(\phi)$

\textbf{Riesz theorem:} There is an one to one correspondence
between the linear functionals $F \in V'$ and the vectors $|\psi
\rangle \in V$ of an inner product space $V$, such that
\begin{equation}
\begin{array}{lcl}
F( \phi )~=~ ( \psi,~ \phi ), ~~\forall~ |\phi \rangle \in V
\end{array}
\end{equation}

\textbf{Use of Dirac's Bra and Ket notation:} Dirac introduced the
bra and ket notation (by splitting the word `bracket') in quantum
mechanics. The vectors in the linear space are called ket vectors,
denoted by $\mid\psi\rangle$ (which we have considered here from
beginning) and the linear functionals in the dual space are called
bra vectors and are denoted by $\langle F\mid$ whose numerical value
is determined as $F(\psi)~=~\langle F\mid\psi\rangle$.

Now according to Riesz theorem there is an one-to-one correspondence
between bras and kets. Therefore it is traditional to use same
algebraic characters for the functional and the particular vector to
which it corresponds differing only in bra and ket sign which
indicates from which space it belongs to. The linear vector spaces
where the rule of Riesz theorem always valid, are Hilbert spaces.
They are the building blocks of quantum systems. We now formally
state the notion of a Hilbert space.

\textbf{Hilbert space:} A Hilbert space is a linear vector space $H$
over the complex field $C$, such that an inner product
$\langle\cdot\mid\cdot\rangle$ is defined on the linear space and
the space is complete in its norm defined in the manner
\begin{equation}
\|\psi\|~=~\sqrt{\langle\psi\mid\psi\rangle},~~~~~\forall~\mid\psi\rangle
\in H
\end{equation}

\section{Logical formalism of quantum mechanics}
Every physical theory involves some basic physical concepts, a
mathematical formalism, and set of correspondence rules which maps
the physical concepts onto the mathematical objects that will
represent them. Those correspondence rules express a physical
problem in mathematical terms, so that the problem may be solved by
purely mathematical techniques that need not have any physical
interpretation. In this context we first recall some basic
concepts \cite{NielsenChuang, Preskill} regarding mathematical
description of any quantum mechanical system.

\textbf{Basic postulates:}
\begin{description}
    \item[\emph{Postulate-1:}] Every physical system is associated with a
separable complex Hilbert space $H$.
    \item[\emph{Postulate-2:}] Every state of a physical system
corresponds to an unique state operator, known as density operator,
which is hermitian, non-negative and of unit trace.
    \item[\emph{Postulate-3:}]  To each dynamical
variable (which is a physical quantity) there corresponds a linear
operator(which is a mathematical object), and the possible values of
the dynamical variable are the eigenvalues of the operator.

Usually, in logical formalism all physical quantities are known as
observables. Every observable $A$ of the physical system is
associated with a self-adjoint (i.e., hermitian) operator
$\Gamma~:~H \longrightarrow H$. Outcomes of any measurements of the
observable $A$ is one of the eigenvalues of the corresponding
operator $\Gamma$.

A hermitian or self-adjoint operator in a Hilbert space has a
spectral representation (by spectral decomposition theorem) as
$\Gamma=\sum_n a_n P_n$. Each $a_n$ is an eigenvalue of $\Gamma$ and
$P_n~\equiv~|n\rangle \langle n|$ is the corresponding orthogonal
projection onto the space of eigenvectors with eigenvalue $a_n$. The
normalized eigenvectors of a hermitian operator forms a complete
orthonormal basis in $H$.
    \item[\emph{Postulate-4:}] If, the outcome of the measurement of an
observable $A$ on a physical system be $a_n$, then just after the
measurement, the system jumps into one of the normalized states
belonging to the support of the corresponding projector $P_n$. This
rule is often designated as the \emph{Collapse Postulate}.
    \item[\emph{Postulate-5:}] If the initial state of a quantum mechanical
system is $\rho$ and a measurement of the observable $A$ is
performed on the system, then the outcome $a_n$ is obtained with the
probability
$$Prob(a_n)~=~tr(P_n \rho)$$
If outcome of the measurement is $a_n$, then the final quantum state
of the system will be $\frac{ P_n \rho P_n^{\dagger}}{tr(P_n
\rho)}$.
This is known as the \emph{Born's Rule}.
    \item[\emph{Postulate-6:}]  The time evolution of any state of a quantum
system is governed by a unitary operator, defined by the Hamiltonian
${\text{H}}$ acting on the system. The dynamics of the system, i.e.,
the evolution of the system described by the state vector $|\psi(t)
\rangle$ with time, is governed by the \emph{Schrodinger equation,}
$$\frac{d}{dt}~| \psi(t)
\rangle~=~\frac{-i}{\hbar}{\text{H}} | \psi(t) \rangle$$
\end{description}

\subsection{Density operator}
Density operator \cite{Preskill} is a linear operator $\rho$ acting on the Hilbert
space that corresponds to the state of the physical system. It obeys
the following rules:

(1) $\rho$ is non-negative,

(2) $\rho$ is self-adjoint,

(3) $\Tr(\rho)=1.$

The set of all density operators acting on a Hilbert space forms a
convex set. Furthermore, they satisfy the relation, $\rho^2 ~\leq~
\rho$. According to this relation we have the following two
classifications of states in a physical system.

\textbf{1. Pure state:} If the density matrix $\rho$ corresponding
to the state satisfies the relation $\rho^2 = \rho$, then the states
are known as pure states. In this case, there is a nice
correspondence between each density operator $\rho$ represents a
pure state and the vectors of the Hilbert space $\emph{H}$, viz.,
$\rho^2 = \rho$ if and only if $\rho~=~\mid \psi \rangle \langle
\psi \mid$ where $\mid \psi \rangle \in \emph{H}$. Thus pure states
are described by rays of the Hilbert space, whether normalized or
not. Pure states can not
be written as a convex combination of the other states. They are the
extreme points of the convex set of all density operators.

\textbf{Phase invariance:} Two pure states $\mid \varphi \rangle$
and $e^{i\alpha} \mid \varphi \rangle$, differing only in some
relative phase $e^{i\alpha}$, where $\alpha$ is a real number, are
said to be physically equivalent up to a global phase factor.
Interesting to note that statistical predictions of any allowable
quantum measurement for these two states are all same. As the
observed properties are equal so states are same from physical point
of view. In this sense global phases are not physically important.

\textbf{2. Mixed state:} All other states which are not pure, called
mixed states. In other words, for every mixed state $\rho$, we have
$\rho^2 ~ < ~ \rho$. Any mixed state can be represented as a convex
combination of some pure states as, $\rho~=~\sum_i \mid \psi_i
\rangle \langle \psi_i \mid$, where each $\mid \psi_i \rangle \in
\emph{H}$.

\emph{ A criterion to detect mixedness of a density operator:} If
$\rho$ be any density operator then $\Tr(\rho^2)\leq 1.$ This
density operator will represent a pure state if and only if
$\Tr(\rho^2)= 1.$ Now we consider the simplest quantum system
described by qubits.

\subsection{Bloch sphere representation of Qubits}

\textbf{Qubits:}  In accordance with cbits as unit of classical
information, the unit of quantum information is denoted by quantum
bits, in short \emph{`qubit'}. This is the simplest possible state
of a quantum system corresponding to the smallest possible
non-trivial Hilbert space of two dimension. An orthonormal basis of
this space can be denoted by $\{ \mid 0 \rangle,~ \mid 1 \rangle
\}$. Since, superposition of pure states are also states of the
system. Therefore, any pure state of a qubit system can be
represented as
\begin{equation}
\mid \varphi \rangle~=~ \alpha |0 \rangle ~+~ \beta \mid 1
\rangle~~; ~~~|\alpha|^2~+~ |\beta|^2~=~1
\end{equation}

\textbf{Bloch sphere representation:} Bloch sphere \cite{Preskill} is a geometrical
way of representing the state of a single qubit system. This is a
unit ball with the principle axes along the direction of spin
polarization of the three Pauli operators. The four orthogonal
operator $I,~\sigma_x, ~\sigma_y, ~\sigma_z$ forms an alternative
basis of the density operators acting on the two-dimensional Hilbert
space for qubits. Any density matrix of the qubit state can be
expressed as,
\begin{equation}
\rho~=~\frac{I+\overrightarrow{n}.\overrightarrow{\sigma}}{2} \ \ \
;\ \ \ \ | \overrightarrow{n} | \leq 1
\end{equation}
where $\overrightarrow{\sigma}~=~(\sigma_x, ~\sigma_y, ~\sigma_z)$
and $\overrightarrow{n}$ denotes the vector of the direction of the
spin polarization of the state $\rho$, and known as \emph{Bloch
vector} for $\rho$. The case, $| \overrightarrow{n} | = 1$
represents pure qubits. If we assume the form of the Bloch vector as
$\overrightarrow{n}=\{n_x,n_y,n_z\}$, then the explicit form of the
qubit state will be
\begin{equation}
\rho~=~\frac{1}{2}(I+n_x\sigma_x+n_y\sigma_y+n_z\sigma_z) =
\frac{1}{2}\left(
\begin{array}{cc}
1+n_z & n_x-in_y \\
n_x+in_y & 1-n_z \\
\end{array}
\right)
\end{equation}
As a geometrical figure a sphere is a convex set. A point on the
surface of the sphere represents a pure state which can not be
written as mixture of ant other qubit. And the points inside the
sphere represent mixed states, that can be expressed in an infinite
number of ways as convex combination of pure states.

For higher dimensional (dim $\geq 3$) single quantum system, however
there is no such geometrical configuration. States in a $d$-
dimensional system are generally known as `qudits'.

\section{Composite system} A composite system consists of more
than one distinct physical subsystems. The state space of a
composite system is the tensor product of Hilbert spaces
corresponding to its subsystems. Consider a composite system
consists of two subsystems, i.e., a bipartite system. Suppose $H_A$
be the Hilbert of the subsystem of the first party that has an
orthonormal basis $\{|\mu_1\rangle,|\mu_2\rangle,\cdots,
|\mu_n\rangle\}$ and the Hilbert space corresponding to the second
party is $H_B$ which has another orthonormal basis $\{|\nu_1\rangle,
|\nu_2\rangle,\cdots, |\nu_m\rangle\}$. Then the combined system of
these two subsystems is $H_A \otimes H_B$ with dimension $n\times m$
and it has a basis given by, $\{|\mu_1\rangle
\otimes|\nu_1\rangle,|\mu_2\rangle \otimes |\nu_1\rangle,\cdots,
|\mu_n\rangle \otimes |\nu_1\rangle, |\mu_1\rangle \otimes
|\nu_2\rangle, |\mu_2\rangle \otimes |\nu_2\rangle,\cdots,
|\mu_n\rangle \otimes |\nu_2\rangle,\cdots, |\mu_1\rangle \otimes
|\nu_m\rangle, |\mu_2\rangle \otimes |\nu_m\rangle, \cdots,
|\mu_n\rangle \otimes |\nu_m\rangle\}$.

In general if the composite systems consist of $k (>2)$ number of
subsystems, we call them usually  multipartite systems. Further if
the states of the $i^{th}$ local subsystem will be $|\psi_i \rangle$
for all $i=1,2,\cdots ,k$, then the joint pure product state of the
composite system will be of the form $|\psi_1 \rangle \otimes
|\psi_2 \rangle \otimes \cdots \otimes |\psi_n \rangle.$ Here after,
for simplicity, in most of the cases, we shall drop the tensor
product notations in writing composite states.

\subsection{Schmidt decomposition of pure bipartite states}

Pure bipartite states has a standard form, known as Schmidt
decomposition. This form is used largely in all kind of information
theoretic tasks performed on pure states including various
entanglement manipulation procedure.

Let $\mid \psi \rangle_{AB}$ be any pure state shared between two
separated parties, Alice and Bob. If the Hilbert spaces
corresponding to the local system of Alice and Bob be $H_A$ and
$H_B$, then there exists orthonormal bases $\{ \mid i_A \rangle \}$
and $\{ \mid i'_B \rangle \}$ of $H_A$ and $H_B$ respectively, such
that $\mid \psi \rangle_{AB}$ can be expressed as
\begin{equation}
\mid \psi \rangle_{AB}~=~ \sum_{i=1}^k a_{i} \mid i_A \rangle \mid
i'_B \rangle
\end{equation}
Where the number of non-zero coefficients $a_i$ in Schmidt
decomposition is a unique number, satisfying $k \leq \min
\{\dim{H_A},\dim{H_B}\}$. It is called Schmidt number of the state
and is constant for every pure state. The Schmidt coefficients can
be taken non-negative real numbers satisfying $0\leq a_i \leq 1$
with $\sum_{i=1}^{k} a_i = 1$. A classification of pure bipartite
states immediately follows from the number of Schmidt coefficients.
If the number of Schmidt terms in a decomposition is one then it is
a pure product state and if it is greater than one, the state is
called an entangled one. It is well-known fact that all fundamental
characteristics of any bipartite pure state is completely determined
by its Schmidt coefficients. Study of all possible kind of local
evolutions of the states are possible by using only the Schmidt
vector, a vector formed by Schmidt coefficients taken in decreasing
order. Thus it is traditional to specify a pure bipartite state by
its Schmidt vector (sometimes it is even unnecessary to specify the
Schmidt bases $\{ \mid i_A \rangle \},~ \{ \mid i'_B \rangle \}$ of
the local subsystems) as,
\begin{equation}
\mid \psi \rangle_{AB}~\equiv~ (a_{1},a_2,\cdots, a_k ),
\end{equation}
where Schmidt coefficients are taken in decreasing order.

Now a more complicated case arises when the state of all the local
subsystems are not pure. Rather we may think that it may be a matter
of search, whether the state of the subsystems are pure or mixed.
Given the most general form of a composite system in terms of
density matrix representation, the corresponding state of any local
subsystem is obtained by computing the reduced density matrices. To
describe the process of finding the reduced density operator, we
first describe the calculation of partial traces.

\subsection{Partial trace and Reduced density matrices}
Consider a bipartite system in the space $H_A \otimes H_B$. Suppose
the density operator representing the state of the joint system is
$\rho_{AB}$ and  $\{|\psi^i_A \rangle,~ i=1, \cdots, m \}, ~
\{|\phi^j_B \rangle,~ j= 1, \cdots, n \},$ be any orthonormal bases
of the subsystems $A$ and $B$ respectively. Then the partial trace
on the joint system is nothing but tracing over any of the subsystem
with respect to an orthonormal basis and after tracing over, the
remaining subsystem is the reduced density matrices of that
subsystem. Usually they are denoted by,
\begin{equation}
\begin{array}{lcl}
\rho_{A}~=~ \Tr_B (\rho_{AB})~=~ \sum_j \langle\phi^j_B | \rho_{AB}
|\phi^j_B \rangle,\\
\rho_{B}~=~ \Tr_A (\rho_{AB})~=~ \sum_i \langle\psi^i_A | \rho_{AB}
|\psi^i_A \rangle.
\end{array}
\end{equation}
In particular, if we consider the state of the joint system is a
pure bipartite state $|\Psi \rangle_{AB}$. Then there exists
orthonormal bases $\{|i_A \rangle \}$ and $\{|i'_B \rangle \}$ of
$H_A$ and $H_B$ respectively, such that $|\Psi \rangle_{AB}$ can be
expressed in its Schmidt form as,
\begin{equation}
|\Psi \rangle_{AB}~=~ \sum^{\kappa}_{i=1} a_{i} |i_A \rangle |i'_B
\rangle~~;~~~~~\kappa\leq \min\{dim(H_A), dim(H_B)\}
\end{equation}
The reduced density matrices of the subsystems $A$ and $B$
respectively are,
\begin{equation}
\begin{array}{lcl}
\rho_{A}~=~ \Tr_B (|\Psi \rangle_{AB}\langle\Psi |)~=~
\sum_{i=1}^{\kappa} |a_i|^{2}|i_A \rangle \langle i_A |,\\
\rho_{B}~=~ \Tr_A (|\Psi \rangle_{AB}\langle\Psi |)~=~
\sum_{i=1}^{\kappa} |a_i|^{2}|i'_B \rangle \langle i'_B |,
\end{array}
\end{equation}
where partial traces are taken over Schmidt bases. Clearly, the
reduced density matrices are in general mixed for a pure bipartite
composite states, except when the joint state is a product one.

Naturally, there arises the cases when the state of all the local
subsystems are not pure. Rather it may be a matter of search to
determine whether the state of the subsystem is pure or mixed. Given
the most general form of a composite system (i.e., a multipartite
system) in terms of density matrix representation, the corresponding
state of any local subsystem is obtained by computing the reduced
density matrices. This is done by tracing out the parts
corresponding to all other subsystems from the density matrix of the
joint composite system. It is interesting to note that the partial
trace operation is the unique operation that provides correct
description of the observable quantities for subsystems of a
composite system. However, the reduced density matrices reveals only
a very limited characteristics of the joint system, as the
correlation between the disjoint subsystems are totally lost in
computing the reduced density matrices.

\section{Quantum operations}

\subsection{POVM measurements:} Let us consider a physical observable
described by the complete set of general measurements $\{M_i\}$.
This observable acts on the system which is originally in the state
$| \psi \rangle$. The probability of the outcome $i$ is, from Born
rule, $p_i=\langle \psi \mid M_i^\dagger M_i \mid \psi \rangle$.
Thus the set of operators $E_i=M_i^\dagger M_i $ is a complete set
of positive operators, sufficient to determine the probability of
different measurement outcomes.

This may also be characterized as a partition of the unity operator
into some non-negative operators. The positivity is required to
ensure the existence of the probability representation given by the
Born rule. To elaborate the concept of performing a generalized
measurement in a joint space $H$, we express it as an extension of
the space $H_A$, on which its action is observed.  Now orthogonal
measurements on the larger space $H= H_A \otimes H_B$, a direct
product space, is not orthogonal on the system of $A$.

Thus the measurement prepare the system in one of the non-orthogonal
states. We may express the orthogonal measurements on the joint
space $H$ as a set of one-dimensional projectors $E_{\mu}
=|\mu\rangle \langle\mu|$, where $|\mu\rangle$ be a normalized state
vector in $H$. We denote the orthogonal projection operator $M$,
that takes the states from the space $H$ to its subspace $H_A$. Then
we construct the POVM as $F_{\mu} = ME_{\mu}M$, each of the
$F_{\mu}$ is a hermitian, non-negative operator \cite{48,Preskill}.
Next we consider symmetry of the transformations and the
corresponding rules for certain class of operations.

\subsection{Transformation symmetry:} The laws of nature are believed
to be invariant under certain space-time symmetry operations
including displacements, rotations and transformations between frame
of references in uniform relative motion. Then some specific
properties of nature expressed as relations corresponding the
observable and states of the system, must be preserved under these
transformations. For example, the eigenvalues of an observable must
be same with that of the equivalent observable in the new reference
frame. Similarly, the inner product of two state vectors must be
always remain invariant after such a symmetry transformation, which
is actually reflected by Wigner's theorem.

\subsubsection{Wigner Theorem:} Any mapping on the inner product space $V$
onto itself that preserves the absolute value of the inner product
$\langle \phi | \psi \rangle$ of the two vectors $|\psi
\rangle,~|\phi \rangle\in V$, may be implemented by an operator $U$
as,
\begin{equation}
\begin{array}{lcl}
|\psi \rangle &\longrightarrow& | \psi' \rangle~=~U| \psi \rangle\\
|\phi \rangle &\longrightarrow& | \phi' \rangle~=~U| \phi \rangle
\label{II}
\end{array}
\end{equation}
Then, the operator $U$ is either unitary or anti-unitary.

Case-1 : When $U$ is an unitary operator, then by definition,
$UU^\dagger=U^\dagger U=I$, the identity operator. Thus inner
product between transformed and original vectors remain same. i.e.,
$\langle \phi' | \psi' \rangle~=~(\langle \phi |U^\dagger)~( U|\psi
\rangle)~=~\langle \phi |(U^\dagger U)| \psi \rangle~=~\langle \phi
| \psi \rangle.$ Thus an unitary transformation preserves the
complex value of an inner product, not merely its absolute value.

Case-2 : If $U$ is anti-unitary then $U(c|\psi \rangle )= c^*U|\psi
\rangle$, for any complex number $c$. Thus in this case the change
in inner product of two vectors will be: $\langle \phi' | \psi'
\rangle=(\langle \phi |U^\dagger)( U|\psi \rangle )= (\langle \phi |
\psi \rangle)^* = \langle \psi | \phi \rangle$, that preserves the
absolute value of inner product.

\subsubsection{Transformation of operators:} Now, any transformation of
state vectors  of the form \ref{II} will be accompanied by a
transformation $A ~\rightarrow~A'$ of the operator corresponding to
some physical observable. The eigenvalues of the operator must
remain invariant. Thus if originally $A| \phi_n \rangle~=~a_n |
\phi_n \rangle$, then finally the eigenvalue equation changes to,
$A'| \phi'_n \rangle~=~a_n | \phi'_n \rangle$. Assuming the
evolution of the state vector to be unitary, i.e., $| \phi'_n
\rangle~=~U| \phi_n \rangle$, the corresponding change in the
observable will be $A ~\rightarrow~A'~=~UAU^{-1}$.

\subsection{Physical operation}

Any physical operation will correspond to a \emph{Superoperator}
$\Im$ acting on the state $\rho$ of the physical system associated
with the Hilbert space $H$, and can be realized by an unitary
evolution on a larger space.

\subsubsection{Superoperator:} A superoperator can be regarded as a
linear map that takes density operators to density operators. To
fulfill the requirement of preserving density operator, such
mappings must have the operator sum representation (or Kraus
representation), given by
\begin{equation}
\rho \longrightarrow \Im(\rho ) \equiv \rho'~=~\sum_{\mu} M_\mu \rho
M_{\mu}^\dagger
\end{equation}
where $\sum_{\mu}
 M_{\mu}^\dagger M_\mu~=~1$.

The properties of this linear map are as follows:\\
(1) $\rho' $ is hermitian, if $\rho$ is so,
\begin{equation}
{\rho'}^\dagger~=~\sum_{\mu} M_\mu \rho^\dagger M_{\mu}^\dagger
~=~\rho'
\end{equation}
(2) $\rho' $ has unit trace, as
\begin{equation}
\Tr {\rho'}~=~\sum_{\mu} \Tr(\rho M_{\mu}^\dagger M_\mu)
~=~\Tr{\rho}~=~1
\end{equation}
(3) $\rho' $ is a positive operator, as for any state $\mid \psi
\rangle$,
\begin{equation}
\langle \psi \mid \rho' \mid \psi \rangle~=~\sum_{\mu} \Tr ( \langle
\psi \mid M_\mu \rho M_{\mu}^\dagger \mid \psi \rangle) ~\geq~0
\end{equation}

It is to be noted that any linear mapping $\Im ~:~\rho \rightarrow
\rho'$, satisfying the above three conditions (1), (2), (3) does not
have a operator sum representation. For obtaining that
representation we need an additional requirement that \emph{$\Im$ is
completely positive}.

\textbf{Kraus Representation Theorem:} Any linear mapping $\Im ~:~
\rho \rightarrow \rho'$ on the states space, that takes a density
$\rho$ matrix to another density matrix $\rho'$ has a operator-sum
representation in the following form,
\begin{equation}
\Im (\rho) = \sum_\mu M_\mu \rho M_\mu^\dagger~ ~;~~ \sum_\mu M_\mu
M_\mu^\dagger~=~1
\end{equation} for any density matrix $\rho$.

In general a quantum operation $\Im$ need not to be trace
preserving. It may be successful with some probability that does not
increase the trace. For that we require only projective measurements
possibly with postselection. Thus, in an alternate way, a quantum
operation $\Im$, in general, is a physical operation that transforms
a state $\rho$ on $H$ into another state $\rho'$ on another Hilbert
space $K$, consists of four elementary operations \cite{Preskill} on
quantum states: (i) unitary transformations, (ii) adding an
uncorrelated ancilla, (iii) tracing out part of the system and (iv)
projective measurements possibly with postselection. It is clear
that the operations (i)-(iii) can be represented by trace preserving
completely positive maps. Now in the next section, we describe some
fundamental concepts of quantum information theory alongwith some
no-go theorems.

\section{Distinguishability and related issues}
In quantum information theory every information about the physical
system, will be encoded in some quantum states. Suppose, there are
$n$ number of results of a random experiment, expressed as
$1,2,3,\cdots,n$. The $i^{th}$ result is encoded in the state
$|\phi_i\rangle$. If it is then required to reveal the result,
having in hand its encoded version, one has to construct a process
of distinguishing the states or simply, to know which state is
given. Thus discrimination of a set of states is directly related
with revealing and processing of the information \cite{38} about the
associated physical system.

\subsection{Existence of non-orthogonal states}

If the dimension of the Hilbert space associated with the system be
$d$, then the system has exactly $d$ number of linearly independent
states, orthogonal to each other. Superposition principle states
that any superposition of these states will also represent a
physical state of the system. In any physical system $(d\geq 2)$
there exists at least two linearly independent non-orthogonal
states. This is a very special and entirely quantum mechanical
phenomenon not seen in classical physics. In classical physics any
two different states of the system are orthogonal which is
equivalent to the statement that any two classical objects are
always distinguishable. In quantum physics the peculiarity of
non-orthogonal states arise and in principle there is no allowable
quantum operation which can perfectly (with probability one)
distinguish any two non-orthogonal states \cite{chefles2}. The
immediate consequences of the existence of non-orthogonal states are
as follows:

The three statements that \emph{two states are non-orthogonal} or,
\emph{two states can not be distinguished with probability one} or,
\emph{two states can not be perfectly cloned by any quantum
machine}, are equivalent.

\subsection{Distinguishability of quantum states}
Suppose a state is chosen at random from a known set of states $\mid
\psi_i \rangle~; ~ ~i=1,2,\cdots,n$. Then the task of determining
the state given, is equivalent with the idea of discriminating the
whole set of $n$ states \cite{nonOrthogonal}. It is obvious that the
task is realizable with certainty, only when the set of $n$ states
are orthonormal. If the states are orthonormal, then we can extend
this set to a basis of the associated Hilbert space $H$, by
adjoining $m-n$ orthonormal states $\mid \psi_i \rangle~; ~
~i=n+1,2,\cdots,m$ where $m(\geq n)$ is the dimension of $H$. So, we
can construct a projective measurement
\begin{equation}
\aleph_n ~\equiv ~ \sum_{i=1}^m \lambda_{i} \mid \psi_i \rangle
\langle \psi_i \mid
\end{equation} And after operating this measurement on the given unknown
state, if we find the outcome of the measurement be $k$, then we can
determine with certainty that the given state is $\mid \psi_k
\rangle$. Since the aspect of distinguishing a set of states is
closely related with ability of copying quantum information encoded
in quantum states, therefore, in the next section we review in brief
the aspect of quantum cloning.

\subsection{Cloning of quantum states}
The quantum cloning operation is performed for copying the
information of a given quantum state on some suitably chosen blank
state \cite{BrussClone,Zurek}. Operating the cloning machine on the
system of input state and the blank state, it will produce two
copies of the input state. Operationally, we may describe the
situation as,
\begin{equation}
\Gamma(|i\rangle|b\rangle|M\rangle) ~=~|i\rangle|i\rangle|M_i\rangle
\end{equation}
where $\Gamma$ is the cloning operation, $\{|i\rangle ;~\forall i\}$
be the set of all input states, $|b\rangle$ is a suitably chosen
blank state, $|M\rangle$ is the initial machine state and
$|M_i\rangle$ is the final states of the machine after the cloning
operation, corresponding to the different input states $|i\rangle$.

As every state represents some information encoded in itself, and
larger number of copies of same state will provide the ability of
recovering more information from the state, therefore cloning
operation is directly related with the aspect of state estimation
\cite{acin}. Having one copy of the input state and sufficient
number of copies of the blank state, if one repeatedly apply a
properly designed cloning machine, then it produces a large number
of copies of the input state. Thus, it is practically equivalent
with estimating the input state \cite{ScaraniGisin}. Now naturally
one may ask that whether there exists any quantum cloning machine
which can copy exactly arbitrary quantum information or not? The
answer is no for deterministic \cite{cloneWootter} as well as
probabilistic \cite{ProbClone} exact cloning. However, universal
inexact cloning machine exists \cite{BrussClone}. Here we now
describe only the famous no-cloning theorem alongwith another famous
no-go theorem, the no-deletion theorem which may be viewed as
somewhat reverse kind of operation than that of cloning. In quantum
deletion, the task is whether it is possible to delete arbitrary
quantum information or not \cite{delet}?

\subsubsection{No-Cloning and No-Deleting Theorems}

\textbf{No-Cloning theorem:} \emph{Universal Exact Cloning is not
possible} \cite{ProbClone,Zurek}.

\noindent\textbf{No-Deletion theorem:} \emph{Exact Deletion of
arbitrary quantum state is not possible} \cite{delet,DeletSignal}.

Both the no-cloning and no-deleting theorems are closely related
with the aspect of indistinguishability of states. Though it is
found that being equivalent with the indistinguishability criteria,
we should note that both the exact cloning and exact deletion of a
set of orthogonal states are quantum mechanically allowable
operations and are always achieved by some unitary machines. Further
more stronger versions of these two no-go theorems are proposed as,

\emph{No-Cloning theorem: Any two non-orthogonal states can not be
cloned exactly by any quantum operation.}

\emph{No-Deletion theorem: Any two non-orthogonal states can not be
deleted exactly by any quantum operation.}

The word '\emph{exactly}' is very much crucial here. Those
restrictions are only imposed for deterministic performance of the
cloning operation. Thus the condition for calling a machine to be
faithful is that, in anyway, it is possible to distinguish each of
the output states by any statistical tests, for any preparation of
the input states. As a probabilistically exact machine, the set of
input states, on which a single quantum exact cloning machine can be
defined, will be enlarged from set of orthogonal states to the set
of linearly independent states.

The above two theorems have some interesting connections with
different directions of the mathematical construction of quantum
mechanics. It is shown that linear dynamics provides these
restrictions over the allowable operations performed on the system.
Unitary evolutions of the systems also restrict such impossible
operations. More fundamental is that if someone assumes the
possibility of such impossible operations, then it is also possible
to send signals faster than the speed of
light \cite{gisinclone,DeletSignal}.

\section{Some ideas from Classical Information Theory}

Subject of Information theory evolves with the very practical
requirement of protecting and conveying information \cite{15g}. In a
more lucid way one may define study of sending, exchanging or
expressing some data or messages as information theory. Classical
information is thus the information that may be represented by some
classical system. For example, result of any random classical
experiment (such as, tossing a coin, or results of a football match)
are always represented by some classical system and they are
essentially known as classical information.

\subsection{Shannon Entropy} Claude Shannon establishes the Shannon entropy as a
quantification rule for data compression, known as `Noiseless coding
theorem'. This expresses, up to what extent a given message can be
compressed. The entropy function is described as,
\begin{equation}
H(p)~=~-\sum_{i=1}^k~(p_i\log p_i)~\leq~ \log k,
\end{equation}
with respect to a probability distribution
$p\equiv\{p_i~;~~~i=1,2,\cdots ,k\}$, where logarithms are taken
with base $2$.

For $k=2$, the entropy of the probability distribution $p_x\equiv \{
x,1-x \}$, is called binary entropy and is given by,
\begin{equation}
h(x)~=~-x\log x -(1-x)\log(1-x)
\end{equation}

\subsection{Mutual information:} It quantifies the amount of
common information between two messages. From Bayes probability rule
this is given by,
\begin{equation}
\begin{array}{lcl}
I(X;Y)& \equiv & ~H(X)-H(X \mid Y) \\
&=&H(X)+H(Y)-H(X, Y) \\
&=&H(Y)-H(Y \mid X),
\end{array}
\end{equation}
where $ H(X, Y)$, is the joint entropy function of a pair of random
variables $X$ and $Y$. Mutual information is symmetric under
interchanges of $X$ and $Y$. This expresses a classical correlation
between classical messages. When $I(X;Y)=0$, then the there is no
classical correlation between $X$ and $Y$. Thus in this situation
learning about one can never provide some extra knowledge about the
other.

\subsection{Relative Entropy of Information}
It expresses the closeness between two probability distributions
$p\equiv \{p_i \}_{i=1}^k$ and $q \equiv \{q_i \}_{i=1}^k$ in the
form,
\begin{equation}
\begin{array}{lcl}
D(p\parallel q)~=~-\sum_{i=1}^k~\{p_i \log(\frac{p_i}{q_i})\}~\geq~
0
\end{array}
\end{equation}
where the equality $D(p\parallel q)~=~0$ holds only when $p=q$. This
is a measure of the classical information, done by comparing the
difference between two probability distributions. The relative
entropy $D(p\parallel q)$ can be viewed as a measure of inefficiency
of assuming that the distribution is $q$, when the actual
distribution is $p$.

{\large\textbf{Conclusion:}} The research of quantum information and
computation has the background of quantum foundations, information
theory and quantum logic. After the ideas of Bennett
\emph{et.al.} \cite{crypto,densecoding,teleport}, the use of quantum
states in computational schemes are enhanced. Thus the research in
the field of quantum information theory started with the goal to
exploit various quantum states in different computational tasks. As
expected some states having purely quantum nature, found to give
better result in developing schemes that are powerful than classical
computational schemes. In composite systems, the invention of
entangled states further enhanced the rapid progress in the field of
quantum information and computation. In recent days some people also
designate quantum information theory as the theory of entanglement.
In the next chapter we will discuss about entanglement in detail.

\chapter{Entanglement: Characterization and Quantification}

\section{EPR paradox and Bell's resolution}
Nature often shows contradiction with our common intuitions. Quantum
mechanics evolved through the study of such apparently peculiar features
of nature explored beyond the predictability of classical world. It
happens to be a stochastic theory rather than a deterministic one.
Afterwards it is found to be something more unpredictable than just
a probabilistic theory. This unpredictability evolved through some
special kind of nonlocal-correlation, that is absent in any
classical system. More specifically, there are quantum correlations
between subsystems that can not be described or interpreted by the
laws of classical mechanics. The correlation, known as entanglement,
causes many nonlocal phenomena. It usually described as non-locality
of quantum systems, but as we will discuss later, there are other
resources of non-locality in a quantum system.

In \emph{1935} Einstein, Podolsky and Rosen proposed the famous EPR
paradox \cite{EPR} which raises a question on the completeness of
quantum mechanics as a physical theory. EPR argument is very much
concerned about the term defined as elements of reality. In that
paper \emph{elements of reality} is described in the following
manner:\\

\emph{"If, without in any way disturbing a system we can predict
with certainty $\ldots$ the values of a physical quantity, then
there exists an element of physical reality corresponding to this
physical quantity".}\\

They considered a pure state describing two particles, say $S_1$ and
$S_2$ which are well-separated by distance but correlated in some
manner. Years after this correlation was termed as
\emph{Entanglement} by Schr\"{o}dinger. The state has the property
that, if one measures the position of the particle $S_1$, one could
determine with arbitrary accuracy, the position of the other
particle $S_2$ and also for the case of momenta. Thus a distant
observer can find out the position of the particle $S_2$ by a
measurement on $S_1$. Locality argument says that this measurement
will not disturb in any way the subsystem of $S_1$. Thus there
exists an element of physical reality corresponding to the position
of $S_2$. Similarly one could conclude that there exists also an
element of physical reality corresponding to the momentum of $S_2$.
Thus from EPR state one could ascribe both the position and momentum
with arbitrarily large probability. This contradicts
\emph{Copenhegen's interpretation of quantum mechanics} that
position and momenta of a particle can not be determined
simultaneously. Thus EPR paradox interprets that quantum mechanics
(rather Copenhegen's quantum mechanics) is an incomplete theory.

This counter-intuitive ideas and many attempts to pose some answers
to it, from the viewpoint of \emph{Hidden variable models} that runs
for years. Then in \emph{1964} comes the Bell's theorem. It
successfully shows the non-locality of all realistic quantum theory.
Bell proved that \cite{BellEPR} even considering statistical quantum
mechanics to be incomplete, it will violate local realism. Thus any
realistic hidden variable theory must be non-local in nature.

In short, the results of Bell experiments concludes with only two
possibilities that either, (1) the world is nonlocal:- which implies
that events happen in nature, violating the principle of reality or,
(2) objective reality does not exist:- there is no matter of fact
about distant events.

However, in essence, with the EPR paradox modern quantum mechanics
observes the birth of a new concept for composite system, known as
entanglement which is responsible for all queers and counter
intuitive ideas in quantum foundations, quantum information and
computation theory \cite{BellBook}.

\section{Separability and concept of entanglement}

Consider a multipartite system shared between distinct parties $A,
~B,~C, ~\cdots$ etc. Each party is situated at a distant place, and
suppose, $H_A$ denote the Hilbert space of the local system of party
$A$, and similarly, $H_B$, $H_C$, etc. The state of the joint system
corresponds with the joint Hilbert space $H_A \otimes H_B\otimes
H_C\otimes \cdots$. The first basic classification of quantum states
is proposed as,

\textbf{1. Separable state:} Suppose the general form of a
multipartite state is represented by convex combination of product
states as follows \cite{werner}:
\begin{equation}
\rho~=~ \sum_i \omega_i~ \rho_A^i\otimes \rho _B^i \otimes \rho_C^i
\otimes \cdots~ ;~ 0\leq \omega_i \leq 1,~\sum_i \omega_i
=1, \label{setI}%
\end{equation}
then, we call it a multipartite separable state.

Pure separable states have the simplest form of product states,
i.e., tensor product of local pure states associated with the
different parties, as stated below,
\begin{equation}
\mid \psi \rangle~=~ |\chi \rangle_A \otimes |\phi \rangle_B
\otimes |\varphi\rangle_C \otimes \cdots
\end{equation}

The physical interpretation of composite systems of two parties, in
product state $| \psi \rangle~=~ |\chi \rangle_A \otimes |\phi
\rangle_B$ is that when the first system is in the state $|\chi
\rangle_A $, the state of the second system is certainly determined
by $|\phi \rangle_B$. So the systems in some separable state always
specifies deterministically the state of one subsystem corresponding
to the state of another subsystem. While for an entangled composite
system, as we will see below, it is really complicated to assign the
properties of individual subsystems.

\textbf{2. Entangled state:} A state of a multipartite system which
can not be expressed in anyway (i.e., considering any basis of the
concerned local Hilbert spaces) in the general form of a separable
state given in Eqs.(\ref{setI}), is defined to be an entangled
state. There is always a special type of correlation between the
different subsystems in an entangled state that can not be
interpreted classically.

One of the basic reason for the above classification scheme is that
the separable states shared between distant parties can be prepared
locally. In other words, it is possible for some distant persons,
holding separate local systems in their own places, to prepare any
separable state of a composite system, by a sequence of local
operations done on their local subsystems and communications between
them through some classical channels shared among the concerned
parties (i.e., by LOCC). This can never be possible for sharing any
entangled state if initially the parties are separated \cite{rsp}.
This phenomena may be described as impossibility of creating
non-local feature in a global system by some local manipulation. In
the next section we will show some power of this non-local feature.

\section{Entanglement as a resource of the system}

Recent development of quantum information theory inspired us to
manipulate entangled systems for implementing various computational
and communication tasks \cite{Benenti}. Emphasis is given on the use
of entanglement for performing such tasks with better precision
comparing to any kind of classical schemes proposed for the same
purposes. Entanglement is now-a-days considered as a valuable
resource of the quantum systems. Quantum teleportation, dense coding
and quantum key distributions are the some successful applications
of this powerful correlation.

\subsection{Quantum Teleportation:} The basic task in quantum
teleportation is to send an unknown information encoded in a quantum
state, say qubit, from one place to another through some quantum
channel, without actually sending the qubit itself, using only local
operations along with classical communications on the subsystems
concerned. The protocol is given by Bennett group
\cite{teleport,purifi} for the qubit system as well as for the qudit
systems. The success relies on the quantum channel considered, taken
usually the maximally entangled states. The interesting fact about
quantum teleportation is that here both the sender as well as the
receiver do not know the state transferred between them. Now, an
unknown qubit information is equivalent with an infinite amount of
classical bits. However, the process of teleporting a qubit
information requires only two cbits of classical information
together with the use of one e-bit of entanglement (this notion of
e-bit will be described later). The situation is different when the
sender knows the information encoded in the qubit and require to
prepare a qubit in receiver's side that contains the same
information. This task is known as remote state
preparation \cite{rsp}. Such kind of tasks are impossible if the
joint state (i.e., the quantum channel) is taken to be separable.
The non-local character of entangled states made the tasks possible.

\subsection{Dense Coding:} Dense coding is one of the most simple
and elegant example to show the power of entangled state to transmit
classical information encoded in quantum states. It is shown that,
by sharing one Bell state, two distant parties can communicate with
certainty, two classical bits of information from one place to
another, by sending a single quantum bit and of course by destroying
the entanglement shared between them \cite{densecoding}. From these
two protocols of Teleportation and Dense Coding in contrast of one
another, an equivalence in quantum information theory is stated as,
\begin{equation}
1 ~Qubit ~ ~ \mathop{\longleftrightarrow}_{1 ~e-bit}~ ~ 2~ cbit
\end{equation}

\subsection{Quantum Key Distribution:} To prepare and share a
secret communication channel between two distant labs so that any
unwanted third party can not eavesdrop the secret, entanglement is
used very efficiently. In BB84 protocol, proposed by Bennett and
Brassard in 1984 \cite{crypto}, two distant parties, say Alice and
Bob generate a secret key by sharing one Classical public channel
(Noiseless and Unjammable) together with Quantum communication
channel, which is assumed to be insecure in terms of an
eavesdropper, who can manipulate the quantum signals in some senses.
The protocol is proposed in such a way that the transmission of
secret information is completely protected against Eave's success of
knowing the secret data without being detected. Alice and Bob, by
their recurring random tests can always detect the eavesdropping and
therefore just by destroying the corrupted channel and starting
refreshed with a new set of data, they would able to communicate
secret information. The protocol, given by Bennett and Brassard, is
based on another invention of modern quantum mechanics, i.e., the
existence of non-orthogonal states. After this, Ekert gave a
protocol \cite{cryp} based on the use of entangled states (i.e.,
Bell states). All the above schemes show how the shared entanglement
may efficiently perform various computational tasks.

\section{LOCC and Entanglement}

Entanglement of quantum system is a peculiar kind of quantum
mechanical correlation, that can be sensed through some operational
criteria. It is generally seen in composite quantum systems shared
between more than one party who are at distant positions. For this
reason, it is sometimes termed as quantum non-locality. By sharing
entanglement between several parties we usually mean some quantum
communications, as there is an equivalence between perfect
entanglement distribution and perfect quantum communication. For
example, if we can transport any qubit without any decoherence
(i.e., disturbance or error formation), then any entanglement shared
by that qubit as a part of a joint system, will also be distributed
perfectly. Thus distribution of entanglement is closely related with
the achievement of a teleportation protocol.

Now, in implementing any kind of computational tasks or information
processing tasks using entangled systems, \emph{Local Operations}
are often very important to perform. By allowable local operations
on a joint system we define ensemble of operations that are
expressed as tensor product of quantum mechanically allowable
operations, i.e., POVM done on local subsystems of single parties.
Local operations are largely used, and encouraged in performing any
tasks as they can be performed in well-controlled environments
without the decoherence induced by communication over long
distances.  Though almost all kind of tasks performed on joint
system, requires interdependence of operations performed in separate
places. Thus the results of operations performed on local
laboratories are communicated through some classical channels (by
any standard telecom technology which can communicate a classical
data perfectly). Studying the status and ability of this general
kind of \textbf{\emph{Local Operations and Classical
Communications}}, in short \emph{LOCC}, is thus an important
motivation  of quantum information theory \cite{chefles3, fan, 36,
37, majorization, nielkemp}. The relation between LOCC and
entanglement can be expressed as laws of quantum information
processing tasks.\\

\textbf{Fundamental laws of quantum information processing}

\emph{Two parties, say, Alice and Bob can not create any amount of
entanglement between them by some local operations on their own
subsystems and communicating through some classical channel, if
initially they are disentangled} \cite{VedralPlenioRippin,
UniqueMeasure}.

\emph{By local operations and classical communications, two distant
parties, say, Alice and Bob, can not increase the total amount of
entanglement shared between them.}

In quantum information theory, a question arises due to some closely
related operations with the LOCC. Those operations are known as
\emph{separable superoperators} \cite{Preskill} and the question is
whether all separable superoperators  can be represented by some
LOCC or not. All LOCC is a kind of separable superoperator but the
reverse is not yet clear \cite{nwe}. In Kraus form, a physical
operation (may or may not be trace preserving) $\Im ~:~
\rho_{ABCD\ldots} \rightarrow \rho'_{ABCD\ldots}$ on the states
space, that takes a density $\rho_{ABCD\ldots}$ matrix shared
between different parties A, B, C, D, $\ldots$ to another density
matrix $\rho'_{ABCD\ldots}$ is known as a separable superoperator,
if it has the operator-sum representation in the following form,
\begin{equation}
\rho'_{ABCD\ldots} = \Im (\rho_{ABCD\ldots}) = \sum_\mu M_\mu
\rho_{ABCD\ldots}M_\mu^\dagger~ ~;
\end{equation}

where $\sum_\mu M_\mu M_\mu^\dagger~\leq~1$ and each $ M_\mu $ is of
the form $M_\mu = A_\mu \otimes B_\mu \otimes C_\mu \otimes D_\mu
\otimes \ldots$ with  $ A_\mu, B_\mu, C_\mu, D_\mu, \ldots$ are
positive linear operators. Now, before going to analyze further the
connections between LOCC and entanglement, we first discuss some
detection procedures and quantification methods of entanglement.

\section{Detection of entanglement}

Detection of a state to be entangled or not is not only a
classification task but it is extremely useful for many practical
purposes.

Here we will mention only some of the good detectors of entanglement
used largely in information theory, like \emph{Bell violation,
Partial transposition, Reduction criteria, Maximal entangled
fraction}, etc.

\textbf{Bell violation:} If the state $\rho$ of a bipartite quantum
system will violate Bell-CHSH inequality,
\begin{equation}
\begin{array}{lcl}
\Tr~(\varrho\emph{B})~\leq~2 \label{belline}%
\end{array}
\end{equation}
where the Bell-CHSH observable $\emph{B}$ is defined as
\begin{equation}
\emph{B}~=~\widehat{a}\overrightarrow{\sigma}\otimes(\widehat{b}+\widehat{b'})\overrightarrow{\sigma}
~+~\widehat{a'}\overrightarrow{\sigma}\otimes(\widehat{b}-\widehat{b'})\overrightarrow{\sigma}
\end{equation}then it is necessarily entangled \cite{Alber}.
$\widehat{a},\widehat{a'},\widehat{b},\widehat{b'}$ are arbitrary
unit vectors of $\Re^3$ and
$\overrightarrow{\sigma}=\{\sigma_x,\sigma_y,\sigma_z\}$.

For $2\times2$ states $\varrho$, the above mentioned Bell inequality
(\ref{belline}) takes the simplified form \cite{bellHorodecki},
\begin{equation}
M(\varrho)~\leq~1
\end{equation}
where $M(\varrho)$ is the sum of the two larger eigenvalues of
$\Gamma^\dagger \Gamma$ and the $3\times3$ real matrix $\Gamma$ is
constructed for the state $\varrho$ by the following prescription,
\begin{equation}
\Gamma = ~\{\Gamma_{ij}\equiv \Tr(\varrho\sigma_i \otimes
\sigma_j)\}_{3\times 3}
\end{equation}
Here we use the notations, $\sigma_1=\sigma_x, \sigma_2=\sigma_y,
\sigma_3=\sigma_z$, for the Pauli operators.

The Bell-inequalities are not sufficient to detect all entangled
states. All pure entangled states violet some Bell-inequalities.
However, there are mixed entangled states that satisfy all the
standard Bell inequalities. In the simplest dimension, i.e., in
$2\times 2$ the so-called Werner states ($U\otimes U$ invariant
states) provide us an example that satisfies Bell-CHSH inequalities
but entangled for a large region of a parameter \cite{werner}. The
entanglement is found initially by the flip-operator. The Werner
states in $2\times 2$, are given by,
\begin{equation}
\rho_W = p |\psi^- \rangle \langle \Psi^- | + \frac{1-p}{4}I
\end{equation}
where $\frac{-1}{3} \leq p \leq 1$ and $|\Psi^- \rangle , I$ are
respectively the singlet state and the identity operator in $2\times
2$. It is easy to check that, for $p> \frac{1}{\sqrt{2}}$, $\rho_W$
violets Bell-inequalities. However, for $\frac{1}{3}< p \leq
\frac{1}{\sqrt{2}}$, they remain entangled. We now present the most
useful detector of entanglement via partial transposition.

\textbf{Partial transposition criteria:} This separability criteria
connected with the partial transposition operation on composite
systems. Let us consider the general form of a bipartite state as
\begin{equation}
\rho_{AB}~=~\sum_{i,j = 1}^m {\sum_{k,l = 1}^n} \alpha_{ij kl} |i
\rangle_A \langle j | \otimes |k \rangle_B \langle l |
\end{equation}
where  $\{|i \rangle_A, ~ i= 1,2,\ldots, m \}$ and $\{|k \rangle_B,
~k=1,2,\ldots, n \}$ are orthonormal bases corresponding to the
subsystems A and B respectively. Then the partial transposition of
this density matrix with respect to the subsystem B, denoted by
$\rho_{AB}^{T_B}$, is defined as
\begin{equation}
\rho_{AB}^{T_B}~=~\sum_{i,j = 1}^m {\sum_{k,l = 1}^n} \alpha_{ij kl}
|i \rangle_A \langle j | \otimes |l \rangle_B \langle k |
\end{equation}

In linear operator theory, the operation transposition is a positive
operator but not a completely positive one. The consequence of not a
completely positive map \cite{Alber}, partial transposition is unable
to preserve positivity of density matrices/operators, but preserves
hermiticity. So, after partial transposition a state cannot be
remain in general a state, it may have some negative eigenvalues. We
then classify the system of bipartite states into two distinct
classes:

(i) {\it states with positive partial transposition} or, PPT states,
i.e., after partial transposition they remain positive operator; and

(ii) {\it states with negative partial transposition} or, NPT
states, i.e., after partial transposition they have at least one
negative eigenvalues.

\emph{Separability criteria:} Now partial transposition of a density
matrix is proved to be a good detector of entanglement \cite{29,31,peres}.
In lower dimensional bipartite states (in
$2\times 2$ and $2\times 3$), it is found to be necessary as well as
sufficient condition for separability \cite{29,31,peres}. Peres
\cite{peres} shows that for every separable state $\rho_{AB}$, both
of its partial transpose are positive  (i.e., $\rho_{AB}^{T_A}>0$,
$\rho_{AB}^{T_B}>0$). Later, for $2\times 2$ and $2\times 3$ system
of states, Horodecki group \cite{29} found that it is also sufficient
one. The result thus known in literature as Peres-Horodecki theorem.
Though for higher dimensional systems it is not sufficient in
general \cite{Alber}. The following are precisely the results
concerning separability and partial transposition.

Theorem-1: Let $\varrho_{AB}$ be a state of the composite system
described by the Hilbert space $H_A \otimes H_B$. Then
$\varrho_{AB}$ is separable if and only if for any positive map
$\Lambda~:~\textit{H}_B~\rightarrow~\textit{H}_B$ the operator $(I
\otimes \Lambda)\varrho_{AB}$ is positive.

Theorem-2: A state $\varrho_{AB}$ of a $2\times 2$ or $2\times 3$
system is separable if and only if its partial transposition is a
positive operator \cite{29,31,peres}.

Immediate consequences of the above results are as follows:

(a) all NPT states are entangled;

(b) all separable states are PPT states.

However, there are entangled states which are PPT states. They have
another interesting property of bound entanglement, for which we
require some notions of quantification procedures for entanglement.
One can now check that the Werner states in $2\times 2$ are
entangled if $p>\frac{1}{3}$ and they are separable for $p \leq
\frac{1}{3}$. Thus partial transposition criteria is found to be a
good detector of entanglement in some cases.

Here we should note that the finding of good entanglement witness is
closely related with the problem of characterizing positive
operators acting on Hilbert spaces. Next, we just mention other two
good detectors of entanglement for a large class of bipartite
states.

\textbf{Reduction criteria:} Separable states in bipartite systems
shared between two parties A and B, must satisfy the following two
inequalities \cite{reduction}
\begin{equation}
I \otimes \varrho_B -\varrho_{AB}~\geq~0~,~~~~\varrho_A \otimes I -
\varrho_{AB}~\geq~0
\end{equation}If any one of the above two conditions is violated
then the state $\varrho_{AB}$ will be entangled. This criteria is
very much useful to detect entanglement. However, there are mixed
entangled states (Werner class of states in $d\times d$) that also
satisfy reduction criteria \cite{werner}. So, the inequality is not
tight enough.

\textbf{Maximal entangled fraction:} The maximally entangled
fraction of a state $\rho_{AB}$ of $d\times d$ system is given by
\begin{equation}
F_{\max}~=~ \max_{\Psi} \langle\Psi|\rho_{AB}|\Psi\rangle
\end{equation}where $|\Psi\rangle$ is any maximally entangled state
of the $d\times d$ system shared between A and B.
Now, it is found that if $F_{\max} > \frac{1}{d}$, then $\rho_{AB}$
is certainly entangled \cite{Alber}.

Both the above detection criteria are not only detects entanglement,
but they are useful in characterizing another aspect of entanglement
related with quantification, i.e., whether distillable or not. In
the next section, we shall describe in brief some of the
quantification procedures of entanglement.

\section{Measure of entanglement}
\textbf{Entanglement is a quantifiable property of the system}

Entanglement is a property of the quantum system, that can be used
to perform various task which are otherwise impossible, or can
enhance the performance of some tasks in quantum information theory.
In practical cases, to apply this resource perfectly the
quantification is very much necessary. The subjects like,
amplification, purification \cite{Vedral2}, concentration \cite{41},
distillation and manipulation are associated with the use of
entanglement. Before going to describe some measures of
entanglement, we first describe the properties of being a good
measure of entanglement \cite{hormeas,plenio1}.

 \textbf{Basic properties of being a measure of
entanglement:}
\begin{enumerate}
    \item  A bipartite entanglement measure denoted by $E(\rho_{AB})$, is a
mapping from density matrices into non-negative real numbers.
    \item  For any separable state $\sigma_{AB}$, the measure should give the
value zero, i.e.,
\begin{equation}
E(\sigma_{AB})~=~ 0
\end{equation}
    \item For any bipartite state $\sigma_{AB}$, its entanglement should
remain unchanged by the action of any local unitary transformation
of the form $U_A \otimes U_B$, i.e.,
\begin{equation}
E(\sigma_{AB})~=~ E(U_A \otimes U_B \sigma_{AB} U_A^\dagger \otimes
U_B^\dagger)
\end{equation}
    \item Local operations, classical communications and
sub-selections cannot increase the expected value of the
entanglement, i.e., if we start with an ensemble in a state
$\sigma_{AB}$ and end up with probability $p_i$ in sub-ensembles in
states $\sigma_i$ then we have,
\begin{equation}
E(\sigma_{AB})~\geq~ \sum_i p_i E( \sigma_i)
\end{equation}
\end{enumerate}

Instead of those above basic requirements, there are some additional
properties which are, though not necessary criteria but sometimes
imposed on entanglement measures for better physical implications.
Such as,

\begin{description}
    \item[Additivity.] Given two pairs of entangled particles in the
total state $\sigma=\sigma_1 \otimes \sigma_2$, where $ \sigma_1,
\sigma_2$ are bipartite states, by additivity, we mean $E(\sigma)~=~
E(\sigma_1)+E(\sigma_2)$. Sometimes instead of additivity we find an
entangled measure $E$ satisfies partial additivity, i.e., for any
bipartite entangled state $\rho$, $E(\rho^{\otimes n})~=~n E(\rho)$.
Also, in asymptotic region we want to observe whether the following
limit exists or not;
\begin{equation}
E^{\infty}(\rho)~=~\lim_{n\rightarrow \infty}\frac{ E( \rho^{\otimes
n})}{n}
\end{equation}
for any bipartite entangled state $\rho$.

    \item[Convexity.] For a set of bipartite density matrices $\{\rho_i\}$
    the entanglement of any convex combination of the states will satisfy,
\begin{equation}
E(\sum_i p_i \rho_i)~\leq~ \sum_i p_i E(\rho_i)
\end{equation}
    \item[Continuity.] Suppose $H_n = H^A_n \otimes H^B_n$ be a
    sequence of bipartite Hilbert spaces and $\rho_n$ and $\sigma_n$
    be sequences of states from $H_n$ corresponding to each positive
    integer $n$. Then we have,
\begin{equation}
\|\rho_n - \sigma_n \| \rightarrow 0~\Rightarrow~\lim_{n\rightarrow
\infty}\frac{ E(\rho_n)- E(\sigma_n)}{1+\log_{2}dim(H_n)}\rightarrow
0
\end{equation}
Sometimes instead of arbitrary Hilbert spaces $H_n = H^A_n \otimes
H^B_n$, we consider the continuity behavior of the measure $E$ only
on $H_n= (H_A)^{\otimes n}\otimes(H_B)^{\otimes n}$.

\end{description}

Now we will discuss about some of the important measures of
entanglement \cite{Vedral1}.

\subsection{Von-Neumann Entropy and Entropy of Entanglement}

The Von-Neumann entropy for any state whose density matrix is
$\rho$, is given by
\begin{equation}
\begin{array}{lcl}
S(\rho)~=~-tr(\rho \log_2 \rho)
\end{array}
\end{equation}Thus if $\lambda_1, \lambda_2,
\cdots, \lambda_n$ are the $n$ eigenvalues of the state $\rho$
(including multiplicity), then we can express the Von-Neumann
entropy as, $S(\rho)~=~-\sum_i \lambda_i \log_2{\lambda_i}$.

Concerning physical importance of this measure, some properties of
the Von-Neumann entropy are discussed below:

(1) \textbf{Purity and Upper Bound:} Von-Neumann entropy is bounded
by the relation $0\leq S(\rho)\leq\log_2 d$, where $d$ is the
dimension of the Hilbert space H. The lowest bound is achievable
(i.e., $S(\rho)=0$) if and only if $\rho$ is a pure state so that we
may express the state as, $\rho~=~\mid \psi \rangle \langle \psi
\mid$. Also, the upper bound $S(\rho)=\log_2 d$ is achievable if and
only if $\rho=\frac{1}{d}I$, i.e., all the non-zero eigenvalues of
the state are equal. Thus, the entropy of a state is maximum when
the state chosen is completely random.

(2) \textbf{Basis Invariance:} Entropy remains unchanged by any
change in the basis of the system, attained by some unitary
transformation over the system; i.e.,
\begin{equation}
\begin{array}{lcl}
S(U\rho U^\dagger)~=~S(\rho)
\end{array}
\end{equation}
This is because entropy depends only on the eigenvalues of the
density matrix.

(3) \textbf{Concavity:} For any real numbers $\alpha_1, \alpha_2,
\cdots, \alpha_n \geq 0$ such that $\sum_{i=1}^n \alpha_i~=~1$
(i.e., $0\leq \alpha_i \leq 1,~ \forall i$), we have
\begin{equation}
S(\sum_{i=1}^n \alpha_i \rho_i )= -\sum_{i=1}^n  \alpha_i
\log_2{\alpha_i}+\sum_{i=1}^n \alpha_i S(\rho_i)~\geq~ \sum_{i=1}^n
\alpha_i S(\rho_i)
\end{equation}
This result can be described as a relation between entropy of a
system and its preparation. Here we observe that the Von-Neumann
entropy increases when the information about the state preparation
decreases.

(4) \textbf{Entropy of measurement:} If we measure the observable
$\textsl{\textsf{A}}=\sum^{n}_{i=1} a_i |\psi_i \rangle \langle
\psi_i|$ in the input state $\rho$, then the outcome $a_i$ occurs
with probability
\begin{equation}
p_i~=~\langle \psi_i |\rho|\psi_i \rangle
\end{equation}

Now the Shannon entropy defined by, $H(Y)~=~-\sum_{i=1}^n  p_i
\log_2{p_i}$, corresponding to the ensemble of measurement outcomes
$ Y= \{p_i ;~i=1,2,\cdots n\}$, will satisfy $H(Y)\geq S(\rho)$. The
equality holds only when, the operator $\textsl{\textsf{A}}$ and
$\rho$ will commute. This can be interpreted physically as the
randomness of the measurement outcomes will be minimized if we
choose to measure an observable that commutes with the density
matrix of the state.

(5) \textbf{Subadditivity:} For any bipartite system in the state
$\rho_{AB}$,
\begin{equation}
S(\rho_{AB}) \leq S(\rho_{A})+S(\rho_{B})
\end{equation}

(6) \textbf{Strong subadditivity:} For any state $\rho_{ABC}$ of a
tripartite system,
\begin{equation}
S(\rho_{ABC})+S(\rho_{B})\leq S(\rho_{AB})+S(\rho_{BC})
\end{equation}

(7) \textbf{Triangle inequality:} For a bipartite system in a state
$\rho_{AB}$,
\begin{equation}
S(\rho_{AB})\geq \mid S(\rho_{A})- S(\rho_{B}) \mid
\end{equation}

Now, for a pure bipartite state $|\Psi \rangle_{AB}$, the {\bf
entanglement or entropy of entanglement} $E(|\Psi \rangle_{AB})$ of
the state is defined by the Von-Neumann entropy of any of its
reduced density matrices. i.e.,

\begin{equation}
E(|\Psi \rangle_{AB}) = S(\rho_{A}) = S(\rho_{B})
\end{equation} where $\rho_i = \Tr_i(|\Psi \rangle_{AB}\langle\Psi
|),~ i=A,B$.

If $|\Psi \rangle_{AB}$ has the Schmidt decomposition, $|\Psi
\rangle_{AB}= \sum^{k}_{i=1} \sqrt{\lambda_i}\mid i_A \rangle \mid
i'_B \rangle$ where $k \leq \min \{\dim{H_A},\dim{H_B}\}, ~ 0\leq
\lambda_i \leq 1,~ \sum^{k}_{i=1}\lambda_i =1$, then we can express
the entanglement of $|\Psi \rangle_{AB}$ as,

\begin{equation}
E(|\Psi \rangle_{AB}) = S(\rho_{A}) = S(\rho_{B})=-\sum_{i=1}^k
\lambda_i \log_2{\lambda_i}
\end{equation}

For a pure product state $|\psi \rangle_{AB}= |\phi \rangle_{A}
|\chi \rangle_{B}$, it is easy to calculate that $E(|\psi
\rangle_{AB})=0$. Also, for a state of the form, $|\Phi
\rangle_{AB}= \frac{1}{\sqrt{d}}\sum^{d}_{i=1}\mid i_A \rangle \mid
i_B \rangle$ where $\{\mid i_A , ~ \mid i_B \rangle, ~i=0,1,2,\ldots
d-1 \}$ are two orthonormal bases of subsystems A and B
respectively, $E(|\Phi \rangle_{AB}) = \log_2{d}$, $d=
\dim{H_A}=\dim{H_B}$. This is the maximal possible value of
entanglement of an entangled state in a two qudit system. For that
reason the states of the kind $|\Phi \rangle_{AB}$ are called
maximally entangled states. All the states that are locally
unitarily connected with $|\Phi \rangle_{AB}$ have also the same
entanglement. Clearly, for two qubit system maximum possible value
of the entanglement is $1$. All the Bell states in two qubit system,
\begin{equation}
|\Phi^{\pm} \rangle =\frac{|00\rangle \pm |11\rangle}{\sqrt{2}}~,~~
|\Psi^{\pm} \rangle =\frac{|01\rangle \pm |10\rangle}{\sqrt{2}}
\end{equation} where first qubit is for system A and second for B,
are pure maximally entangled states with entanglement $1$. Thus,
this value of the entanglement is known as the unit of entanglement
and we usually denote it by 1-ebit. Interestingly, \textbf{entropy
of entanglement is the unique measure of pure state
entanglement} \cite{UniqueMeasure}.

Quantification of mixed entangled states is a very hard tasks in
quantum information theory. There is no single good measure of
entanglement for mixed states. Next, we will proceed to some
measures of entanglement that are proposed from some physically
relevant processes. First, we describe the measure connected with
the preparation process and then turn up to measures related with
extraction or purification processes.

\subsection{Entanglement of formation}
In historical sense, this is the first measure proposed for
quantifying entanglement of a system, by the observation that in
asymptotic limit, the entanglement cost of any pure state in terms
of Bell states is the entropy of entanglement \cite{wooter1,55}. It
gives an upper bound of efficiency of purification process.

{\bf Entanglement of Formation} $E_F$ of a bipartite mixed state
$\rho_{AB}$ is defined to be the minimum value of convex sum of pure
state entanglements over all possible ensembles of pure states which
realizes the mixed state $\rho_{AB}$. If $\rho_{AB}$ can be prepared
from a mixture of the ensemble $\xi\equiv\{p_i,|\Psi_i\rangle\}$, as
$\rho_{AB}~=~ \sum_i p_i |\Psi_i \rangle \langle \Psi_i |$, then
$E_F(\rho_{AB})~=~\min_\xi \{ E(\xi)\}$; where $E(\xi) = \sum_i p_i
E(|\Psi_i \rangle)$.

As this minimization is taken over all possible decompositions of
the density matrix $\rho_{AB}$ into pure states, thus it is
extremely difficult to compute numerically.\\

\textbf{Entanglement Cost:} The entanglement cost of a mixed state
$\rho_{AB}$ is described as \cite{Vidal3},
\begin{equation}
E_C(\rho_{AB})~=~ \lim_{n\rightarrow \infty}
\frac{E_{F}(\rho_{AB}^{\otimes n})}{n}
\end{equation}For pure states it is equal to entropy of entanglement.

\subsection{Distillability} The problem of distillation is
associated with the aspect of extracting entanglement from the
system. Thus distillability is tested only for entangled systems
\cite{AcinDistil,purifi,distil,ErrorCorr,15b1,18a,Shor1}. The subject is
based on the idea of using entanglement as a resource of the system.
Idea for using entanglement as a resource of the system is build out
of the quantum computational protocols, proposed on the maximally
entangled states like, teleportation, dense coding, etc. In
particular, for a number of spatially separated parties,
entanglement is a physical resource with which we can overcome the
practical restrictions of allowing only local operations and
classical communications (i.e., LOCC).

Consider the situation where a source produces some quantum system
in a particular state $\rho_{AB}$. Then the state $\rho_{AB}$ is
distillable if using a large number of copies of (say $N$ copies of
the state, it is possible to prepare a smaller number (say, $M$) of
copies of a maximally entangled state by performing LOCC on the
system. Formally, the distillable entanglement is defined as:

\textbf{Distillable entanglement:} The distillable entanglement
$E_D(\rho_{AB})$ of a bipartite state $\rho_{AB}$ is given by,
\begin{equation}
E_D(\rho_{AB})~=~ \sup_{\it P} \zeta_{\it P}~ ~\text{with}~~
\zeta_{\it P}\equiv \lim_{n \rightarrow \infty} ~ \frac{m}{n}
\end{equation}
where $m$ copies of Bell state $|\Phi^{+} \rangle \equiv
\frac{|00\rangle+|11\rangle}{\sqrt{2}}$ can be extracted from $n$
copies of $\rho_{AB}$ and ${\it P}$ be any LOCC protocol.\\
Now this definition is also in asymptotic sense and it is really
hard to calculate. For product or separable states, distillable
entanglement is zero. But, there are also mixed entangled states
\cite{horb} with zero distillable entanglement. This result of
distillability give birth of a fundamental classification scheme for
entangled system. (i) States that have non-zero distillable
entanglement (i.e., having free entanglement) and (ii), the states
with zero distillable entanglement (known as bound entangled
states). This classification is not complete yet. There are examples
of PPT bound entangled states, but it is not known whether there
exists NPT bound entangled states or not \cite{npt,reduction}.
However, to tackle this problem we need to define distillability of
any bipartite mixed entangled state in an alternate way. A bipartite
mixed entangled state $\rho_{AB}$ is distillable iff there exists a
positive integer $k$ and a Schmidt rank $2$ pure entangled state
$|\psi\rangle$, such that
\begin{equation}
\langle\psi| (\rho^{T_A})^{\otimes k}|\psi\rangle~<~0
\end{equation}
Thus to check whether a bipartite mixed state is distillable, one
has to start from $k=1$ and search for the existence any rank $2$
pure state $|\psi\rangle$, so that the above condition is satisfied.
If one fails for $k=1$, then it requires to check for
$k=2,3,\cdots.$ This definition is given on the basis that if a
quantum state is distillable then projecting it onto a $2\times 2$
subspace, we may able to check whether it is entangled or not. By
Peres-Horodecki criteria, the $2\times 2$ states are entangled if it
has negative partial transpose.

Regarding the relation between distillable entanglement $E_D$,
entanglement cost $E_C$ and any other measure of entanglement $E$ we
have \cite{UniqueMeasure},
\begin{equation}
E_D~\leq~E~\leq~E_C
\end{equation}
It is found that there are zero distillable entanglement states with
non-zero entanglement cost \cite{Vidal3}, an irreversibility in
quantum information processing.

The definition of distillable entanglement proposed only for
bipartite states and but there is no immediate extension for
multipartite entangled states. There are some other well known
measures like, \emph{Relative entropy of
Entanglement} \cite{VedralPlenioRippin}, \emph{Logarithmic
Negativity} \cite{logneg}, \emph{Squashed Entanglement} \cite{15f},
\emph{Concurrence} \cite{55}, etc. For some specific
tasks they are very much usable.\\

\textbf{Concurrence:} The entanglement of formation for two-qubit
system is exactly calculable by a special method. Wootters
provide \cite{55} a new measure of entanglement known as concurrence,
defined by,
\begin{equation}
C_{\rho}=\max\{0, \lambda_1 - \lambda_2- \lambda_3- \lambda_4\}
\end{equation}
where $\{\lambda_i ~; ~~i=1,2,3,4\}$ are the eigenvalues of the matrix
$R\equiv\sqrt{\sqrt{\rho}\widetilde{\rho}\sqrt{\rho}}$ arranged in
decreasing order (i.e., ${\lambda_1}$ is the maximum eigenvalue).
The matrix $\widetilde{\rho}$ is formed as,
\begin{equation}
\widetilde{\rho}=~(\sigma_y \otimes \sigma_y) \rho^* (\sigma_y
\otimes \sigma_y)
\end{equation} where $\rho^*$ is formed by taking complex conjugation
of each elements of the density matrix $\rho$.

Then the entanglement of formation $E_F(\rho)$ for the $2\times 2$
state $\rho$ is given by,
\begin{equation}
E_F(\rho) ~=~ h(\Theta)~;
~~\Theta=\frac{1+\sqrt{1-{C_{\rho}}^2}}{2}
\end{equation}
where the binary entropy function $h(\cdot)$ is defined by
\begin{equation}
h(x)= -x\log_2 x -(1-x)\log_2 (1-x)
\end{equation}

Thus, for two-qubit system $E_F(\rho)$ is a monotonically increasing
function of ${C_{\rho}}$. The function ${C_{\rho}}$ is an
entanglement monotone, ranges from $0$ to $1$. Also, for pure
two-qubit states $E_F(\rho)$ is exactly equal to the entropy of
reduced density matrices. The general formula of concurrence for any
pure bipartite state $|\Psi\rangle_{AB}~\in~H_A\otimes H_B$ is given
by \cite{chen,min,min1,55},
\begin{equation}
C(|\Psi\rangle_{AB})~=~ \sqrt{2(1-\Tr(\rho_S^2))}
\end{equation}where
$\rho_S=\Tr_S |\Psi\rangle_{AB}\langle\Psi|~;~~S=A$ or, $B$. For
mixed bipartite states $\rho_{AB}$, concurrence is given by the
convex roof,
\begin{equation}
C(\rho_{AB})~=~\inf\sum_i p_i C(|\Psi_i\rangle_{AB})
\end{equation} for all possible decomposition of $\rho_{AB}=\sum_i p_i |\Psi_i\rangle_{AB}\langle
\Psi_i|$ where $0\leq p_i \leq 1, \sum_i p_i =1$. For separable
states it gives exactly the value zero. However, for any general
mixed states the calculation requires some optimization
procedures.\\

\textbf{Relative entropy of entanglement:} This is generalization of
the concept of classical relative entropy. The relative entropy
between two states $\sigma$ and $\rho$ is given by,
\begin{equation}
\begin{array}{lcl}
D(\sigma \| \rho) &=& \Tr~ [\sigma (\log_2 \sigma~-~\log_2 \rho)] ~,~if~ Supp~ \rho \subset Supp ~ \sigma\\
&=& \infty ~,~ otherwise
\end{array}
\end{equation}
The relative entropy $D(\sigma \| \rho) \geq 0~\forall ~ \rho,
\sigma $ where equality holds if and only if $\rho = \sigma$. The
relative entropy of entanglement of a bipartite state $\rho_{AB}$ is
then defined by,
\begin{equation}
E(\rho_{AB})  ~=~ \inf_{\sigma_{AB} \in S} D(\rho_{AB} \|
\sigma_{AB})
\end{equation} where $S$ is the set of all separable states \cite{VedralPlenioRippin}.
The motivation of this measure of entanglement is to find the
distance between the entangled state $\rho_{AB}$ with the set of all
separable states. In other words, the amount of entanglement is
quantified by the distinguishability between $\rho_{AB}$ and the
closest possible separable state. It is an upper bound of distillable entanglement.\\

\textbf{Logarithmic negativity:} The quantity known as
\emph{Negativity} is the first one that attempts to find a
computable measure of entanglement, and is defined by,
\begin{equation}
\emph{N} (\rho_{AB})  ~=~ \ \frac{\| \rho_{AB}^{T_B} \| ~-~1}{2}
\end{equation} where $\| X \| ~=~ tr\sqrt{X^\dagger X}$ is the
trace norm.

This definition suffer from the deficiency that it is not an
additive measure. Thus an improvised form of this measure is
proposed as \emph{Logarithmic Negativity} \cite{logneg,vidallog} and defined
by,
\begin{equation}
E_N (\rho_{AB})  ~=~ \ \ \log_2 \| \rho_{AB}^{T_B} \|
\end{equation} It is an additive measure. Interestingly, logarithmic negativity
is a full entanglement monotone that is not convex.

We have discussed before some of the important characteristics of a
good measure of entanglement. Certainly, additivity is
one of them. In almost all aspect of using entanglement as a
resource, the additivity is a natural requirement. However, there
are only very few measures of entanglement with additivity property.
Logarithmic negativity is one and the next measure of entanglement
also satisfies the additivity requirement.

\textbf{Squashed Entanglement:} Squashed entanglement \cite{15f} for
a bipartite state $\rho_{AB}$ is defined by,
\begin{equation}
E_S (\rho_{AB})  ~=~ \inf \Big[\ \ \frac{1}{2} I(\rho_{ABE})
~:~\Tr_E\{\rho_{ABE}\}=\rho_{AB} \Big]
\end{equation} where the $\rho_{ABE}$ is any possible extension of
$\rho_{AB}$ with the third subsystem $E$ (may be considered as the
environment) and $I(\rho_{ABE})$ is the quantum conditional mutual
information of $\rho_{ABE}$, defined as
\begin{equation}
I(\rho_{ABE})  ~=~ S(\rho_{AE}) ~+~S(\rho_{BE}) ~-~S(\rho_{ABE})
~-~S(\rho_{E})
\end{equation}Squashed entanglement has nice properties like, it
is zero for all separable states and coincides with the entropy of
entanglement for the pure states. It is an entanglement monotone
with superadditivity in general and additive on tensor products.
Also, it is convex and continuous almost everywhere and bounded by
$E_D$ and $E_F$.

\section{Thermodynamical aspect of entanglement} Physically it is
much more interesting to deal with open systems, i.e., quantum
systems in contact with environment. In that case, we are largely
concerned with the evolution of the open systems without monitoring
the environment \cite{32a}. Now, like energy, entanglement is
considered as a resource of the system. Thus for proper utilization
of it, the quantification of this resource is required for
information processing. Related with the quantification problem,
there is a fundamental restriction on the evolution of a physical
system consists of different non-interacting distant subsystems. The
laws of nature itself evoke an irreversibility on the joint system
\cite{Vidal1}. It is found that in any process performed on any
entangled system the amount of entanglement invested for the process
is always greater than or equal to the amount of entanglement
recovered from the process. For example, the amount of entanglement
required to prepare a state (by any LOCC protocol) is always greater than
or equal to the amount of entanglement that can be extracted from it
\cite{UniqueMeasure}. Simply, we may describe it as, "resource
cannot be created out of nothing". The purification process of
entangled systems also exhibit such irreversibility \cite{Vidal2}.
Again, as observed, entanglement of a joint system has a relation
with the concept of entropy \cite{32a}. Even in quantification
process, we find, entanglement is directly connected with the idea
of classical Shannon's entropy. And the concept of entropy had
evolved through the knowledge of thermodynamics
\cite{UniqueMeasure}. Interestingly, there is always an underlying
irreversibility that occurs in quantum information processing tasks.

\subsection{Characteristics  of Entanglement}
Thus we have discussed in this chapter some of the fundamental
properties of entangled system evolved directly through the
definition of entanglement \cite{Bruss,VidalMonotone,Vidal1}. In
short, we enlist them as follows:

\begin{itemize}
    \item Separable states contain no entanglement.
    \item All non-separable states allow some tasks to achieve
    better results by some global operations rather than by LOCC alone,
    hence all non-separable states are defined to be entangled.
    \item The entanglement of the states does not increase under any
    sort of LOCC.
    \item In particular, entanglement of a system does not change under Local
    Unitary Operations.
    \item There are maximally entangled states in every dimension.
    In a $d\times d$ pure bipartite system, any state, locally
    unitarily connected with a state of the form
    \begin{equation}
\mid \Phi_d^{+} \rangle_{AB}~=~ \frac{1}{\sqrt{d}}\sum_i  \mid i_A
\rangle \otimes \mid i_B \rangle
\end{equation} is maximally entangled.

\end{itemize}

\section{Discussions}

Entangled states are used for performing various information
processing tasks. Thus it is very important to investigate the
properties of those states together with the process of preparing
and sharing them among some distant parties. Different measures of
entanglement are proposed in view of either how much amount of some
known entangled state (like, Bell states having one e-bit of
entanglement) is sufficient to obtain the required state or how many
copies of a known entangled state are recoverable from the state, or
sometimes how good the state is, for performing some information
processing tasks relative to other known states. This is a very
recent and growing field. A vast area of research is wide open. One
of the interesting topic is manipulating entanglement by local
means. In the next chapter, some ideas of manipulating bipartite
pure entangled states by local operations on the subsystems with
classical communications between them, will be discussed.

\chapter{Majorization and Incomparability}\footnote{Some portions of this chapter is
published in Quantum Information and Computation, 5(3), 247-257
(2005).}

\section{Concept of majorization}
Given two sets of real numbers, i.e., two vectors of equal length as
$x\equiv (x_1, x_2, \cdots, x_d)$ and $y\equiv (y_1, y_2, \cdots,
y_d)$, to measure the disorder between these vectors we need a
specific mathematical modeling. Majorization is a mathematical
process of comparing the vectors in the most general and elegant way
\cite{Bhatia,maj1,maj,vidalmaj}. It is widely applicable in various fields
like, computer science, information theory, etc.

If the lengths of both the vectors are not same, we equate them by
assuming the other elements of the short length vector to be all
zero. To describe the process, we formulate two new vectors
$x^{\downarrow}\equiv (x^{\downarrow}_1, x^{\downarrow}_2, \cdots,
x^{\downarrow}_d)$ and $y^{\downarrow}\equiv (y^{\downarrow}_1,
y^{\downarrow}_2, \cdots, y^{\downarrow}_d)$ by re-arranging the
components of the $d$-dimensional vectors $x$ and $y$ in a
decreasing way such as,
\begin{equation}
\begin{array}{lcl}
x^{\downarrow}_1 \geq x^{\downarrow}_2 \geq \cdots \geq
x^{\downarrow}_d , ~~y^{\downarrow}_1 \geq y^{\downarrow}_2 \geq
\cdots \geq y^{\downarrow}_d
\end{array}
\end{equation}

Then the vector $x$ is said to be majorized by the vector $y$ (or,
alternately, $y$ majorizes $x$), denoted by $x\prec y$, if and only
if
\begin{equation}
\sum_{i=1}^k x^{\downarrow}_i ~\leq ~\sum_{i=1}^k
y^{\downarrow}_i,~~\forall~k=1,2,\cdots,d-1  \label{maj2}%
\end{equation} and
\begin{equation}
\sum_{i=1}^d x^{\downarrow}_i ~=~\sum_{i=1}^d y^{\downarrow}_i \label{maj1}%
\end{equation}

Example: If $x_i~\geq~0$ with $\sum_{i=1}^d x_i ~=~1$, then
\begin{equation}
\begin{array}{lcl}
(\frac{1}{d},\frac{1}{d},\cdots, \frac{1}{d})\prec
(x^{\downarrow}_1, x^{\downarrow}_2, \cdots, x^{\downarrow}_d) \prec
(1,0,\cdots ,0)
\end{array}
\end{equation}

\textbf{An alternative definition:} If $x^{\uparrow}$ be the vector
obtained from $x$ by rearranging the elements in an increasing order
then, $x^{\uparrow}_j~=~x^{\downarrow}_{d-j+1},~~\forall~~j$ such
that $1\leq j \leq d$. Thus we can write, $\sum_{i=1}^k
x^{\uparrow}_i ~=~\sum_{i=1}^d x_i~-~\sum_{i=1}^{d-k}
x^{\downarrow}_i.$

So the majorization relation in equation (\ref{maj2}) can also be
expressed as, $x\prec y$ if and only if
\begin{equation}
\sum_{i=1}^k x^{\uparrow}_i ~\geq ~\sum_{i=1}^k
y^{\uparrow}_i,~~\forall~k=1,2,\cdots,d-1  \label{maj3}%
\end{equation} and
\begin{equation}
\sum_{i=1}^d x^{\uparrow}_i ~=~\sum_{i=1}^d y^{\uparrow}_i  \label{maj4}%
\end{equation}

Equations (\ref{maj1}) and (\ref{maj4}) are both equivalent with the
condition $\sum_{i=1}^d x_i= \sum_{i=1}^d y_i$.

\textbf{Trace condition of majorization:} Let us consider the vector
$|e\rangle~=~(1,1,\cdots,1)$ of dimension $n$. Then given any vector
$|x\rangle \in \Re^n$, we can define
\begin{equation}
tr(x) ~=~\sum_{j=1}^n ~x_{j}=\langle x,e \rangle,
\end{equation}
where $\langle \cdot,\cdot \rangle$ denotes the inner product of two
vectors in $\Re^n$.

Now, for any subset $I$ of $\{1,2,\cdots,n\}$, the vector
$|e^I\rangle \in \Re^n$ is defined as, $|e^I\rangle = (
e_1,e_2,\cdots,e_n)$, where $e_j = 1$ for all $j\in I $ and $e_j=
0$, otherwise. Then we have a relation which provide us another
definition of majorization between two vectors:

\emph{For any vectors $x,y \in \Re^n$, $x\prec y$ if and only if for
each subset $I$ of $\{1,2,\cdots,n \}$ there exists another subset
$J$ of $\{1,2,\cdots,n\}$, with $\mid I \mid~=~ \mid J\mid$ such
that}
\begin{equation}
\begin{array}{lcl}
\langle x,e^I \rangle ~\leq~\langle y,e^J \rangle
\end{array}
\end{equation} and
\begin{equation}
\begin{array}{lcl}
tr(x) ~=~tr(y).
\end{array}
\end{equation}

\section{Doubly stochasticity and majorization}

\textbf{Linear mapping on vector spaces:}  Let $\Im(V,W)$ be the
space of all linear mappings from a vector space $V$ to another
vector space $W$. Then corresponding to every basis of $V,W$, each
linear mapping belonging to $\Im(V,W)$ has a unique matrix
representation.

If $\emph{H, K}$ are Hilbert spaces with dimensions $n,~m$
respectively and the bases of the Hilbert spaces $\{|e_1\rangle,
|e_2\rangle, \cdots, |e_n\rangle \}$ and $\{|f_1\rangle,
|f_2\rangle, \cdots, |f_m\rangle \}$ are chosen to be orthonormal,
then the matrix representation of each operator $A\in \Im (H,K)$ is
obtained by, $A~=~(a_{ij})_{m\times n}$ where $a_{ij}~=~\langle
f_i | Ae_j \rangle$.

Now we define some characteristics of linear maps $A$ defined on
$C^n$.

\emph{A linear maps $A$ on $C^n$ is called
\textbf{positivity-preserving} if it transforms any vector with
non-negative coordinates to some vector with non-negative
coordinates.}

\emph{A linear maps $A$ on $C^n$ is called \textbf{trace-preserving}
if $ tr (Ax) ~=~ tr (x)$ for all $x \in C^n$.}

\emph{A linear maps $A$ on $C^n$ is called \textbf{unital} if
$Ae~=~e$ where $e$ is a unit vector.}

The majorization relation between two vectors is associated with a
very special kind of matrix defined below.

\textbf{Doubly stochastic matrix related with majorization:} An
$n\times n$ matrix $A=(a_{ij})_{n\times n}$ is said to be doubly
stochastic, iff
\begin{equation}
\begin{array}{lcl}
~~~~~~~~a_{ij}\geq  0,~~~~\forall~~i,j,\\
\sum_{i=1}^n ~a_{ij}= 1,~~~\forall~~j,\\
\sum_{j=1}^n ~a_{ij}= 1,~~~\forall~~i.
\end{array}
\end{equation}Alternately, we can say that a matrix $A$ is
\emph{doubly-stochastic} if and only if the corresponding linear
operator $A$ is \emph{positivity-preserving, trace-preserving} and
\emph{unital}.

Then, the results about the majorization process are as follows:

\textbf{Theorem-1:} A matrix $A$ is \emph{doubly-stochastic} if and
only if $Ax~\prec~x$, for all vector $x \in \Re^n$.

\textbf{Theorem-2:} If $x,~y\in \Re^n$, then $x\prec y$ if and only
if $x~=~Ay$, for some doubly-stochastic matrix $A$.

\subsection{Matrix/Operator majorization} Quantum mechanics is a
probabilistic theory where states of a physical system are
represented by density matrices. Any density matrix is hermitian,
positive semi-definite, trace preserving. The evolution of the
system has some relation with majorization. For this reason, we
require the idea of majorization of matrices, preferably for
hermitian matrices.

\emph{If $\Omega_1$ and $\Omega_2$ are hermitian matrices and
$\lambda_i$ be the vector of eigenvalues arranged in non-increasing
order, corresponding to the matrix $\Omega_i$, for $i=1,2$, then we
define $\Omega_1 \prec \Omega_2$ if and only if $\lambda_1 \prec
\lambda_2$.}

Now, quantum operators are largely connected with the notion of
majorization. Briefly, they are as follows:

\emph{Lemma-} A quantum operation is doubly stochastic if it is both
trace preserving and unital.

Thus doubly stochastic operators are of great importance in quantum
information theory. They have many properties including that, they
will never increase the entropy of a system.

\emph{Theorem-1:} Suppose $A$ is hermitian and $\xi$ is a doubly
stochastic operator, then $\xi(A) \prec A$.

\emph{Theorem-2:} If $\xi$ is a trace-preserving quantum operation
that is not doubly stochastic, then there exists a hermitian
operator $A$ such that $A \prec \xi (A)$ but $\xi(A) \not\prec A$.

\emph{Schur's theorem-} Let $A$ be a $d \times d$ hermitian matrix.
Let $diag(A)$ denote the vector whose components are the diagonal
entries of $A$ and $\lambda_A$ denote the vector whose components
are the eigenvalues of the operator $A$. Then $diag(A) \prec
\lambda_A$.

The set of all doubly-stochastic matrices have a nice geometric
configuration alongwith the permutation matrices.

\textbf{\emph{Birkhoff's theorem:}} The set of all $n \times n$
doubly-stochastic matrices is a convex set whose extreme points are
the  $n\times n$ permutation matrices.

As there are $n!$ number of permutation matrices of size $n$, so
from Birkhoff's theorem, every $n\times n$ doubly stochastic matrix
is a convex combination of these $n!$ number of matrices.

\subsection{Random unitary evolution}
Unitary evolution is a very important and general kind of physical
operation. To study the nature of any physical system it is
necessary to explore its behavior under every possible unitary
operation performed on it. Transformation of the state of a system
under any unitary operation is termed as a unitary evolution of the
system.

\textbf{Random unitary evolution:} Any random unitary evolution, of
a physical system in an initial state $\rho$, will be denoted by
$\xi(\rho)$ and is described as a set of unitary matrices $U_i$
together with the probabilities $p_i$ which evolve the physical
system to the final state $\rho^f$ such that
$$\rho^f~=~\xi(\rho)~=~\sum_i p_i U_i \rho U_i^\dagger$$

If we represent the evolution operator $\xi$, as a set of operations
$\{\xi_i~;~i=1,2,\cdots n\}$ then the elements of this set can be
expressed as $\xi_i~=~\sqrt{p_i} U_i$, where
$$\sum_i\xi_i~\xi_i^\dagger~=~\sum_i p_i U_i U_i^\dagger~=~\sum_i
p_i~=~1$$Thus we see that every random unitary evolution is unital
in nature. Next, if we compute the trace of the final state, then we
have,
$$\emph{\Tr}(\rho^f)~=~\emph{\Tr}(\sum_i p_i U_i \rho
U_i^\dagger)~=~\sum_i p_i \emph{\Tr}(U_i \rho U_i^\dagger)~=~\sum_i
p_i \emph{\Tr}(\rho )~=~\emph{\Tr}(\rho ).$$Thus any unitary
evolution is trace-preserving in nature.

It is interesting to observe that every random unitary evolution is
a doubly stochastic operation, as $\xi(\rho)$ is both unital and
trace-preserving. And we have the following theorem:

\textbf{Uhlmann's theorem:} For any pair of hermitian matrices $A$
and $B$, the following conditions are equivalent:
\begin{enumerate}
  \item $A \prec B$
  \item There exists a random unitary evolution characterized by a set of
  unitary matrices $U_i$ and probabilities $p_i$, connecting $A$ and
  $B$ such that $A~=~\sum_i p_i U_i B U_i^\dagger$.
  \item There exists a doubly stochastic operation $\xi$, such that
  $A=\xi(B)$.
\end{enumerate}

We previously seen that every random unitary evolution is a doubly
stochastic matrix, but the converse is not true. For our purpose we
confine ourselves with the evolution of hermitian matrices.
Uhlmann's theorem implies that in this case, every doubly stochastic
matrix is also a random unitary evolution.

\textbf{Change in entropy of the system:} For a single quantum
system in state $\rho$, we measure the disorder by the von-Neumann
entropy of the system, i.e., $S(\rho)~=~-\Tr(\rho \log_2 \rho)$. It
is interesting to note the behavior of entropy of two systems, when
one of them is majorized by the other. For any pair of density
matrices $\rho$ and $\sigma$, the relation $\rho\prec\sigma$ will
then imply $S(\rho)\geq S(\sigma)$. This may be also described as a
natural equivalency between two processes of measuring disorder of a
system.

\section{Majorization in quantum mechanics}

The algebraic relation via majorization between two vectors is
connected very nicely with the evolution of quantum system. In
different areas of quantum mechanics, idea of majorization is used
very successfully. It reveals the characteristics of physical system
and its possible physical evolutions. Majorization criteria also
detects possible measurement outcomes and operational outputs of a
physical system. We mention some of those results \cite{43} here.
\begin{description}
\item[Probability of any Measurement outcome:]

\emph{Probabilities of any measurement in an orthonormal basis are
connected with the eigenvalues of the initial state by the
majorization relation.}

Suppose the state corresponding to a physical system is represented
in its spectral decomposition form as, $\rho= \sum_i \lambda_i |i
\rangle \langle i|$, where $\{\lambda_i ~;~ i=1,2,\cdots d\}$ are
the eigenvalues of the density matrix $\rho$ and $\{|i\rangle ~;~
i=1,2,\cdots d\}$ are the corresponding eigenvectors. If a
measurement in an orthonormal basis $\{|\phi_k \rangle ~;~
k=1,2,\cdots d\}$ is performed on this system, then the $k^{th}$
result will occur with the probability,
\begin{equation}
\begin{array}{lcl}
p_k~&=&~ \langle \phi_k | \rho | \phi_k \rangle \\
~&=&~\sum_{i=1}^{d} \lambda_i |\langle\phi_k |i\rangle|^2
\end{array}
\end{equation}

Now, for any density matrix $\rho$ characterizing the state of a
system, the necessary and sufficient condition for the existence of
any measurement basis of the system, for which a given set of values
$\lambda_\rho =~\{\lambda_i ~;~ i=1,2,\cdots d\}$ will serve as the
probabilities of that measurement is described by;

\item[Quantum Measurement without Post-Selection:]

\emph{If $\rho$ be the density matrix corresponding to the initial
state of a physical system then measuring the system in any complete
set of orthonormal projectors, the final state of the system will be
$\rho^f$ if and only if  $\rho^f \prec \rho$.}

The measurement associated with any complete set of projection
operators, represented by, $\{P_j;~~~j=1,2,\cdots,d\}$, is usually
called von Neumann measurement. If such a measurement is performed
on the physical system, then the measurement outcomes will be one of
the state corresponding to a projector $P_j$. Without knowing the
actual outcome obtained, the final state after measurement will be
described by,  $$\rho^f~=~\sum_j P_j \rho P_j$$ And then we will
find the evolution of the system, characterized by some unitary
operations, as follows,
$$\rho^f~=~\sum_k U_k \rho U_k$$ Therefore, by Uhlmann's theorem, $\rho^f
\prec \rho$.

\item[Possible Pure State Decomposition of any Density matrix:]

\emph{For any given density matrix $\rho$ with vector of eigenvalues
$\lambda \equiv (\lambda_1, \lambda_2,\cdots,\lambda_d)$ there
exists a pure state ensemble $\{p_j, |\phi_j \rangle\}$
corresponding to some fixed probability vector $p\equiv
(p_1,p_2,\cdots,p_d)$ if and only if $ p \prec \lambda$.}

In many tasks of information processing it is often necessary to
decompose some given density matrix in terms of different pure state
ensembles $\{p_j, |\phi_j \rangle\}$. It is then important to decide
whether for a given probability vector, there exists some set of
pure states that may be represented by the density matrix, $$\rho~=~
\sum_j p_j |\phi_j \rangle \langle \phi_j |$$ Clearly, it is
determined by the majorization criteria.

\item[Measurement increases disorder of the system:]

\emph{If $\lambda \equiv (\lambda_1, \lambda_2,\cdots,\lambda_d)$ be
the vector of eigenvalues of some physical system then there exists
a measurement with possible set of outcomes $\{\rho_j ~;~
j=1,2,\cdots,d\}$ occurring with probabilities $\{p_j ~;~
j=1,2,\cdots,d\}$, if and only if}$$\lambda~\prec~\sum_j p_j
\rho_j$$ This result implies, measurements acquire some information
about the system being measured.
\end{description}

\section{Bipartite state transformation by LOCC}

The simplest kind of non-local states are the states shared between
two distant parties. In such bipartite cases, many properties of the
joint system is revealed by the states of the local subsystems,
i.e., by the reduced density matrices. Bipartite entangled states
are required for performing different kinds of computational tasks
like, Teleportation, Dense Coding, etc. Now, sometimes for
performing specific kind of information processing tasks, one
requires some specific entangled states. But the specific state may
not be available always. Then further task is that, given a known
state shared between two parties, whether it is possible to convert
the given state according to their own choice. For such purpose, it
is interesting to find out the set of states that may obtained from
a particular entangled state. And, it is much preferable to perform
the transformation of entangled states by LOCC \cite{34,36}. This
LOCC restriction is not only imposed due to practical usefulness,
but also due to the cause that conversion of entangled states by
global operations is physically equivalent of transforming states of
a single system by valid physical operations. Thus properties of
entangled systems may not be reflected if we allow global operations
to perform on a physical system.

Various ideas are prescribed for making possible the transformation
of pure bipartite states $|\Psi\rangle_{AB}$ to $|\Phi\rangle_{AB}$,
shared between two parties A and B, under LOCC in both the
deterministic and probabilistic manner. Main idea of most of those
schemes can be described as a scheme for manipulating entangled
states to enhance the required tasks. The most important result is
due to M.A. Nielsen \cite{majorization,vidalmaj}, where we find a necessary
and sufficient condition for deterministic conversion of pure
bipartite entangled states under LOCC.

\subsection{Nielsen's criteria}

Nielsen provided a criteria \cite{majorization}, related with
majorization, for transforming one pure bipartite state to another
by any sort of allowed local operations and classical communications
with certainty.

\textbf{Nielsen's majorization criteria:} Suppose the pure bipartite
state $|\Psi\rangle_{AB}$ is shared between two parties, say, Alice
and Bob situated at two distant places. The task is to determine the
possibility of converting the joint state $|\Psi\rangle_{AB}$ with
certainty, to another pure bipartite state $|\Phi\rangle_{AB}$ under
LOCC performed by Alice and Bob. First, let us represent
$|\Psi\rangle_{AB}$, $|\Phi\rangle_{AB}$ in the Schmidt form with
decreasing order of Schmidt coefficients:
\begin{equation}
\begin{array}{lcl}
|\Psi\rangle_{AB}&=& \sum_{i=1}^{d} \sqrt{\alpha_{i}}
|\mu_i\rangle_A |\nu_i\rangle_B ~~;~ ~ \alpha_{1}\geq \alpha_{2}\geq
\cdots
\alpha_{d}\geq 0,\\& &~~~~~~~~~~~~~~~~~~~~~~~~~~~~~~~~~~~~~~~~~~~~\sum_{k=1}^{d} \alpha_{k} = 1\\
|\Phi\rangle_{AB}&=& \sum_{j=1}^{d} \sqrt{\beta_{j}}
|\eta_j\rangle_A |\zeta_j\rangle_B\rangle ~~;~ ~\beta_{1}\geq
\beta_{2}\geq \cdots\beta_{d}\geq 0,\\&
&~~~~~~~~~~~~~~~~~~~~~~~~~~~~~~~~~~~~~~~~~~~~~~ \sum_{k=1}^{d}
\beta_{k}=1
\end{array}
\end{equation}
Here $\{|\mu_i\rangle_A |\nu_i\rangle_B ~;~ i=1,2,\cdots,d \}$ is
the Schmidt basis for the state $|\Psi\rangle_{AB}$ and
$\{|\eta_j\rangle_A |\zeta_j\rangle_B  ~;~ j=1,2,\cdots,d \}$ is the
same for state $|\Phi\rangle_{AB}$. We denote, the Schmidt vectors
corresponding to the states $|\Psi\rangle_{AB}$ and
$|\Phi\rangle_{AB}$ as:
$\lambda_\Psi\equiv(\alpha_1,\alpha_2,\cdots,\alpha_d),$
$\lambda_\Phi\equiv(\beta_1,\beta_2,\cdots,\beta_d)$. Then according
to Nielsen's criterion, the necessary and sufficient condition for
the conversion of $|\Psi\rangle_{AB}$ to $|\Phi\rangle_{AB}$
(expressed as, $|\Psi\rangle_{AB}\rightarrow| \Phi\rangle_{AB}$)
with certainty under LOCC is, $\lambda_\Psi\prec\lambda_\Phi$, i.e.,

\begin{equation}
  \sum_{i=1}^{k}\alpha_{i}\leq \sum_{i=1}^{k}\beta_{i},~ ~\forall~
~k=1,2,\cdots,d \label{nielsen}
\end{equation}

\textbf{Change in entanglement:} Let us now recall the
thermodynamical law of entanglement, i.e., the amount of
entanglement shared between some spatially separated parties can not
be increased by LOCC. Therefore, as a consequence of thermodynamical
law of entanglement, if $|\Psi\rangle_{AB}\rightarrow
|\Phi\rangle_{AB}$ is possible under LOCC with certainty, then
$E(|\Psi\rangle_{AB})\geq E(|\Phi\rangle_{AB})$; where $E(\cdot)$
denotes the entropy of entanglement. Now, from Nielsen's criteria we
have the following corollary.

\emph{Corollary:} Let $|\psi\rangle_{AB}$ be any pure entangled
state of a $m \times n$ system. The number of Schmidt coefficients
of $|\psi \rangle_{AB}$ is $d\leq~\min \{ m,n \}$. Then,
$|\Psi\rangle_{AB}\rightarrow |\psi\rangle_{AB}$ is always possible,
where $|\Psi\rangle_{AB}$ is any $d\times d$ maximally entangled
state. This is obvious, as the Schmidt vector of $|\Psi\rangle_{AB}$
is $(\frac{1}{d},\frac{1}{d},\cdots,\frac{1}{d})$. Also, for any
pure bipartite product state $|\Upsilon \rangle_{AB}$ we have,
$|\psi\rangle_{AB}\rightarrow |\Upsilon \rangle_{AB}$, as the
Schmidt vector corresponding to any pure product state is
$(1,0,\cdots,0)$.

\section{Incomparable pair of states}

If the above criterion for transforming
$|\Psi\rangle_{AB}\rightarrow| \Phi\rangle_{AB}$ with certainty
under LOCC, is not satisfied, then that pair of states is usually
denoted by, $|\Psi \rangle_{AB} \not \rightarrow |\Phi
\rangle_{AB}$. Though, in such cases, it may happen that
$|\Phi\rangle_{AB}\rightarrow |\Psi\rangle_{AB}$ under LOCC. If for
some pair of pure bipartite states $(~|\Psi\rangle,~|\Phi\rangle)$
shared between, say, Alice and Bob, both
$|\Psi\rangle\not\rightarrow |\Phi\rangle$ and
$|\Phi\rangle\not\rightarrow |\Psi\rangle$ happen together, then we
denote it by, $|\Psi\rangle\not\leftrightarrow |\Phi\rangle$ and
describe $(|\Psi\rangle, |\Phi\rangle)$, as a pair of incomparable
states. Also, $E(|\Psi\rangle)\geq E(|\Phi\rangle)$ does not
guarantee $|\Psi\rangle\rightarrow |\Phi\rangle$ under LOCC with
certainty. One of the most interesting feature of the existence of
an incomparable pair of states is that we are unable to say which
state has a greater amount of entanglement than the other.

For every pair of pure states in $2\times 2$ system it is always be
the case that either $|\Psi\rangle\rightarrow |\Phi\rangle$ (when
$E(|\Psi\rangle)\geq E(|\Phi\rangle)$) or $|\Phi\rangle\rightarrow
|\Psi\rangle$ (when $E(|\Psi\rangle)\leq E(|\Phi\rangle)$). So, in
$2\times 2$, there is no pair of pure entangled states which are
incomparable as described above. Therefore, we look beyond the
$2\times2$ system of states.

In $3 \times 3$ system, there exists incomparable pair of states.
The incomparability of Schmidt rank $3$ states has some strange
character which we will discuss latter. Now, we explicitly mention
the criterion of incomparability for a pair of pure entangled states
$(|\Psi\rangle, |\Phi\rangle)$ of $m\times n$ system, where $\min \{
m,n \}=3$.

Suppose, the Schmidt vectors corresponding to the states
$|\Psi\rangle,~ |\Phi\rangle$ are $(a_1, a_2, a_3)$ and $(b_1, b_2,
b_3)$ respectively, where $a_1> a_2> a_3~,~b_1> b_2> b_3~,~a_1+ a_2+
a_3=1=b_1+ b_2+ b_3$. Then it follows from Nielsen's criterion that
$|\Psi\rangle, |\Phi\rangle$ are incomparable if and only if either
of the two relations,
\begin{equation}
\begin{array}{lcl}
a_1 > b_1 >  b_2 > a_2 > a_3 > b_3,\\
b_1 > a_1 >  a_2 > b_2
> b_3 > a_3 \label{incom}%
\end{array}
\end{equation} hold.

For higher dimensional states, incomparable pair of states appear in
several ways. However, it is interesting to note that for infinite
dimensional states incomparability is generic \cite{16,owari}.

\subsection{Catalysis}

Now, it is natural to ask that what we could do with an incomparable
pair of states, by means of LOCC? If we require a pure bipartite
state $|\phi\rangle$ and we have a finite but sufficient number of
copies of a pure bipartite state $|\psi\rangle$, then Vidal's
theorem \cite{Vidalpure} provided us the way of converting
probabilistically $|\psi\rangle$ to $|\phi\rangle$ by LOCC. This is
of course of no use for deterministic conversion. For this purpose,
Jonathan and Plenio~ \cite{jpcat}, found that sometimes collective
operation may be useful to convert deterministically $|\psi\rangle$
to $|\phi\rangle$ by LOCC, where $|\psi\rangle\not\rightarrow
|\phi\rangle$. They showed that if we assist the conversion by
another pure bipartite entangled state $|\chi\rangle$, say, a
catalyst, then the conversion $|\psi\rangle \otimes |\chi\rangle
\longrightarrow |\phi\rangle\otimes|\chi\rangle$ may be possible by
collective LOCC deterministically. Let us illustrate it by an
example in detail.

\emph{Example:} Consider two pure bipartite entangled states
$|\psi\rangle, |\phi\rangle $ of $4\times 4$ system, having the
following Schmidt vectors,
\begin{equation}
\begin{array}{lcl}
\lambda_{|\psi\rangle} &=& (.4, .4, .1, .1)\\
\lambda_{|\phi\rangle} &=& (.5, .25, .25, 0)\label{cat}%
\end{array}
\end{equation}This two Schmidt vectors violates Nielsen's criteria
for transforming the state $|\psi\rangle$ to $|\phi\rangle$ by LOCC
with certainty. Calculating entanglement of the two states we found
that $E(|\psi\rangle)>E(|\phi\rangle)$. Thus there is no restriction
for local conversion of $|\psi\rangle$ to $|\phi\rangle$ from the
constraint of non-increase of entanglement by local manipulation.
Therefore, one may ask for a scheme to implement the task of
transforming the pure state shared among two parties by allowing
some extra entanglement shared between them without exchange of
quantum communication through some quantum channel, i.e., whether
there exists any catalyst or not? Suppose, two parties share a pure
bipartite state $|\chi\rangle$, having Schmidt vector
$\lambda_{|\chi\rangle} = (.6, .4)$. In this new situation, the
local system of each party has two parts corresponding to the two
entangled states $|\psi\rangle$ and $|\chi\rangle$. The Schmidt
vectors corresponding to the initial and final joint states
$|\psi\rangle \otimes |\chi\rangle$ and
$|\phi\rangle\otimes|\chi\rangle$ are respectively,
$\lambda_{|\psi\rangle \otimes |\chi\rangle} = (.24, .24, .16, .16,
.06, .06, .04, .04)$ and $\lambda_{|\phi\rangle\otimes|\chi\rangle}
= (.3, .2, .15, .15, .1, .1, 0, 0)$. Then, it is easy to check from
Nielsen's criteria that by performing LOCC collectively on the joint
entangled system it becomes possible to transform $|\psi\rangle
\otimes |\chi\rangle \longrightarrow
|\phi\rangle\otimes|\chi\rangle$ with certainty. Here, it is
interesting to note that the extra amount of entanglement shared
between the parties, preserved by the process. Also, the entangled
state $|\chi\rangle$ is not used as a channel for transforming
quantum information.

But what type of pairs are really catalyzable? It is really hard to
categorize. Consider two pure bipartite states $|\psi\rangle,
|\phi\rangle$ of Schmidt rank $d$ with Schmidt vectors,
\begin{equation}
\begin{array}{lcl}
\lambda_{|\psi\rangle} &=& (\alpha_1, \alpha_2,\cdots,
\alpha_d)\\
\lambda_{|\phi\rangle} &=& (\beta_1, \beta_2,\cdots, \beta_d)
\end{array}
\end{equation}
respectively. Let $|\chi\rangle$ be a pure bipartite state acting as
a catalyst for the local conversion of $|\psi\rangle$ to
$|\phi\rangle$ with certainty.

Jonathan and Plenio~\cite{jpcat} showed that, if the form of Schmidt
vector for the catalytic state $|\chi\rangle$ is assumed to be
$\lambda_{|\chi\rangle} = (\gamma_1, \gamma_2, \cdots, \gamma_d)$,
then from Nielsen's criteria, the first and last conditions for
$|\psi\rangle \otimes |\chi\rangle \longrightarrow
|\phi\rangle\otimes|\chi\rangle$ by LOCC with certainty are
$\alpha_1 \gamma_1 \leq \beta_1 \gamma_1$ and $\alpha_d \gamma_d
\geq \beta_d \gamma_d$. And it is equivalent with the conditions,
\begin{equation}
\begin{array}{lcl}
&\alpha_1 \leq \beta_1& ~\verb"and"~ \ \ \ \alpha_d \geq \beta_d
\end{array}
\end{equation}
This is of course a necessary condition, but not sufficient, for
catalytic transformation. Thus if it is violated for some pair of
states $(|\psi\rangle, |\phi\rangle)$, then there does not exist any
pure bipartite entangled state, that may be used as catalyst. Also,
if $|\psi\rangle\rightarrow |\phi\rangle$ is possible by catalysis,
then $E(|\psi\rangle)\geq E(|\phi\rangle)$. For $3\times3$ system of
states, violation of Nielsen's criteria for the pair
$(|\psi\rangle,~|\phi\rangle)$ implies either both the conditions
$\alpha_1 > \beta_1$ and $\alpha_3 > \beta_3$ or the conditions
$\beta_1 > \alpha_1$ and $\beta_3 > \alpha_3$ must hold
simultaneously. This implies the violation of necessary condition
for catalysis. So, for any incomparable pair of states of $3\times
3$ system, the transformation by LOCC with certainty could not be
possible by any catalytic state. The existence of catalytic state is
first observed in $4\times4$ incomparable pairs and only in this
level a necessary and sufficient condition for the existence of $2
\times 2$ system of catalytic state is found until now~\cite{1a}.
Therefore, except of some numerical evidences, it is really hard to
determine whether an incomparable pair is catalyzable or not.
Investigation in this direction is going on by several
groups~\cite{trump,cat1,VidalNielsen,Vidal4}.

\subsection{Multi-copy transformation}

Another interesting feature of local conversion of non-local states
shown by Bandyopadhyay\emph{ et.al.} \cite{5}. Sometimes if we
increase the number of copies of the states, then deterministic
conversion of incomparable states under LOCC may be possible. There
exists some pair of bipartite entangled states $(|\psi\rangle,
|\phi\rangle)$ such that, $|\psi\rangle\not\rightarrow |\phi\rangle$
where $E(|\psi\rangle)>E(|\phi\rangle)$, but $|\psi\rangle^{\otimes
k}\rightarrow {|\phi\rangle}^ {\otimes k}$ is possible for some
positive integer k. This phenomena is called multi-copy
transformation. A sufficient condition for an incomparable pair to
remain incomparable even if we increase the number of copies as
large as possible is that, either $\alpha_1 < \beta_1$ and $\alpha_d
< \beta_d$, or, $\alpha_1 > \beta_1$ and $\alpha_d > \beta_d$ must
hold simultaneously \cite{4,3,5}. Existence of this kind of
incomparable pair of states are known as strongly incomparable.

Let us reconsider the example in catalytic transformation. It is
shown by Duan \emph{et.al.} \cite{CatMulticopy} that for the incomparable pair of
states of $4\times 4$ system represented by,
\begin{equation}
\begin{array}{lcl}
\lambda_{|\psi\rangle} &=& (.4, .4, .1, .1),\\
\lambda_{|\phi\rangle} &=& (.5, .25, .25, 0)
\end{array}
\end{equation}
the conversion is possible if three copies of the initial state are
available to produce the three copies of the final state under
deterministic LOCC, i.e.,
\begin{equation}
\begin{array}{lcl}
\mid\psi\rangle^{\otimes 3}\rightarrow{\mid\phi\rangle}^{\otimes 3}
\end{array}
\end{equation} is possible under deterministic LOCC.
It must be noted here that the condition for strong incomparability
implies a violation of the necessary condition for catalytic
transformation. Thus when multi-copy transformation is not possible
for a pair of pure bipartite states, then so also the catalytic
transformation. Later it has been shown that catalytic
transformation is asymptotically equivalent with multi-copy
transformation \cite{CatMulticopy}. This equivalency is established
in two steps. Firstly, if for a pair of incomparable states
$(\mid\psi\rangle, \mid\phi\rangle)$, multi-copy transformation is
possible, then there exists some finite dimensional bipartite
entangled state, which may assist this transformation catalytically.
Conversely, if there exist some catalytic state $|\chi\rangle$, for
which $|\psi\rangle \otimes |\chi\rangle \longrightarrow
|\phi\rangle\otimes|\chi\rangle$ by LOCC, then
$\mid\psi\rangle^{\otimes k}\rightarrow {\mid\phi\rangle}^ {\otimes
k}$ is possible, if $k\rightarrow \infty$.

\subsection{Mutual catalysis}

Now, another interesting process is provided by Feng \emph{et.al.},
and other groups \cite{mutualcat,4,3,40}, known as mutual catalysis.
The basic objective in this process is: given two pairs of
incomparable states, say, $|\psi_1\rangle\not\rightarrow
|\phi_1\rangle, |\psi_2\rangle\not\rightarrow |\phi_2\rangle,$
whether $|\psi_1\rangle \otimes |\psi_2\rangle
\longrightarrow|\phi_1\rangle\otimes|\phi_2\rangle$ is possible
under LOCC with certainty or not. Emphasis is given on the special
kind of mutual catalysis (in~\cite{4}, it is defined as super
catalysis), where in the conversion process, we recover not only the
entanglement assisted in the process but more than that, i.e., for
some incomparable pair $\mid\psi\rangle\not\rightarrow
\mid\phi\rangle$ there exists $( \mid\chi\rangle, \mid\eta\rangle)$
with $E(|\eta\rangle)\geq E(|\chi\rangle)$ and
$\mid\eta\rangle\rightarrow \mid\chi\rangle$ such that by collective
local operation $|\psi\rangle \otimes |\chi\rangle
\longrightarrow|\phi\rangle\otimes|\eta\rangle$ is possible
deterministically. It is interesting that the necessary condition
for the existence of such special kind of mutual catalytic pair
$(\mid\chi\rangle, \mid\eta\rangle)$, is the same as that for
catalyst. Hence this type of mutual catalysis is not possible for
$3\times3$ system of incomparability. It is shown, not analytically,
but by some numerical examples, that there are system of states for
which catalyst does not exist but mutual catalysis works. Trivially
it is always possible that $|\psi\rangle \otimes |\phi\rangle
\longrightarrow|\phi\rangle\otimes|\psi\rangle$ under LOCC with
certainty. So, existence of mutual catalytic state is always
possible. But it is not of use, as our target state $|\phi\rangle$
is not in our hand and in the process of trivial mutual catalysis we
have to use it.

We have investigated possibility of deterministic conversion by LOCC
on the ground of mutual catalysis. We have tried to give analytic
method of searching out some optimal process of finding mutual
catalytic states that may be suitable from some physical viewpoint.
Our work emphasis on the idea of using entanglement of a composite
system as a resource and also on conservation and proper
manipulation of this valuable resource.

\subsection{Probabilistic Conversion by Bound entanglement}

Here comes the question of proper use of entanglement to reach the
target state. By the use of entanglement we mean, forget about
recovering the entanglement used in the process, but to concentrate
on converting the input state to the desired one. Use of bound
entanglement solves the problem of conversion to the desired one
probabilistically. Recently, Ishizaka~\cite{Ishizaka} showed that
using PPT-bound entanglement local conversion of any two pure
entangled states (no matter what the Schmidt rank of the states are)
is always possible, at least with some probability. So, the problem
of finding target state of our interest now remains for the case of
deterministic local conversion only.

\subsection{Deterministically unresolved class}

We have observed that the necessary condition for the existence of
super-catalytic pair $(\mid\chi\rangle, \mid\eta\rangle)$ is the
same as that for catalyst. All pure incomparable states in
$3\times3$ are strongly incomparable. So, all the processes of
catalysis, mutual catalysis with some recovery and multi-copy
transformation will fail for all $3\times3$ pure incomparable pair
of states and also for all pure strongly incomparable bipartite
classes. Therefore one may ask, is it not possible to get the target
state under collective LOCC with certainty for such incomparable
pairs? The answer is yes. Two possible ways for resolving
incomparability of such classes are discussed here \cite{catlys}. In
both the cases we use free unrecoverable entanglement for
deterministic local transformation.

\section{Assistance by entanglement}

Suppose we have a pair of incomparable pure bipartite states
$|\psi\rangle_{AB}, |\phi\rangle_{AB}$ of $d\times d$ system, where
the source state $|\psi\rangle_{AB}$ and the target state
$|\phi\rangle_{AB}$ are taken in the following form:
\begin{equation}
\begin{array}{lcl}
|\psi\rangle_{AB} = \sum_{i=1}^{d} \sqrt{a_{i}}|i\rangle_{A}
|i\rangle_{B},~
~a_{i} \geq a_{i+1} \geq 0,~\forall~i=1,2,\cdots,(d-1), ~ ~\sum_{i=1}^{d} a_{i}= 1\\
|\phi\rangle_{AB} = \sum_{i=1}^{d} \sqrt{b_{i}}| i\rangle_{A}
|i\rangle_{B},~ ~b_{i} \geq b_{i+1} \geq 0, \ \forall~
i=1,2,\cdots,(d-1), ~ ~\sum_{i=1}^{d} b_{i} = 1
\end{array}
\end{equation}
Now, consider the $(d-1)\times (d-1)$ maximally entangled state
$|\Psi_{max}^{d-1}\rangle_{AB} =
\frac{1}{\sqrt{d-1}}\sum_{i=d+1}^{2d-1}| i\rangle_{A}
|i\rangle_{B},$ and the product state $|P\rangle_{AB} =
|d+1\rangle_{A}~|d+1\rangle_{B}.$ We want to make possible the joint
transformation $|\psi\rangle_{AB} \otimes
|\Psi_{max}^{d-1}\rangle_{AB} \rightarrow |\phi\rangle_{AB} \otimes
|P\rangle_{AB}$ under LOCC with certainty. If the Schmidt
coefficients of $|\psi\rangle_{AB} \otimes
|\Psi_{max}^{d-1}\rangle_{AB}$ is arranged in a decreasing order,
then all the first $d$ coefficients are equal to $\frac{a_1}{d-1}$.
Again the Schmidt vector of $|\phi\rangle_{AB} \otimes
|P\rangle_{AB}$ is $\{b_1,b_2,\cdots,b_d,0,0,\cdots \}$. Thus by
Nielsen's criteria, one has to check the first $d$ conditions. We
may write those conditions in the simplified form as,
\begin{equation}
\frac{k a_1}{d-1}\leq\sum_{i=1}^{k}b_i~;~~~ \forall \
k=1,2,\ldots,d-1.
\end{equation}
To prove this, we first state a theorem which shows an intricate
relation holds between first and last Schmidt coefficients of any
pair of incomparable states.

{\bf Theorem.} For any pair of  incomparable states
$|\psi\rangle_{AB} \not\leftrightarrow|\phi\rangle_{AB}$ in $d\times
d$ system, where $|\psi\rangle_{AB} = \sum_{i=1}^{d} \sqrt{a_{i}} \
|i\rangle_{A} |i\rangle_{B}$ and $|\phi\rangle_{AB} = \sum_{i=1}^{d}
\sqrt{b_{i}} \ | i\rangle_{A} |i\rangle_{B},$ with $a_{i} \geq
a_{i+1} \geq 0,~\forall ~~i, \sum_{i=1}^{d} a_{i} = 1$ and $b_{i}
\geq b_{i+1} \geq 0,~\forall ~~i,~ \sum_{i=1}^{d} b_{i} = 1,$ the
following always holds:
\begin{equation}
a_1+b_d<1, \ b_1+a_d<1.
\end{equation}
Proof: From Nielsen's criteria, $|\psi\rangle_{AB}
\not\leftrightarrow|\phi\rangle_{AB}$
implies either of the following two cases must be hold.\\
Case-1. When, $a_1\leq b_1$ and $\sum_{i=1}^{k} a_{i} >
\sum_{i=1}^{k} b_{i}$, for some $k = 2,3,\cdots, d-1$. Then\\
\begin{equation}
\begin{array}{lcl}
a_1+b_d  \leq   b_1+ b_{d}< 1
\end{array}
\end{equation}and
\begin{equation}
\begin{array}{lcl}
b_1+a_d  &=&  b_1+ 1-\sum_{i=1}^{d-1} a_{i} \\ &=&  b_1+
1-\sum_{i=1}^{k} a_{i} -\sum_{i=k+1}^{d-1} a_{i}  \\ &<&  b_1+
1-\sum_{i=1}^{k} b_{i} -\sum_{i=k+1}^{d-1} a_{i}  \\ &=&
1-\sum_{i=2}^{k} b_{i} -\sum_{i=k+1}^{d-1} a_{i}  \\ &<& 1
\end{array}
\end{equation}Case-2. When, $a_1\geq b_1$ and $\sum_{i=1}^{k} a_{i} <
\sum_{i=1}^{k} b_{i}$, for some $k = 2,3,\cdots, d-1$; the proof is similar to the previous case.

This proves the theorem.

So, the incomparability condition itself implies that
$|\psi\rangle_{AB} \otimes |\Psi_{max}^{d-1}\rangle_{AB} \rightarrow
|\phi\rangle_{AB} \otimes | P\rangle_{AB}$ is possible under LOCC
with certainty. Therefore, for any pair of incomparable states with
a given Schmidt rank the maximally entangled state of the next lower
rank is sufficient to assist the
joint transformation under LOCC.\\

Next, we show that instead of using lower rank maximally entangled
state, the conversion may be possible under LOCC if we use lower
rank non-maximally entangled states so that we need as minimum use
of the resource as possible. In explicit form the minimum amount of
entanglement that is required for the local transformation of any
$3\times 3$ incomparable pair is

$\min_c E(|\chi\rangle)= - c\log_{2} c -(1-c)\log_{2} (1-c)$ where
$|\chi\rangle= \sqrt{c}|00\rangle+\sqrt{(1-c)}|11\rangle$. Let the
Schmidt vectors of the pure bipartite states $|\psi\rangle$ and
$|\phi\rangle$ are respectively,
$\lambda_{|\psi\rangle}=(a_{1},a_{2},a_{3})$ and
$\lambda_{|\phi\rangle}=(b_{1},b_{2},b_{3})$. The joint
transformation $|\psi\rangle\otimes|\chi\rangle\rightarrow
|\phi\rangle\otimes|P\rangle$ will be possible if the following
conditions,
\begin{equation}
\begin{array}{lcl}
a_1 c&\leq & b_1\\
a_1 c+ \max\{a_2 c, ~a_1 (1-c)\}&\leq & b_{1}+b_{2}
\end{array}
\end{equation}
hold(all other conditions are obvious). Now, $a_2 c \gtrless ~a_1
(1-c)$ if and only if $c \gtrless\frac{a_1}{a_1+a_2}$. Here we want
to choose the maximum possible value of $c$ for which the required
transformation will occur. Thus, if we assume $c >
\frac{a_1}{a_1+a_2}$, i.e., $\max\{a_2 c, ~a_1 (1-c)\} = a_2 c$.
Then, the second condition becomes $c \leq \frac{b_{1} + b_{2}}{
a_{1}+ a_{2}} < 1$. Again, if $\max\{a_2 c, ~a_1 (1-c)\}=a_1 (1-c)$,
then the second condition is obvious. So, we only need to check the
possibility of the transformation when $c > \frac{a_1}{a_1+a_2}$. We
discuss separately the cases of incomparability of the two states,
as follows:

{\bf{Type-1:}} When, $a_{1} < b_{1}, a_{1}+ a_{2} > b_{1} + b_{2},$
for all values of $c\in [\frac{1}{2},1)$, we have $a_1 c <  b_1$.
So, we have to select the state $|\chi\rangle$ in such a way that
the second condition $c \leq \frac{b_{1} + b_{2}}{ a_{1}+ a_{2}} <
1$, will be satisfied. Thus in this case the only condition to make
possible the joint transformation
$|\psi\rangle\otimes|\chi\rangle\rightarrow
|\phi\rangle\otimes|P\rangle$ will be, $$c \leq \frac{b_{1} +
b_{2}}{ a_{1}+ a_{2}} < 1$$ As the maximum possible value of $c$ is,
$c_0 = \frac{b_{1} + b_{2}}{a_{1}+ a_{2}} < 1$, so the minimum
amount of entanglement required in this process is,
$E_0=-c_0\log_{2} c_0-(1-c_0)\log_{2}(1-c_0)$.

{\bf{Type-2:}} If, $a_{1} > b_{1}, a_{1}+ a_{2} < b_{1} + b_{2},$
the second condition is obvious, as for all $c\in[\frac{1}{2},1)$, $
c (a_{1}+ a_{2}) < a_{1}+ a_{2} < b_{1} + b_{2}.$ The condition for
the above transformation to be possible is given by $c\leq
\frac{b_{1}}{ a_{1}} < 1$. The minimum amount of entanglement
required in the process is, $E=E_0$ corresponding to $c=c_0=
\frac{b_{1}}{a_{1}}.$

We conclude this section with an interesting result that in any
$d\times d$ system, from a non-maximally pure entangled state
$|\psi^d\rangle_{AB}$ of $d\times d$ system $(d \geq 3)$, we are
able to reach the maximally entangled state
$|\phi\rangle_{AB}=|\Psi_{max}^d\rangle_{AB}$ of the same $d\times
d$ system by the use of the next lower rank maximally entangled
state $|\Psi_{max}^{d-1}\rangle_{AB}$ through collective local
operation with certainty.

{\bf Corollary-1.} In any $d\times d$ system for $(d \geq 3)$, a
non-maximally pure entangled state $|\psi^d\rangle_{AB}~=~
\sum_{i=1}^{d}{\sqrt{a_{i}}|i\rangle_{A} |i\rangle_{B}}$ can be
transformed to the maximally entangled state
$|\Psi_{max}^d\rangle_{AB}~=~
\frac{1}{\sqrt{d}}\sum_{i=1}^{d}|i\rangle_{A} |i\rangle_{B}$ of same
dimension by using the next lower rank maximally entangled state
$|\Psi_{max}^{d-1}\rangle_{AB}$ through collective local operations
with certainty. i.e.,~~ $|\psi^d\rangle_{AB} \otimes
|\Psi_{max}^{d-1}\rangle_{AB} \rightarrow
|\Psi_{max}^{d}\rangle_{AB} \otimes | P\rangle_{AB},$ where
$|P\rangle_{AB}$ is a product state, is possible under LOCC with
certainty, if the largest Schmidt coefficient, $a_1$ of
$|\psi^d\rangle_{AB}$ satisfies the relation,
$a_1\leq\frac{d-1}{d}.$

We have from Nielsen criteria,
\begin{equation}
\begin{array}{lcl}
\frac{a_1}{d-1}&\leq & \frac{1}{d}\\
2.~\frac{a_1}{d-1}&\leq & 2.~\frac{1}{d}\\
&\vdots &\\
&\vdots &\\
(d-1)~\frac{a_1}{d-1}&\leq & (d-1)~\frac{1}{d}\\
a_1+\frac{a_2}{d-1}&\leq & 1\\
&\vdots &\\
&\vdots &
\end{array}
\end{equation}

We observe that each of the first (d-1) conditions are equivalent
with the relation, $a_1\leq\frac{d-1}{d}$ and all the next
conditions are obvious, as the right-hand is always 1.

Therefore, the result of Corollary-1, follows directly from the
theorem above. In fact, instead of using the state
$|\psi^d\rangle_{AB}$, the above transformation is possible by a
$2\times 2$ state only.

{\bf Corollary-2.} The transformation $|\psi\rangle_{AB} \otimes
|\Psi_{max}^{d-1}\rangle_{AB} \rightarrow$ $
|\Psi_{max}^{d}\rangle_{AB} \otimes | P\rangle_{AB},$ is possible
under LOCC with certainty, if we take $|\psi\rangle_{AB}$, as a
$2\times 2$ state with Schmidt coefficients $(\frac{d-1}{d},
\frac{1}{d}).$

By a recursive way, Corollary-2 immediately suggests that it is
possible to achieve a maximally entangled state of any arbitrary
Schmidt rank $d, \ d \geq 3$ by using a finite number (to be exact,
$d-1$) of suitably chosen $2\times 2$ states only. This will be
presented in the next corollary, given below.

{\bf Corollary-3.} The transformation, $|\psi_1\rangle_{AB} \otimes
|\psi_2\rangle_{AB} \otimes \cdots|\psi_{d-1}\rangle_{AB}
\rightarrow |\Psi_{max}^{d}\rangle_{AB} \otimes | P\rangle_{AB},$ is
possible under LOCC with certainty, where $\{|\psi_i\rangle_{AB}, \
\forall i=1,2,\cdots,d-1\}$ are $2\times 2$ states with Schmidt
coefficients $(\frac{d-i}{d-i+1}, \frac{1}{d-i+1})$, respectively.

\section{Mutual co-operation}
In this section our main goal is to provide an auxiliary
incomparable pair so that the collective operation enables us to
find the desired states; i.e., given a pair  $|\psi\rangle
\not\leftrightarrow |\phi\rangle$, we want to find an auxiliary pair
$|\chi\rangle \not\leftrightarrow |\eta\rangle$ such that
$|\psi\rangle \otimes |\chi\rangle \rightarrow |\phi\rangle \otimes
|\eta\rangle,$ is possible under LOCC with certainty. There are
several ways to find nontrivial $(|\chi\rangle,|\eta\rangle)$. We
explicitly provide the form of the auxiliary pair for all possible
incomparable pair $(|\psi\rangle,|\phi\rangle)$ in $3\times 3$
system. One of the interesting feature of such incomparable pairs is
that we are unable to say that which state has a greater amount of
entanglement than the other. So in this way we may resolve the
incomparability of $(|\psi\rangle,|\phi\rangle)$ with
$E(|\psi\rangle)<E(|\phi\rangle)$ by mutual co-operation which
obviously claims that $E(|\chi\rangle)>E(|\eta\rangle)$. To remove
the trivial case of joint transformation through the existence of
single crosswise transformations $|\psi\rangle \rightarrow
|\eta\rangle$ or $|\chi\rangle \rightarrow |\phi\rangle$, we further
impose the restriction that $|\psi\rangle \not \rightarrow
|\eta\rangle$ and $|\chi\rangle \not\rightarrow |\phi\rangle$.

We have also studied analytically that in $3\times 3$ system, from
two copies of a pure entangled state we are able to find two
different pure entangled states, both of which are incomparable with
the source state.

We first provide some examples that will show such features and then
in two subsections we shall give analytical results for $3\times 3$
system of incomparable states. We explicitly provide the form of the
auxiliary pair for all possible incomparable pairs
$(|\psi\rangle,|\phi\rangle)$ in $3\times 3$ system.

\noindent{\em \textbf{Example 1}}.-- Consider a pair of pure
entangled states of the form

$$|\psi\rangle = \sqrt{0.4}| 00 \rangle +\sqrt{0.4}| 11
\rangle + \sqrt{0.2}| 22 \rangle,$$

$$|\phi\rangle = \sqrt{0.48}| 00 \rangle +\sqrt{0.26}| 11
\rangle + \sqrt{0.26}| 22 \rangle,$$

$$|\chi\rangle = \sqrt{0.49}| 33 \rangle +\sqrt{0.255}| 44
\rangle + \sqrt{0.255}| 55 \rangle,$$

$$|\eta\rangle = \sqrt{0.41}| 33 \rangle +\sqrt{0.41}| 44
\rangle + \sqrt{0.18}| 55 \rangle.$$

It is easy to check that $|\psi\rangle \not\leftrightarrow
|\phi\rangle$ and $|\chi\rangle \not\leftrightarrow |\eta\rangle$;
whereas, $E(|\psi\rangle)\approx 1.5219
> E(|\phi\rangle) \approx 1.5188,  E(|\chi\rangle) \approx 1.5097 >
E(|\eta\rangle) \approx 1.5001$; and if we allow collective
operations locally on the joint system, then the transformation
$|\psi\rangle \otimes|\chi\rangle\rightarrow|\phi\rangle
\otimes|\eta\rangle$ is possible with certainty, i.e., the two
incomparable pairs, after mutual co-operation, are able to make the
joint transformation possible.

If we review the whole process critically, then we find something
more than we have discussed earlier. Here we see that the
comparability of the joint operation actually evolved through the
co-operation with the comparable class of states, i.e., the four
states chosen are related in such a way that $|\psi\rangle
\rightarrow |\eta\rangle$ and $|\phi\rangle \rightarrow
|\chi\rangle$. So here we reduce the incomparability of two states
by choosing some class of states comparable with them. It is obvious
that such a pair of states always exist for any incomparable pair,
i.e., incomparable pairs can always be made to compare by collective
LOCC. Without going into the details of the proof, we just mention
here that this approach resolves the incomparability of the
$3\times3$ states in one way. However, someone may think that only
by the help of comparable classes we may able to reach the desired
states which are incomparable initially. Obviously, the answer is in
the negative. The next example is given in support of this last
remark.

\noindent{\em \textbf{Example 2}}.-- Consider two pairs of pure
entangled states $(|\psi\rangle,|\phi\rangle)$ and
$(|\chi\rangle,|\eta\rangle)$ of the form

$$|\psi\rangle = \sqrt{0.41}| 00 \rangle +\sqrt{0.38}| 11
\rangle + \sqrt{0.21}| 22 \rangle,$$

$$|\phi\rangle = \sqrt{0.4}| 00 \rangle +\sqrt{0.4}| 11
\rangle + \sqrt{0.2}| 22 \rangle,$$

$$|\chi\rangle = \sqrt{0.45}| 33 \rangle +\sqrt{0.34}| 44
\rangle + \sqrt{0.21}| 55 \rangle,$$

$$|\eta\rangle = \sqrt{0.48}| 33 \rangle +\sqrt{0.309}| 44
\rangle + \sqrt{0.211}| 55 \rangle.$$

It is quite surprising to see that not only $|\psi\rangle
\not\leftrightarrow |\phi\rangle$ and $|\chi\rangle
\not\leftrightarrow |\eta\rangle$ but also $|\psi\rangle
\not\leftrightarrow |\eta\rangle$, $|\chi\rangle \not\leftrightarrow
|\phi\rangle$. Beside this we also get the extra facility to prepare
$|\chi\rangle$ from $|\psi\rangle$ as $|\psi\rangle \rightarrow
|\chi\rangle$. From the informative point of view the picture is
although, $E(|\psi\rangle)\approx 1.5307
> E(|\phi\rangle) \approx 1.5219,$ and  $E(|\chi\rangle) \approx
1.5204 > E(|\eta\rangle) \approx 1.50544,$ but still independently
we can not convert  $|\psi\rangle$ to either one of $|\phi\rangle$
or $|\eta\rangle$ and also $|\chi\rangle$ to either one of
$|\phi\rangle$ or $|\eta\rangle$ with certainty under LOCC.
Therefore, although the resource states have greater information
content in terms of pure state entanglement, the individual pairs
aren't convertible. But considering them together, we are able to
break the overall incomparability. Thus, apart from the usual way in
mutual catalysis, i.e., achieve transformation of the first pair and
recover as much as possible the amount of entanglement from second
pair, here, we have tried to overcome the incomparability of both
the two pairs of incomparable states together, by collective LOCC
with certainty.

To give rise of the fact that mutual co-operation also exists in
other dimensions, we are now providing other two sets of
incomparable pairs in $4 \times 4$ system which are strongly
incomparable so that deterministic local conversions are not
possible by assistance of catalytic states and $2 \times 2$ mutual
catalytic states, but by mutual co-operation the transformation is
possible.

\noindent{\em \textbf{Example 3}}.-- Consider two pairs of pure
entangled states $(|\psi\rangle,|\phi\rangle)$ and
$(|\chi\rangle,|\eta\rangle)$ of the form:

$$|\psi\rangle = \sqrt{0.4}| 00 \rangle +\sqrt{0.3}| 11
\rangle + \sqrt{0.2}| 22 \rangle + \sqrt{0.1}| 33 \rangle,$$

$$|\phi\rangle = \sqrt{0.45}| 00 \rangle +\sqrt{0.29}| 11
\rangle + \sqrt{0.14}| 22 \rangle + \sqrt{0.12}| 33\rangle,$$

$$|\chi\rangle = \sqrt{0.5}| 44 \rangle +\sqrt{0.25}| 55
\rangle + \sqrt{0.2}| 66 \rangle + \sqrt{0.05}| 77 \rangle,$$

$$|\eta\rangle = \sqrt{0.48}| 44 \rangle +\sqrt{0.36}| 55
\rangle + \sqrt{0.12}| 66 \rangle + \sqrt{0.04}| 77\rangle.$$

It is easy to check that $|\psi\rangle \not\leftrightarrow
|\phi\rangle$ and $|\chi\rangle \not\leftrightarrow |\eta\rangle,$
and $E(|\psi\rangle)\approx 1.846
> E(|\phi\rangle) \approx 1.800,$  $E(|\chi\rangle) \approx
1.680 > E(|\eta\rangle) \approx 1.592.$  However, one may check
$|\psi\rangle \otimes|\chi\rangle \longrightarrow \ |\phi\rangle
\otimes|\eta\rangle,$ is possible under LOCC. From crosschecking, we
find that  $|\psi\rangle \rightarrow |\eta\rangle$ but $|\chi\rangle
\not\rightarrow |\phi\rangle$, however, $|\phi\rangle \rightarrow
|\chi\rangle$.

\noindent{\em \textbf{Example 4}}.-- Consider another two pairs of
pure entangled states $(|\psi\rangle,|\phi\rangle)$ and
$(|\chi\rangle,|\eta\rangle)$ of the form:

$$|\psi\rangle = \sqrt{0.4}| 00 \rangle +\sqrt{0.3}| 11
\rangle + \sqrt{0.2}| 22 \rangle + \sqrt{0.1}| 33 \rangle,$$

$$|\phi\rangle = \sqrt{0.45}| 00 \rangle +\sqrt{0.29}| 11
\rangle + \sqrt{0.14}| 22 \rangle + \sqrt{0.12}| 33\rangle,$$

$$|\chi\rangle = \sqrt{0.5}| 44 \rangle +\sqrt{0.23}| 55
\rangle + \sqrt{0.22}| 66 \rangle + \sqrt{0.05}| 77\rangle,$$

$$|\eta\rangle = \sqrt{0.48}| 44 \rangle +\sqrt{0.36}| 55
\rangle + \sqrt{0.12}| 66 \rangle + \sqrt{0.04}| 77\rangle.$$

Here also it is easy to verify that $|\psi\rangle
\not\leftrightarrow |\phi\rangle$, $|\psi\rangle \not\leftrightarrow
|\eta\rangle$, $|\chi\rangle \not\leftrightarrow |\phi\rangle$ and
$|\chi\rangle \not\leftrightarrow |\eta\rangle.$ But surprisingly
$|\psi\rangle \rightarrow$ $|\chi\rangle$. Now it is very
interesting that we can prepare the state of co-operation from the
state in our hand. The relations between the entanglement of those
states are, $ E(|\psi\rangle) \approx 1.846 > E(|\phi\rangle)
\approx 1.800,$ and $E(|\chi\rangle) \approx 1.684 >$
$E(|\eta\rangle) \approx 1.592,$  and $|\psi\rangle \otimes
|\chi\rangle \longrightarrow |\phi\rangle \otimes |\eta\rangle$, is
possible under LOCC with certainty. All the examples we have
provided here, are non-trivial one. Next, we will show some
analytical results for $3\times 3$ system of incomparable states.

\subsection{Conversion by an auxiliary incomparable pair}

Now we concentrate only to the case of incomparable pairs in $3
\times 3$ system of states. We will show that for every pair of
incomparable pure entangled states $(|\psi_1\rangle,
|\phi_1\rangle)$, there is always a pair of incomparable pure
entangled states $(|\psi_2\rangle, |\phi_2\rangle)$ such that
$|\psi_1\rangle \ \otimes |\psi_2\rangle \longrightarrow \
|\phi_1\rangle \otimes |\phi_2\rangle,$ is possible under LOCC with
certainty. The main idea of this portion is, assuming
$|\psi_1\rangle$ as the source state and $|\phi_1\rangle$ as the
target state, we choose the nontrivial auxiliary incomparable pair
$(|\psi_2\rangle, |\phi_2\rangle)$ such that by collective LOCC the
joint transformation of both pairs is possible with certainty.

Consider, $|\psi_1\rangle \equiv (a_1, a_2, a_3), |\phi_1\rangle
\equiv (b_1, b_2, b_3 )$ where $a_1 \geq a_2 \geq a_3 \geq 0, a_1 +
a_2 + a_3 =1,$ $b_1 \geq b_2 \geq b_3 \geq 0, b_1 + b_2 + b_3 =1.$
There are two possible cases of incomparability that exist in this
dimension, which are discussed and treated differently below. Let us
first consider the most arbitrary form of the other pair of states
as: $|\psi_2\rangle \equiv (c_1, c_2, c_3)$ and $|\phi_2\rangle
\equiv (d_1, d_2, d_3 )$, then a necessary condition for the
existence of the joint transformation by LOCC with certainty will be
\begin{equation}
\begin{array}{lcl}
a_1 c_1 &<& b_1 d_1\\
a_3 c_3 &>& b_3 d_3
\end{array}
\end{equation}
Thus, in the case $ a_1 > b_1$, we must have $ c_1 < d_1$ and in
case of $ a_1 < b_1$, we must choose the other pair so that $ c_3 >
d_3$ (as, for the incomparability of the states $|\psi_1\rangle ,~
|\phi_1\rangle$, we have $a_3 < b_3$).

{\bf{Case-1:}}  $ a_1 > b_1, a_1+a_2 < b_1+b_2 .$ As described
above, we have the restriction $ c_1 < d_1$, then imposing the
condition that the pair of states $(|\psi_2\rangle, |\phi_2\rangle)$
are also incomparable in nature, we have $d_1 \geq c_1 \geq c_2 \geq
d_2 \geq d_3 \geq c_3$. Now, without disturbing our target pair, for
simplicity, we consider, $c_1 = c_2$ and $d_2 = d_3$. We rename the
Schmidt vectors of the second pair as $|\psi_2\rangle \equiv
(\beta_1, \beta_1,\beta_2), |\phi_2\rangle \equiv (\alpha_1,
\alpha_2, \alpha_2 )$ where $\beta_1 > \beta_2 >0, 2\beta_ 1
+\beta_2 =1, \alpha_1 > \alpha_2
> 0, \alpha_1 +2 \alpha_2 =1, \beta_1 < \alpha_1, 2\beta_1 >
\alpha_1+\alpha_2$. Therefore, with the choice of the state
$|\phi_2\rangle$ in such a manner that $b_3\alpha_1 > b_1\alpha_2,$
we arrange the Schmidt coefficients of the final joint state in the
decreasing order as:
\begin{equation}
|\phi_1\rangle \otimes |\phi_2\rangle \equiv \{b_1\alpha_1 ,
b_2\alpha_1 , b_3\alpha_1 , b_1\alpha_2 , b_1\alpha_2 , b_2\alpha_2,
b_2\alpha_2, b_3\alpha_2, b_3\alpha_2\}
\end{equation}

We consider separately the different subcases.\\

If $a_3\beta_1>a_1\beta_2$, then
\begin{equation}
|\psi_1\rangle \otimes |\psi_2\rangle \equiv \{a_1\beta_1 ,
a_1\beta_1 , a_2\beta_1 , a_2\beta_1 , a_3\beta_1 , a_3\beta_1 ,
a_1\beta_2 , a_2\beta_2 , a_3\beta_2\}
\end{equation}
Thus the joint transformation is possible by LOCC with certainty, if
the following conditions, using Nielsen's criteria, are satisfied.

\begin{equation}
\begin{array}{lcl}
a_1\beta_1 &<& b_1\alpha_1\\
2a_1\beta_1 &<& (b_1+b_2)\alpha_1\\
(2a_1+a_2)\beta_1 &<& \alpha_1\\
2(a_1+a_2)\beta_1 &<& \alpha_1+b_1\alpha_2\\
(2a_1+2a_2+a_3)\beta_1 &<& \alpha_1+2b_1\alpha_2\\
2\beta_1 &<& \alpha_1+(2b_1+b_2)\alpha_2\\
(a_2+a_3)\beta_2 &>& 2b_3\alpha_2\\
a_3\beta_2 &>& b_3\alpha_2
\end{array}
\end{equation}

Among the above conditions, the second one implies the first (as,
$2a_1\beta_1 < (b_1+b_2)\alpha_1 < 2b_1\alpha_1$) and the 8th
condition would imply the 7th (as, $(a_2+a_3)\beta_2 > 2a_3\beta_2 >
2b_3\alpha_2$). Again by assuming the validity of the second and
fourth conditions we found that the third condition is satisfied
automatically (adding both sides of the second the fourth condition
we have, $2(2a_1+a_2)\beta_1 < 2(b_1+b_2)\alpha_1 + b_3\alpha_1 +
b_1\alpha_2 < 2\alpha_1$). In the same way the fifth condition
follows from the fourth and the sixth conditions. Thus, we are left
with four conditions. We transform them into conditions given on
$\beta_2$, by using $2\beta_ 1 +\beta_2 =1$. We arrive at the final
condition of transformation as:
\begin{equation}
\beta_2>\max\{\frac{\alpha_2b_3}{a_3}, \alpha_2(b_2+2b_3),
\frac{\alpha_2(2-b_1)-a_3}{(1-a_3)}, 1-\frac{\alpha_1
(b_1+b_2)}{a_1}\}
\end{equation}when $a_3\beta_1>a_1\beta_2$. And if,

\begin{equation}
\max\{\frac{a_1}{a_2}, \frac{b_1}{b_3}\}<\frac{\alpha_1}{\alpha_2}
\end{equation}
and
\begin{equation}
\frac{a_3}{(2a_1+a_3)}>\beta_2>\max\{\frac{\alpha_2b_3}{a_3},
\alpha_2(b_2+2b_3), \frac{\alpha_2(2-b_1)-a_3}{(1-a_3)},
1-\frac{\alpha_1 (b_1+b_2)}{a_1}\}
\end{equation}
Under such a choice the required joint transformation is always
possible.

In the above process there may arise a similar condition like our
first example. For this type of choice we have always
$|\psi_1\rangle\rightarrow |\phi_2\rangle.$ Except this choice we
further require that those cross pairs $(|\psi_1\rangle,
|\phi_2\rangle)$ or $(|\psi_2\rangle, |\phi_1\rangle),$ remain
incomparable too. To fulfill this requirement the state
$|\psi_2\rangle$ is chosen slight differently, as $|\psi_2\rangle
\equiv (\beta_1, \beta_2,\beta_3),$ where $\beta_1 > \beta_2 >
\beta_3 >0, \beta_ 1 +\beta_2 +\beta_3=1,$ such that\\
$\alpha_1>a_1>\beta_1>b_1>b_2>\beta_2>a_2>\alpha_2>a_3>\beta_3>b_3$.

We also impose the extra conditions, $a_1\beta_3>\beta_1
a_3>b_1\alpha_2,$ $a_1\beta_1<b_1\alpha_1, a_3\beta_3
> b_3\alpha_2$ and $\{(\beta_1 a_3-a_2\beta_3)-(a_3-b_3)\}<
\min\{0, (\alpha_1 b_3-\alpha_2b_2), (a_2\beta_2-\alpha_2b_2) \}.$

Therefore, from Nielsen's criteria, $|\psi_1\rangle \ \otimes
|\psi_2\rangle \longrightarrow \ |\phi_1\rangle \otimes
|\phi_2\rangle,$ if
\begin{equation}
\begin{array}{lcl}
a_1\beta_1 &<& b_1\alpha_1\\
a_1(\beta_1+\beta_2) &<& (b_1+b_2)\alpha_1\\
a_1(\beta_1+\beta_2)+\max\{a_2\beta_1,a_3\beta_3\}

a_1^2+2a_1a_2 &<& (2b_1+b_2)\alpha_1\\
(a_1+a_2)^2 &<& 2(b_1+b_2)\alpha_1\\
(a_1+a_2)^2+a_1a_3 &<& (2b_1+2b_2+b_3)\alpha_1\\
(a_1+a_2)^2+2a_1a_3 &<& 2\alpha_1\\
(a_1+a_2)^2+(2a_1+a_2)a_3 &<& 2\alpha_1+b_1\alpha_3\\
a_3^2 &>& b_3\alpha_3
\end{array}
\end{equation}

After such a choice, the pair $(|\psi_2\rangle, |\phi_1\rangle)$
became incomparable except when $a_2=a_3.$ But, whenever we face the
case $b_1=b_2$ and $a_2=a_3$, then correspondingly we see that,
$|\psi_1\rangle\rightarrow |\phi_2\rangle$ and
$|\psi_2\rangle\rightarrow |\phi_1\rangle.$

{\bf{Case-2:}} $a_1<b_1, a_1+a_2>b_1+b_2. $ In this case we choose,
$|\psi_2\rangle \equiv (\beta_1, \beta_2,\beta_3), |\phi_2\rangle
\equiv (\alpha_1, \alpha_1, \alpha_2 )$ where $\beta_1 > \beta_2
>\beta_3 >0, \beta_ 1 +\beta_2 +\beta_3=1, \alpha_1 > \alpha_2 > 0,
2\alpha_1 + \alpha_2 =1, \beta_1 > \alpha_1, \beta_1 +\beta_2 >
2\alpha_1.$ Now there arises two different subcases, $a_1\gtrless
\frac{1}{2}$ which must be considered separately as in the case of
$a_1< \frac{1}{2}$ there is a possibility for $a_1=a_2$.

Firstly, when $a_1< \frac{1}{2},$ we choose the state
$(|\psi_2\rangle ,|\phi_2\rangle)$  in such a way that $\alpha_1 b_3
>a_1 \beta_3 >\beta_1 a_3$ and
\begin{equation}
\alpha_1>\max \{\frac{\beta_1 a_1}{b_1}, \frac{\beta_1 (a_1+a_2) a_1
\beta_2}{2b_1+b_2},\frac{ (1-\beta_3)(1-a_3)}{2(1-b_3)} \}
\end{equation}
Secondly, when $a_1 \geq \frac{1}{2},$ we choose the state
$(|\psi_2\rangle,$ $|\phi_2\rangle)$ in such a way that $\beta_1
=\frac{1}{2}$ and $\alpha_1 b_3> a_1 \beta_3
>\beta_1 a_3,$

\begin{equation}
\alpha_1>\max \{\frac{a_1}{2 b_1}, \frac{a_1+a_2+ 2 a_1
\beta_2}{2(2b_1+b_2)}, \frac{(0.5+\beta_2)(1-a_3)}{2(1-b_3)},\frac{2
a_1+a_2}{4(b_1+b_2)}, \frac{a_1+a_2-a_2 \beta_3}{2-b_3} \}
\end{equation}
It is interesting to note that in the first subcase when $a_1= a_2$
we have, $|\psi_1\rangle\rightarrow |\phi_2\rangle.$ Except this
case, our choice maintains $|\psi_i\rangle\not\leftrightarrow
|\phi_j\rangle, \ \forall i,j=1,2.$

\subsection{Joint transformation of pairs with the same initial state}

At the beginning of this subsection we want to present the special
result for $3\times 3$ system of pure entangled states as follows:\\

\textbf{Theorem:} \emph{For any source state $|\psi\rangle$ in
$3\times 3$ system, with distinct Schmidt coefficients there always
exist two states $(|\chi\rangle,$ $|\eta\rangle)$ such that both of
them are incomparable with $|\psi\rangle$, but from two copy of
$|\psi\rangle$ we are able to get them by collective LOCC with
certainty.}

Suppose the source state is $|\psi\rangle\equiv(a_1, a_2, a_3)$ with
$a_1> a_2> a_3>0, \ a_1+ a_2+ a_3 =1$ and the other states are
expressed with the most arbitrary form as: $|\chi\rangle\equiv(b_1,
b_2, b_3)$ and $|\eta\rangle\equiv(c_1, c_2, c_3)$ with $b_1 \geq
b_2 \geq b_3>0$, $b_1+ b_2+ b_3 =1$ and $c_1 \geq c_2 \geq c_3 >0$,
$ c_1+ c_2+ c_3 =1.$ For incomparability of the two pair of states
$(|\psi\rangle,~|\chi\rangle)$ and $(|\psi\rangle,~|\eta\rangle)$,
we need to impose either of the two relations between their Schmidt
coefficients, as specified in Eq.(\ref{incom}). Then it follows from
Nielsen's condition that there is always a possible range of
$(|\chi\rangle,|\eta\rangle)$ such that $|\psi\rangle^{ \otimes 2}
\longrightarrow |\chi\rangle\otimes|\eta\rangle,$ under LOCC with
certainty. It should be noted that the cases of failure of this
general result is only the small number of cases where irrespective
of the incomparability condition, the Schmidt coefficients of the
source state are not all distinct, i.e., either $a_1=a_2$ or
$a_2=a_3.$

This result is very important because we must keep in our mind the
fact, that multiple copy transformation is not possible for states
in $3 \times 3$ system. Thus, whenever we require the joint
transformation to be possible under deterministic LOCC, while both
of the final states are incomparable with the initial state, we are
constrained to keep the pair of final states $(|\chi\rangle,
|\eta\rangle)$, to be incomparable too. Now with this result in our
hand, let us try to fix $|\chi\rangle$ as our target state and find
the possible range (if exists at all) of $|\eta\rangle;$ i.e., given
two copies of the source state $|\psi\rangle$, our aim is to find a
$|\eta\rangle;$ incomparable with the source state, so that
$|\psi\rangle ^{\otimes 2} \longrightarrow |\chi\rangle
\otimes|\eta\rangle,$ is possible under LOCC with certainty.

Like the previous section here also we have two cases of
incomparability of the target pair $(|\psi\rangle,~|\chi\rangle)$,
as given in Eq.(\ref{incom}). From Nielsen's criteria, to perform
the joint transformation by deterministic LOCC, we necessarily
require to satisfy the relation, $a_1^2 \leq b_1 c_1$ and $a_3^2 >
b_3 c_3$. Thus, in case $a_1>b_1$, we are bound to impose the
condition $a_1< c_1$. Otherwise, when $a_1<b_1$, to retain the
incomparability of $(|\psi\rangle,~|\chi\rangle)$, we have
$a_3<b_3$. And, then we must choose $|\chi\rangle$ in such a manner
that $a_3>c_3$ and as, we have to satisfy $|\chi\rangle
\nleftrightarrow |\eta\rangle$, we also require $a_1>c_1$.

{\bf{Case-1:}} When, $a_1<b_1, a_1+a_2>b_1+b_2.$ Here, without any
loss of generality, we may assume, $c_1=c_2$ and rename the Schmidt
coefficients of the pure state $|\eta\rangle$ as, $|\eta\rangle
\equiv (\alpha_1, \alpha_1, \alpha_2 )$ where $\alpha_1 > \alpha_2 >
0, 2\alpha_1 + \alpha_2 =1, a_1 > \alpha_1, a_1+a_2 < 2\alpha_1.$
Thus, we obtain the relation between the Schmidt coefficients as:
$$b_1>a_1>\alpha_1>a_2>b_2\geq b_3
>a_3> \alpha_2.$$ Consider now two subcases separately.

Firstly, we consider the case, when ${a_2}^2>a_1a_3$. We require the
joint transformation $|\psi\rangle ^{\otimes 2} \longrightarrow
|\chi\rangle \otimes|\eta\rangle,$ is possible under LOCC with
certainty. As, ${a_2}^2>a_1a_3$, the Schmidt vector of $|\psi\rangle
^{\otimes 2}$ is:\\
$(a_1^2,a_1a_2,a_1a_2,{a_2}^2,a_1a_3,a_1a_3,a_2a_3,a_2a_3,{a_2}^2)$\\and
the Schmidt vector of $|\chi\rangle
\otimes|\eta\rangle$ is:\\
$(b_1\alpha_1,b_1\alpha_1,b_2\alpha_1,
b_2\alpha_1,b_3\alpha_1,b_3\alpha_1,b_2\alpha_3,b_3 \alpha_3).$\\
Choose, $|\eta\rangle$ in such a way that $\alpha_3<
\frac{b_3}{2b_1+b_3}$.

Then, by Nielsen's criteria, for the transformation $|\psi\rangle
^{\otimes 2} \longrightarrow |\chi\rangle \otimes|\eta\rangle,$ we
must have:
\begin{equation}
\begin{array}{lcl}
a_1^2 &<& b_1\alpha_1\\
a_1^2+a_1a_2 &<& 2b_1\alpha_1\\
a_1^2+2a_1a_2 &<& (2b_1+b_2)\alpha_1\\
(a_1+a_2)^2 &<& 2(b_1+b_2)\alpha_1\\
(a_1+a_2)^2+a_1a_3 &<& (2b_1+2b_2+b_3)\alpha_1\\
(a_1+a_2)^2+2a_1a_3 &<& 2\alpha_1\\
(a_1+a_2)^2+(2a_1+a_2)a_3 &<& 2\alpha_1+b_1\alpha_3\\
a_3^2 &>& b_3\alpha_3\\
\end{array}
\end{equation}
Now, we investigate those conditions in detail. If $a_1^2 <
b_1\alpha_1$, then $a_1^2+a_1a_2 < 2a_1^2< 2b_1\alpha_1$. Again,
adding both sides of the 4th and 6th conditions, we obtain the 5th
condition. Next, we try to prove the 7th condition;

$\{(a_1+a_2)^2+(2a_1+a_2)a_3\} - (2\alpha_1+b_1\alpha_3) =
\{(a_1+a_2+a_3)^2-(a_2a_3+a_3^2)\} - (1-\alpha_3+b_1\alpha_3) =
1-(a_2a_3+a_3^2) - (1-\alpha_3+b_1\alpha_3) =
(1-b_1)\alpha_3-(a_2+a_3)a_3 < (1-b_1)\frac{a_3^2}{b_3}-(a_2+a_3)a_3
= \{(b_2+b_3)a_3-(a_2+a_3)b_3\}\frac{a_3}{b_3} =
\{b_2a_3-a_2b_3\}\frac{a_3}{b_3} <0.$

Also, $a_3^2 - b_3\alpha_3= a_3^2 - b_3(1-2\alpha_1) =  a_3^2 - b_3+
2\alpha_1 b_3 > a_3^2 - b_3+ \frac{(a_1+a_2)^2}{(b_1+b_2)} b_3=
a_3^2 - b_3+ \frac{(1-a_3)^2}{(1- b_3)} b_3>0. $

Then, such a joint transformation occurs if,
$\frac{(a_1+a_2)^2}{2(b_1+b_2)}<a_1$ and the range of $|\eta\rangle$
is specified by the relation:
\begin{equation}
\alpha_1>\max\{ a_1-\frac{{a_1}^2-{a_2}^2}{2},
\frac{(a_1+a_2)^2}{2(b_1+b_2)},
\frac{(a_1)^2}{b_1},\frac{a_1(a_1+2a_2)}{(2b_1+b_2)}\}
\end{equation}
Next, when ${a_2}^2<a_1a_3$, then the condition for such
transformation is; $a_1+2a_2< 2b_1+b_2.$ Under this condition range
of $|\eta\rangle$ is specified by the relation:
\begin{equation}
\alpha_1>\max\{a_1-\frac{{a_1}^2-{a_2}^2}{2},
\frac{a_1(2-a_1)}{(2-b_3)},\frac{(a_1)^2}{b_1},\frac{a_1(a_1+2a_2)}{(2b_1+b_2)}\}
\end{equation}

{\bf{Case-2:}} When $a_1>b_1, a_1+a_2<b_1+b_2,$ we take
$|\eta\rangle \equiv (\alpha_1, \alpha_2, \alpha_2 )$ where
$\alpha_1 > \alpha_2> 0, \alpha_1 +2 \alpha_2 =1, a_1 < \alpha_1,
a_1+a_2 > \alpha_1+\alpha_2.$ Then the condition for such
transformation is, $a_3< \frac{1}{2}(1-\frac{{a_1}^2}{b_1}).$ Under
this condition, we have not only one $|\eta\rangle,$ but a range of
it specified either by the relation:
\begin{equation}
\alpha_2<\min\{\frac{a_1a_3}{b_1},
\frac{a_3^2}{b_3},\frac{a_3(2a_2+a_3)}{(b_2+2b_3)}\}, \ \mbox{for} \
{a_2}^2>a_1a_3,
\end{equation}
or, by the relation:
\begin{equation}
\alpha_2<\min \{ a_3+\frac{{a_2}^2-{a_3}^2}{2}, \frac{a_3^2}{b_3},
\frac{a_3(2a_2+a_3)}{(b_2+2b_3)} \}, \ \mbox{for} \ {a_2}^2<a_1a_3
\end{equation}

Finally, here we must mention that our process works for most of the
cases of incomparability. But, it is not always successful; i.e.,
choosing any arbitrary incomparable pair, we might not be able to
reach the target state by this method. This small range of failure
of the process is possibly due to the fact that we didn't ever
bother about the amount of entanglement contained into the states.
It is possible that $E(|\psi\rangle)\ll E(|\chi\rangle);$ for which
there doesn't exists such a state $|\eta\rangle,$ incomparable with
$|\psi\rangle$ and $ E({|\psi\rangle}^{\otimes2})> E(|\chi\rangle
\otimes |\eta\rangle).$

In conclusion, with this method any incomparable pair of pure
bipartite entangled states in any finite dimension, can be maid to
compare (i.e., transform one to another), under LOCC with certainty,
by providing some pure entanglement. We observe that mutual
co-operation is an useful process to break the incomparability of
two pairs under LOCC. This is not only discussed as an abstract or
rather complicated theory, but we provide the algorithmic structure
by which this goal can be really achieved.

\section{The role of entanglement}
The various methods of transforming a pure bipartite states to other
described above are again examples of the role of entanglement in
performing different information processing tasks. The amount of
entanglement content determines the direction of local
transformation for a particular pair of pure entangled states. The
presence of unperturbed entangled state sometimes enhance the
transformation. When the presence of entanglement is not sufficient
to assist the transformation, then by exploiting entanglement, one
can also able to perform the local conversion. Our schemes are
proposed in this direction. Both of the schemes proposed here are
elaborated analytically, starting from some numerical examples. They
provide us in practical senses, the way of proper utilization of the
resource, for the process of transforming an arbitrary incomparable
pair of states, with extra input of entanglement, by deterministic
LOCC. The results are obviously related with the manipulation and
processing of pure bipartite entangled states.

In conclusion, in this chapter, we have studied incomparability as a
peculiar feature of pure state entanglement that exists in quantum
systems. Incomparability shows some constraint on local conversion
of entangled states. The results also indicate that a kind of
non-locality is associated with the system that may not be fully
describable by the amount of entanglement, but rather it may
concerned with the construction or purification processes which are
connected with the Schmidt vector of the states. It again brings us
to explore a very fundamental area of the quantum information theory
regarding the notion of quantification of pure bipartite entangled
states that all are equal with entropy of entanglement. It is thus
interesting to search for different non-local features of such
systems of incomparable states through different views of quantum
information theory. Apart from the manipulating entanglement, we
will show in the next two chapters  that the notion of
incomparability also plays a role of detection for some classes of
nonphysical operations.

\chapter{Spin-flipping and Incomparability}\footnote{Some portions of this chapter is
published in Physical Review A {\bf 73}, 044303 (2006).}

\section{Spin-flipping of a qubit}

In quantum information theory, one of the main object is to encode some
information in quantum states. In
classical world, if the information of the space direction
$\overrightarrow{n}$ is represented by a classical
spin-$\frac{1}{2}$ particle polarized in the direction
$\overrightarrow{n}$, then it is as good as the spin-$\frac{1}{2}$
particle polarized along $-\overrightarrow{n}$. But, the situation in quantum
systems is quite different. There is no single quantum operation that could
reverse the spin direction in an arbitrary manner. Further, if we use composite
quantum states to encode information, then the state
$|\overrightarrow{n}, \overrightarrow{n}\rangle$ of two
spin-$\frac{1}{2}$ particles both polarized along
$\overrightarrow{n}$, is not equivalent with the state
$|\overrightarrow{n}, -\overrightarrow{n}\rangle$ corresponding to two
spin-$\frac{1}{2}$ particles polarized in opposite directions
$\overrightarrow{n}$ and $-\overrightarrow{n}$. The root of this problem
is not completely solved, however, many interesting results found in this
direction \cite{25,27,37}. Below, we have described one such situation.

There is an intimate relation between discriminating a set of quantum
states and the process of estimating the set. It is shown by
Gisin \emph{et.al.} \cite{27}, the set of anti-parallel spin
states defined by,

$S^a=\{| \overrightarrow{n},
-\overrightarrow{n}\rangle~;~~\overrightarrow{n}\in\Re^3\}$\\can be
better distinguished than the corresponding set of states of
parallel spins describes by,

$S^p=\{| \overrightarrow{n},
\overrightarrow{n}\rangle~;~~\overrightarrow{n}\in\Re^3\}.$

Thus, two set of states are not equivalent while we want to discriminate
its members. Also, it is found that the ensemble $S^a$
of states with antiparallel spins, have a larger entropy. The physical reason
behind such behavior of states of spin-$\frac{1}{2}$ particles may
be the non-universality of the operation that transforms an arbitrary
spin-direction to the opposite spin direction. However, it is clear that
the quantum spin-flipping operation has its own limitations which imposes some
constraints over the system. In this chapter, we will show a connection
between spin-flipping of qubits and the class of pure incomparable states.

\subsection{Flipping operation}

The flipping operation acts on a qubit to reverse the
spin-polarization direction of the qubit. In other words, a flipping
operation transforms a qubit to its orthogonal qubit.
For example, the NOT gate represented by $2\times 2$ matrix in $z$-basis
$\left(
  \begin{array}{cc}
    0 & 1 \\
    1 & 0 \\
  \end{array}
\right)$ flips perfectly the states $|0\rangle$ and $|1\rangle$,
i.e., the spin directions $Z^+$ and $Z^-$, but it cannot flip the spin
direction $X^+$. Mathematically, we define the flipping operation $\Omega$
on a single arbitrary input qubit $| \varphi \rangle= \alpha | 0 \rangle
~+~ \beta | 1 \rangle~~;~~~|\alpha|^2~+~ |\beta|^2~=~1$ as follows:
\begin{equation}
\begin{array}{lcl}
| \varphi \rangle \rightarrow |\overline{\varphi}
\rangle~=~\Omega | \varphi \rangle~=~ \beta^* | 0 \rangle ~-~
\alpha^* | 1 \rangle
\end{array}
\end{equation}
so that the inner product of the input qubit $| \varphi \rangle $ and
the output qubit  $|\overline{\varphi}
\rangle$ will vanish. Thus, if defined on a single qubit, it is
a very natural operation achievable by a simple rotation of the
Bloch sphere about its center. Now if one requires a single flipping
operation defined to act perfectly on a large class of qubits, there
arises some restrictions.

\subsection{No-flipping principle}

Gisin first showed that there is a restriction over general flipping
operation which is similar in some sense with the 'No-Cloning' and
'No-Deleting' principles, but completely different in its
operational status. Non-existence of universal exact flipping
machine is a kind of constraint on the quantum systems that has been
directly observed from the unitary dynamics of quantum evolutions.
One could verify that the universal exact flipper, which if operated
on an arbitrary qubit, will reverse the spin polarization direction,
is not unitary but an anti-unitary operation and thus it is not in
general a physical operation. {\it No-flipping principle says that
exact flipping of an unknown qubit state, is not possible}. Like
no-cloning, no-deleting principles, it is also a fundamental
restriction to the allowable operations on quantum systems. However,
instead of at least two subsystems to describe no-cloning or
no-deleting principles, we require only one single system to
describe flipping operation. Further, there always exist a quantum
machine which can act as an exact flipper for any set of two qubit
states, even if they are non-orthogonal in nature \cite{25,46a}. One
interesting observation is found by Gisin and Popescu \cite{27} and
also by Massar \cite{37} that the two antiparallel-spin state
$|\overrightarrow{n},-\overrightarrow{n}\rangle$ contains more
information than that of two-parallel-spin state
$|\overrightarrow{n},\overrightarrow{n}\rangle.$ It is conjectured
that the origin of this feature is the property that the anti
parallel spin states span the whole four dimensional Hilbert space
of two spin-$\frac{1}{2}$ system, while the parallel
spin-$\frac{1}{2}$ system spans only a three dimensional subspace
constituted by the symmetric states. Apart from such restrictions,
there exists universal optimal flipping machine for qubits (Buzek
\emph{et.al.} \cite{optimalFlip}).

\subsection{Exact flipping of states of one great circle}
The flipping operation on qubit system can be visualized nicely by
the Bloch sphere representation. It is easy to imagine a valid
physical operation on the system which may flip a fixed spin
polarization direction by a $180^\circ$ rotation of the Bloch sphere
about its center. Obviously, all the vectors lying on the circle of
rotation, will be flipped exactly by this operation. Thus, we see
that the operation acts exactly on every qubit represented by a ray
on a great circle. It is further proved that the largest
set of states which can be exactly flipped by a single quantum
machine, is a set of states lying on a great circle of the Bloch
sphere \cite{25}. Interestingly, one could find that any three
states of the Bloch sphere, not lying in one great circle can not
be flipped exactly by a single quantum machine. Thus, a stronger
version of no-flipping principle can be expressed as,

\emph{Any three qubit states, with spin-polarization directions not
lying in one great circle can not be flipped exactly by a single
quantum machine.}

\subsection{Relation with other detectors}
Any physical system would necessarily respect no-signalling
condition of the special theory of relativity. Violation of this
principle of nature is used as a detector for exploring the
impossibilities of some processes to be a physical one
\cite{ProbClone,DeletSignal}. Recently, we proved \cite{flip} that,
both the principles of No-Signaling and Non-increase of entanglement
by LOCC, separately implies the exact flipping operation, defined on
the minimum number (i.e., three) of qubits not lying in one great
circle, is impossible in nature. The work also shows the importance
of the underlying linearity assumption, as by allowing linearity in
pure state superposition level, one could even create entanglement
between separated subsystems. Research along the direction of
finding inter-relations between No-flipping principle and other
constrains of quantum mechanics \cite{22}, has created lots of
interest.

\section{No-flipping verses incomparability}

In this section, we will show the inter-relation between the
no-flipping principle and the notion of incomparable states in
pure bipartite entangled states \cite{incomflip}. We have illustrated
our results firstly with examples.\\

\emph{\textbf{Example-1}:} Consider three states representing the
three axes of the Bloch sphere in usual basis, as $|0_x
\rangle=\frac{|0\rangle+|1\rangle}{\sqrt{2}}~,~|0_y
\rangle=\frac{|0\rangle+i|1\rangle}{\sqrt{2}}~,
|0_z\rangle=|0\rangle$. Beside of not lying in one great circle of
the Bloch sphere, this three states are mutually non-orthogonal and far
separated from each other. We choose the particular setting of a
pure bipartite state in the form:
\begin{equation}
\begin{array}{lcl}
|\Psi^i\rangle_{AB}=
\frac{1}{\sqrt{3}}~\{~|0\rangle_A|0_z0_z\rangle_B +
|1\rangle_A|0_x0_y\rangle_B + |2\rangle_A|0_y0_x\rangle_B~\}
\end{array}
\end{equation}
This is a three particle state, shared between two space-like
separated parties Alice and Bob, so that Alice has a $3$-dimensional
system with the basis $\{|0\rangle,|1\rangle,|2\rangle\}$ and Bob
has two particles, each one belongs to a qubit system. As a
bipartite state it represents a $3\times 4$ system. We also impose
the restriction that the reduced density matrix of Bob's subsystem
admits a representation in terms of the three input states $|0_x
\rangle, |0_y \rangle, |0_z \rangle$. As the operation is performed
on Bob's system and it is defined on the above input states only,
thus it is necessary that the reduced density matrix of Bob's system
must be represented by $|0_x \rangle, |0_y \rangle, |0_z \rangle$,
on which our flipping machine is defined to act. Such as here we
have $\rho_B= Tr_A (|\Psi^i\rangle_{AB}\langle\Psi^i|)=
\frac{1}{3}\{P[|0_z\rangle ]\otimes P[|0_z\rangle ]+P[|0_x\rangle ]
\otimes P[|0_y\rangle ]+P[|0_y\rangle ]\otimes P[|0_x\rangle ]\}$.
The Schmidt vector corresponding to the initial joint state is
$\lambda^i=(\frac{2}{3},\frac{1}{6},\frac{1}{6})$. If the existence
of exact flipping machine for the three states $~|0_x \rangle,~|0_y
\rangle,~|0_z \rangle$ is possible, then by applying this machine to
one of the two particles on Bob's side (say, on the last qubit
system), the joint state between them exactly transforms to the
state:
\begin{equation}
\begin{array}{lcl}
|\Psi^f
\rangle_{AB}=\frac{1}{\sqrt{3}}~\{~|0\rangle_A|0_z\overline{0_z}~\rangle_B
+ e^{i\chi} |1\rangle_A|0_x\overline{0_y}~\rangle_B +
e^{i\eta}|2\rangle_A|0_y\overline{0_x}~\rangle_B~\}
\end{array}
\end{equation}
where $e^{i\chi},e^{i\eta}$ are some arbitrary phase-factors. The
Schmidt vector corresponding to the final joint state is $\lambda^f
=(\frac{1}{3}+\frac{1}{2\sqrt{3}}~, \frac{1}{3}~,
\frac{1}{3}-\frac{1}{2\sqrt{3}})$. From Eq.(\ref{incom}) it is easy
to check that $|\Psi^i \rangle_{AB},~|\Psi^f\rangle_{AB}$ is a pair
of incomparable states. Hence by Nielsen's criterion it is
impossible to locally transform $|\Psi^i \rangle_{AB}$ to $|\Psi^f
\rangle_{AB}$ with certainty. In this example, we see that the
impossibility of an operation in quantum mechanics can be
established from the contradiction that it forces two incomparable
states to become comparable by LOCC with certainty.

The result shows, how a local anti-unitary operation evolves the
system in an unphysical way. To explore this unphysical nature of
anti-unitary operators, we have considered in example-1, a joint system
between some separated parties, because on a single system it is
really difficult to distinguish unitary and anti-unitary operators.
Now, applying an anti-unitary operator (say, $L$) locally on a joint
system of $3 \times (2 \times 2)$ dimension, i.e., applying the
operator $I_A \otimes (I \otimes L)_B$ on the joint system (where
$I$ indicates, identity operator), we find a case of incomparability.
The bi-partite entanglement of the joint system changes and it changes
in such a manner that there is no way to compare the initial and final
joint states locally. Next, we provide another example that shows a
reverse phenomena.\\

\emph{\textbf{Example-2}:} Consider a pair of pure entangled states,
shared between Alice and Bob, in the form:
\begin{equation}
\begin{array}{lcl}
|\Psi\rangle_{AB}= \sqrt{.51}~|0\rangle_A|0\rangle_B +
\sqrt{.30}~|1\rangle_A|1\rangle_B +
\sqrt{.19}~|2\rangle_A|2\rangle_B,\\
|\Phi\rangle_{AB}= \sqrt{.49}~|0\rangle_A|0\rangle_B +
\sqrt{.36}~|1\rangle_A|1\rangle_B +
\sqrt{.15}~|2\rangle_A|2\rangle_B. \label{ex2}%
\end{array}
\end{equation}
The Schmidt vectors corresponding to the states $|\Psi \rangle_{AB}$
and $|\Phi \rangle_{AB}$ are $\lambda^{\Psi}= (.51,.30,.19)$ and
$\lambda^{\Phi}=(.49,.36,.15)$ respectively. Using Eq.(\ref{incom})
we find, $(|\Psi \rangle_{AB}, |\Phi \rangle_{AB})$ is a pair of
incomparable states. Suppose Bob has a two qubit system on his side
and the orthogonal states $\{|0\rangle_B, |1\rangle_B, |2\rangle_B
\}$ have the form,

$|0\rangle_B=|\psi\rangle_{B_1}|\psi\rangle_{B_2}~,~|1\rangle_B=
|\overline{\psi}\rangle_{B_1}|\psi\rangle_{B_2}~,
~|2\rangle_B=|\overline{\psi}\rangle_{B_1}|\overline{\psi}\rangle_{B_2}~,$\\
where $|\psi \rangle$ is an arbitrary qubit state with Bloch vector
$\overrightarrow{n_{\psi}}$, i.e., $|\psi \rangle \langle \psi | =
\frac{1}{2}[I + \overrightarrow{n_{\psi}}\cdot
\overrightarrow{\sigma}]$ and $|\overline{\psi}\rangle$ is
orthogonal to $|\psi \rangle$.

Now tracing out Alice's system and the second qubit of Bob's side
(i.e., system $B_2$), one qubit reduced subsystem of Bob
corresponding to the two joint systems in Eq.(\ref{ex2}) will be
of the form,
\begin{equation}
\begin{array}{lcl}
\rho_{B_1}^{\Psi}= Tr_{A B_2 }\{| \Psi \rangle \langle \Psi |\}=~
\frac{1}{2}~[I + .02~\overrightarrow{n_{\psi}}\cdot
\overrightarrow{\sigma}]\\
\rho_{B_1}^{\Phi}= Tr_{A B_2 }\{| \Phi \rangle \langle \Phi |\}=~
\frac{1}{2}~[I - .02~\overrightarrow{n_{\psi}}\cdot
\overrightarrow{\sigma}].
\end{array}
\end{equation}The states of the above equation are two mixed qubit
states with spin-polarization directions along two exactly opposite
vectors. Here, we have extended the idea of flipping operation
from only pure qubit states to general qubit states. Thus, by
applying any set of local operations, if it is possible to transform the
joint state between Alice and Bob from $|\Psi\rangle_{AB}$ to
$|\Phi\rangle_{AB}$, then the reduced state of one qubit subsystem on
Bob's side will be changed from $\rho_{B_1}^{\Psi}$ to $\rho_{B_1}^{\Phi}$
exactly, by LOCC with certainty. It is clear that the spin direction
$\overrightarrow{n_{\psi}}$ of the arbitrary qubit state
$|\psi\rangle$ is reversed after the operation, i.e., transformed to
$-\overrightarrow{n_{\psi}}$. So, if we extend the LOCC
transformation criterion so that the states $|\Psi\rangle_{AB},
|\Phi\rangle_{AB}$ are interconvertible by some operation then
consequently on a subsystem, the spin-polarization of an arbitrary
qubit state is being reversed. This is quite similar of preparing an
arbitrary spin flipper machine and is an alternative way of
establishing the incomparable nature of the pair of states in
Eq.(\ref{ex2}).

\subsection{Spin-flipping for mixed qubit}
As discussed in the second chapter, the general form of any qubit
system in Bloch sphere representation is given by,
$\rho=\frac{1}{2}(I+\overrightarrow{n}.\overrightarrow{\sigma}) \ \
;\ \  | \overrightarrow{n} | \leq 1$, where $\overrightarrow{n}$
denotes the Bloch vector of the state or the direction of the spin
polarization of the state. Bloch sphere is a three-dimensional
geometrical visualization of the state of the system and it is
reasonable to define the flipped state to be the state, corresponding
to the direction that is exactly opposite to the direction
$\overrightarrow{n}$. The new direction representing
the mixed qubit will obviously be denoted by $-\overrightarrow{n}$,
preserving the previous length of the vector $|\overrightarrow{n}|$.
Thus, mixed qubit flipping may be described as reversal of the direction
of spin polarization of the qubit. The operation can be described as,
$\Gamma:~\rho^i \longrightarrow \rho^f$ where the initial mixed
state is $\rho^i=~\frac{1}{2}\{I+\overrightarrow{n }\cdot
\overrightarrow{\sigma}\}$ and the flipped state will be of the form
$\rho^f=~\frac{1}{2}\{I-\overrightarrow{n }\cdot
\overrightarrow{\sigma}\}$, with $|\overrightarrow{n } | \leq 1$.

\section{Non-existence of Universal exact flipper}
To generalize the main result corresponding to the first example, we
consider three arbitrary states not lying in one great circle in
their simplest form,
\begin{equation}
\begin{array}{lcl}
& & |0\rangle, \\
& &  |\psi\rangle= a|0\rangle + b|1\rangle,  \\
& & |\phi\rangle= c|0\rangle + d~e^{i\theta}|1\rangle \label{qubits}%
\end{array}
\end{equation}
where a, b, c, d are real numbers satisfying the relation $ a^2 +
b^2 =~ 1=~ c^2 + d^2~$ and $0 < \theta < \pi $. We call these three
states as three arbitrary qubits, as by suitable change of basis it
is possible to express any three qubits in this form.

Suppose two spatially separated parties Alice and Bob are sharing
the entangled state,
\begin{equation}
\begin{array}{lcl}
|\Omega\rangle_{AB}= \frac{1}{\sqrt{3}}~\{~|0\rangle_A|00\rangle_B
+|1\rangle_A|\psi\phi\rangle_B +|2\rangle_A|\phi\psi\rangle_B~\}
\end{array}
\end{equation}
where Alice has a 3-dimensional orthogonal local system, having the
basis, $\{|0\rangle,~|1\rangle,~|2\rangle \}$ and Bob has a two
qubit system.

Using Eq.(\ref{qubits}) one can rewrite the joint system shared
between A and B in the usual basis as,
\begin{equation}
\begin{array}{lcl}
|\Omega\rangle_{AB}&=& \frac{1}{\sqrt{3}}~\{~|0\rangle_A|00\rangle_B
+|1\rangle_A((a|0\rangle + b|1\rangle)(c|0\rangle +
d~e^{i\theta}|1\rangle))_B \\&~~& +|2\rangle_A((c|0\rangle +
d~e^{i\theta}|1\rangle)(a|0\rangle + b|1\rangle))_B~\}\\ &=&
\frac{1}{\sqrt{3}}~\{~(|0\rangle+ac|1\rangle+ac|2\rangle)_A|00\rangle_B
+(ad~e^{i\theta}|1\rangle+bc|2\rangle)_A |01\rangle_B\\&~~&
+(bc|1\rangle+ad~e^{i\theta}|2\rangle)_A |10\rangle_B
+bd~e^{i\theta}(|1\rangle+|2\rangle)_A |11\rangle_B ~\}
\end{array}
\end{equation}
Tracing out the part of the state on Bob's side we focus on the
reduced density matrix of Alice's side, as follows:
\begin{equation}
\begin{array}{lcl}
\rho_A^i &=& Tr_B (|\Omega\rangle_{AB} \langle \Omega|) \\
&=& \frac{1}{3}~ \{P[|0\rangle+ac|1\rangle+ac|2\rangle]+
P[ad~e^{i\theta}|1\rangle+bc|2\rangle]+\\ & &
P[bc|1\rangle+ad~e^{i\theta}|2\rangle]
+ b^2 d^2 P[|1\rangle+|2\rangle]\}\\
&=& \frac{1}{3} \{P[|0\rangle]+P[|1\rangle]+P[|2\rangle]+
ac(|0\rangle\langle1|+|1\rangle\langle0|+|0\rangle\langle2|
+|2\rangle\langle0|)\\&+&(a^2 c^2+ abcde^{i\theta}+abcde^{-i\theta}+
b^2 d^2)
(|1\rangle\langle2| +|2\rangle\langle1|)\}\\
&=& \frac{1}{3}~ \{P[|0\rangle]+P[|1\rangle]+P[|2\rangle]+
ac(|0\rangle\langle1|+|1\rangle\langle0|+|0\rangle\langle2|
\\ & &~~+|2\rangle\langle0|)+|\langle \psi |
\phi\rangle|^2(|1\rangle\langle2| +|2\rangle\langle1|) \}
\end{array}
\end{equation}
where to avoid notational complexity, we drop above the suffix $A$
on the R.H.S. for the subsystem of Alice. In the sequel, sometimes we
will also drop suffix $A$ or $B$ where the case may be.

Consider now that Bob has a exact flipping machine defined on just
these three states $|0\rangle,~|\psi\rangle,~|\phi\rangle$. The
flipping operation can be described as:
\begin{equation}
\begin{array}{lcl}
& & |0\rangle \longrightarrow |1\rangle  \\
& & |\psi\rangle \longrightarrow
e^{i\mu}|\overline{\psi}\rangle ~=~ e^{i\mu} (b|0\rangle - a |1\rangle ) \\
& &  |\phi\rangle \longrightarrow e^{i\nu}|\overline{\phi}\rangle
~=~ e^{i\nu} (d~e^{-i\theta}|0\rangle - c|1\rangle )
\end{array}
\end{equation}
where  $|\overline{\psi} \rangle,| \overline{\phi} \rangle$ are the
states orthogonal to $|\psi\rangle,~|\phi\rangle$ respectively and
$e^{i\mu}$ and $e^{i\nu}$ are some arbitrary phase factors. Now
assume that Bob applies the above mentioned flipping machine on any
one of his two subsystems (say, on the second subsystem). After this
local operation on Bob's subsystem, the shared state between Alice
and Bob takes the form,
\begin{equation}
\begin{array}{lcl}|\Omega^{f}\rangle_{AB}&=& \frac{1}{\sqrt{3}} \{|0\rangle_A |01\rangle_B +
e^{i \nu}|1\rangle_A|\psi\overline{\phi}\rangle_B + e^{i
\mu}|2\rangle_A|\phi\overline{\psi}\rangle_B \}\\
 &=& \frac{1}{\sqrt{3}} \{|0\rangle_A |01\rangle_B + e^{i
\nu}|1\rangle_A ((a|0\rangle + b|1\rangle)(de^{-i\theta}|0\rangle -
c|1\rangle))_B \\ & &+ e^{i \mu}|2\rangle_A ((c|0\rangle +
d~e^{i\theta}|1\rangle)(b|0\rangle - a |1\rangle))_B \}\\&=&
\frac{1}{\sqrt{3}}~ \{(ad e^{i(\nu-\theta)}|1\rangle+ bc
e^{i\mu}|2\rangle)_A |00\rangle_B\\ & &+(|0\rangle-ac e^{i
\nu}|1\rangle-ac e^{i \mu}|2\rangle)_A |01\rangle_B \\ &
&+(bde^{i(\nu-\theta)}|1\rangle+ bde^{i(\mu+\theta)}|2\rangle)_A
|10\rangle_B\\ & &+(-bce^{i \nu}|1\rangle+ ade^{i(
\mu+\theta)}|2\rangle)_A |11\rangle_B \}
\end{array}
\end{equation}
The final density matrix of Alice's side is:
\begin{equation}
\begin{array}{lcl}
\rho^{f}_A &=& \frac{1}{3} \{P[ade^{i(\nu-\theta)}|1\rangle+ e^{i
\mu}bc|2\rangle] + P[|0\rangle-ac e^{i \nu}|1\rangle-ac e^{i
\mu}|2\rangle]\\
& & + P[bde^{i(\nu-\theta)}|1\rangle+ e^{i
\mu}bde^{i\theta}|2\rangle] + P[e^{i \nu}bc|1\rangle- e^{i
\mu}ade^{i\theta}|2\rangle] \}\\&=&\frac{1}{3} \{P[|0\rangle] +
P[|1\rangle]+P[|2\rangle]\\
& &- ac(e^{-i\nu} |0 \rangle\langle 1| + e^{i\nu} |1\rangle\langle
0|+e^{-i\mu} |0\rangle\langle2|  + e^{i\mu}
|2\rangle\langle0|)\\
& & +(\langle \phi | \psi\rangle )^2 e^{i(\nu-\mu)} |1\rangle
\langle 2| +(\langle \psi | \phi\rangle )^2 e^{i(\mu-\nu)} |2\rangle
\langle 1|  \}
\end{array}
\end{equation}
The eigenvalue equation for the initial local density matrix
$\rho^i_A$ is,
\begin{equation}
\begin{array}{lcl}
(1-3\lambda)^3 - 3(1-3\lambda)A+ B=0 \label{initial}%
\end{array}
\end{equation}
and that of the final local density matrix $\rho^{f}_A$ is,
\begin{equation}
\begin{array}{lcl}
(1-3\lambda)^3 - 3(1-3\lambda)A + B'=0 \label{final}%
\end{array}
\end{equation}
where $A=\frac{1}{3}[2a^2c^2+{|\langle \psi | \phi\rangle
|}^4],~B=2a^2c^2{|\langle \psi | \phi\rangle |}^2$ and $ B'=2a^2c^2~
Re\{{\langle \phi | \psi\rangle }^2\}.$ It is interesting to
observe that the phase factors $e^{i\mu}, e^{i\nu}$ vanishes in
this stage. So the final result obtained, doesn't care about the
phase factors related with the operation.

We find the roots of the equations (\ref{initial}) and (\ref{final})
(i.e., initial and final eigenvalues) and compare them by using
Cardan's method. This method is applied to find roots of the
cubic equation of the form:
\begin{equation}
\begin{array}{lcl}
x^3 - 3Gx + H=0 \label{Cardan}%
\end{array}
\end{equation}
with $G\geq 0$. The roots of this equation will denoted by $x_1,
x_2, x_3$ and may be expressed in the following form:
\begin{equation}
\begin{array}{lcl}
x_1~&=&~2\sqrt{G}\cos(\frac{2\pi}{3}+\alpha),\\
x_2~&=&~2\sqrt{G}\cos(\alpha), \\
x_3~&=&~2\sqrt{G}\cos(\frac{2\pi}{3}-\alpha)
\end{array}
\end{equation}where $\cos(3\alpha)=~\frac{-H}{2\sqrt{G^3}}$. By
suitable change of the variables, our initial and final eigenvalue
equations can be transformed to the above form of Eq.(\ref{Cardan}).
If we denote the initial and final eigenvectors as
$\widetilde{\lambda^i}= (\alpha_1,~\alpha_2,~\alpha_3 )$ and
$\widetilde{\lambda^f}= (\beta_1,~\beta_2,~\beta_3 )$, then we can
express them as
\begin{equation}
\begin{array}{lcl}
\alpha_1&=&\frac{1}{3}\{1 - 2\sqrt{A}\cos
(\frac{2\pi}{3}+\theta^i)\},\\
\alpha_2&=& \frac{1}{3}\{1 - 2\sqrt{A}\cos{\theta^i}\}\\
\alpha_3&=& \frac{1}{3}\{1 - 2\sqrt{A}
\cos(\frac{2\pi}{3}-\theta^i)\}
\end{array}
\end{equation}
and
\begin{equation}
\begin{array}{lcl}
\beta_1&=& \frac{1}{3}\{1 -2\sqrt{A}\cos
(\frac{2\pi}{3}+\theta^f)\},\\
\beta_2&=& \frac{1}{3}\{1 - 2\sqrt{A} \cos {\theta^f}\},\\
\beta_3&=& \frac{1}{3}\{1 - 2\sqrt{A}
\cos(\frac{2\pi}{3}-\theta^f)\}
\end{array}
\end{equation}
where, $\cos(3\theta^i)~=~\frac{-B}{2\sqrt{A^3}}$, and
$\cos(3\theta^f)~=~\frac{-B'}{2\sqrt{A^3}}$.

The eigenvectors $\widetilde{\lambda^i}$ and $\widetilde{\lambda^f}$
are not the Schmidt vectors as the eigenvalues are not arranged in
decreasing order. However, we will establish the relation
of incomparability as given in Eq.(\ref{incom}), by rearranging
the eigenvalues in decreasing order. Now to compare the states
$|\Omega\rangle_{AB}$ with $|\Omega^f\rangle_{AB}$ in terms of their
local convertibility, we do not need the explicit values of the
Schmidt coefficients but only we have to find the relations between them.
To do this, we first compare the coefficients of the eigenvalue
equations, $B$ and $B'$. Here $B=B'+4a^2 b^2 c^2 d^2 \sin^2 \theta$,
so $B\geq 0, B\geq B'$. Therefore, we have two possibilities,
either, $0<B'<B$ or, $B'<0<B.$\\

If \textbf{$0<B'<B,$} we have $0>\cos(3\theta^f)>\cos(3\theta^i)$.
This imply, $3\theta^i, 3\theta^f \in
(\frac{\pi}{2},\frac{3\pi}{2})$. We find four cases corresponding to
the different regions of $\theta^i$, and  $\theta^f$.

Firstly, we consider the case when $3\theta^i, 3\theta^f \in
(\frac{\pi}{2},\pi).$ We also know previously that
$\cos(3\theta^f)>\cos(3\theta^i)$. Now in the region
$(\frac{\pi}{2},\pi)$, $\cos(3\theta^f)> \cos(3\theta^i) \Rightarrow
3\theta^i
> 3\theta^f$. So, we have,
\begin{equation}
\begin{array}{lcl}
\frac{\pi}{6}< \theta^f< \theta^i< \frac{\pi}{3},\\
\Rightarrow \frac{\sqrt{3}}{2}>\cos(\theta^f)>
\cos(\theta^i)>\frac{1}{2}\\
\Rightarrow\sqrt{3A}>2\sqrt{A}\cos(\theta^f)>
2\sqrt{A}\cos(\theta^i)>\sqrt{A}\\ \Rightarrow
\frac{1}{3}(1-\sqrt{A})> \frac{1}{3}\{1 -
2\sqrt{A}\cos{(\theta^i)}\}>\frac{1}{3}\{1 - 2\sqrt{A} \cos
{(\theta^f)}\}\\~~~~~~~~~~~~~~~~~~~~~~~~~~~~~~~~
> \frac{1}{3}(1-\sqrt{3A})
\end{array}
\end{equation} i.e.,
$\frac{1}{3}(1-\sqrt{A})> \alpha_2> \beta_2
> \frac{1}{3}(1-\sqrt{3A})$.

Proceeding in this way, we find,
\begin{equation}
\begin{array}{lcl}
\frac{\pi}{6}< \theta^f< \theta^i< \frac{\pi}{3},\\
\Rightarrow \frac{2\pi}{3}+\frac{\pi}{6}< \frac{2\pi}{3}+\theta^f<
\frac{2\pi}{3}+\theta^i< \frac{2\pi}{3}+\frac{\pi}{3},\\
\emph{i.e.}, (\pi -\frac{\pi}{6})< \frac{2\pi}{3}+\theta^f<
\frac{2\pi}{3}+\theta^i< \pi\\
\Rightarrow \cos(\pi -\frac{\pi}{6})>\cos(\frac{2\pi}{3}+\theta^f)>
\cos(\frac{2\pi}{3}+\theta^i)>\cos{\pi}\\
\Rightarrow -\frac{\sqrt{3}}{2}>\cos(\frac{2\pi}{3}+\theta^f)>
\cos(\frac{2\pi}{3}+\theta^i)>-1\\
\Rightarrow-\sqrt{3A}>2\sqrt{A}\cos(\frac{2\pi}{3}+\theta^f)>
2\sqrt{A}\cos(\frac{2\pi}{3}+\theta^i)>-2\sqrt{A}\\ \Rightarrow
\frac{1}{3}(1+\sqrt{3A})< \frac{1}{3}\{1 -
2\sqrt{A}\cos{(\frac{2\pi}{3}+\theta^f)}\}<\frac{1}{3}\{1 - 2\sqrt{A} \cos
{(\frac{2\pi}{3}+\theta^i)}\}\\~~~~~~~~~~~~~~~~~~~~~~~~~~~~~~~~ <
\frac{1}{3}(1+2\sqrt{A})
\end{array}
\end{equation}

Thus, we have,
$\frac{1}{3}(1+\sqrt{3A})<\beta_1<\alpha_1<\frac{1}{3}(1+2\sqrt{A})$.

Again,
\begin{equation}
\begin{array}{lcl}
\frac{\pi}{6}< \theta^f< \theta^i< \frac{\pi}{3},\\
\Rightarrow \frac{2\pi}{3}-\frac{\pi}{6}> \frac{2\pi}{3}-\theta^f>
\frac{2\pi}{3}-\theta^i> \frac{2\pi}{3}-\frac{\pi}{3},\\
\emph{i.e.}, \frac{\pi}{2}> \frac{2\pi}{3}-\theta^f>
\frac{2\pi}{3}-\theta^i> \frac{\pi}{3}\\
\Rightarrow \cos{\frac{\pi}{2}}<\cos(\frac{2\pi}{3}-\theta^f)<
\cos(\frac{2\pi}{3}-\theta^i)<\cos{\frac{\pi}{3}}\\
\Rightarrow 0< \cos(\frac{2\pi}{3}-\theta^f)<
\cos(\frac{2\pi}{3}-\theta^i)< \frac{1}{2}\\
\Rightarrow 0< 2\sqrt{A}\cos(\frac{2\pi}{3}-\theta^f)<
2\sqrt{A}\cos(\frac{2\pi}{3}-\theta^i)<\sqrt{A}\\ \Rightarrow
\frac{1}{3}> \frac{1}{3}\{1 -
2\sqrt{A}\cos{(\frac{2\pi}{3}-\theta^f)}\}>\frac{1}{3}\{1 - 2\sqrt{A} \cos
{(\frac{2\pi}{3}-\theta^i)}\}\\~~~~~~~~~~~~~~~~~~~~~~~~~~~~~~~~ >
\frac{1}{3}(1- \sqrt{A})
\end{array}
\end{equation}

i.e., $\frac{1}{3}>\beta_3>\alpha_3>\frac{1}{3}(1-\sqrt{A})$.

Thus, the eigenvalues of $\rho^i_A,~\rho^{f}_A$ are related as:
\begin{equation}
\begin{array}{lcl}
\alpha_1>\beta_1>\beta_3>\alpha_3>\alpha_2>\beta_2.
\end{array}
\end{equation}
So, by Eq.(\ref{incom}), the states $|\Omega\rangle_{AB},$
$|\Omega^f\rangle_{AB}$ are incomparable in this region.\\

In a similar manner we have investigated the other regions and
we find incomparability between the initial and final bipartite
states. Results of the other three possible regions for $0<B'<B,$
will be as follows:\\
When, $3\theta^i, 3\theta^f~\in (\pi,\frac{3\pi}{2})$ then,
$\alpha_1>\beta_1>\beta_2>\alpha_2>\alpha_3>\beta_3.$\\
When, $3\theta^i\in(\frac{\pi}{2},\pi)$ and $3\theta^f\in
(\pi,\frac{3\pi}{2})$ then,
$\alpha_1>\beta_1>\beta_2>\alpha_3>\alpha_2>\beta_3$.\\
When, $3\theta^i\in(\pi,\frac{3\pi}{2})$ and $3\theta^f\in
(\frac{\pi}{2},\pi)$ then,
$\alpha_1>\beta_1>\beta_3>\alpha_2>\alpha_3>\beta_2.$\\

Otherwise, $B'<0<B,$ and we have,
$\cos(3\theta^f)>0>\cos(3\theta^i)$, which implies $3\theta^i~\in
(\frac{\pi}{2},\frac{3\pi}{2})$ and $3\theta^f~\in
\{(0,\frac{\pi}{2})\bigcup(\frac{3\pi}{2},2\pi)\}$. The following
subcases for different regions of $\theta^i, \theta^f$
are considered separately.\\
When, $3\theta^i\in (\frac{\pi}{2},\pi)$ and $3\theta^f\in
(0,\frac{\pi}{2})$ then,
$\alpha_1>\beta_1>\beta_3>\alpha_3>\alpha_2>\beta_2.$\\
When, $3\theta^i\in (\frac{\pi}{2},\pi)$ and $3\theta^f\in
(\frac{3\pi}{2},2\pi)$ then,
$\alpha_1>\beta_1>\beta_2>\alpha_3>\alpha_2>\beta_3.$\\
When, $3\theta^i\in(\pi,\frac{3\pi}{2})$ and $3\theta^f\in
(0,\frac{\pi}{2})$ then,
$\alpha_1>\beta_1>\beta_3>\alpha_2>\alpha_3>\beta_2.$\\
When, $3\theta^i\in(\pi,\frac{3\pi}{2})$ and $3\theta^f\in
(\frac{3\pi}{2},2\pi)$ then,
$\alpha_1>\beta_1>\beta_2>\alpha_2>\alpha_3>\beta_3.$\\
In all the above cases, the results show that the states
$|\Omega\rangle_{AB}$ and $|\Omega^f\rangle_{AB}$ are incomparable
in nature.

Equations (\ref{initial}) and (\ref{final}) will be identical (and
hence the eigen vectors of $\rho^i_A$ and $\rho^{f}_A$) only when
$B=B'$, which imply $abcd \sin\theta=0$, i.e., the three input
states $|0\rangle, |\psi\rangle, |\phi\rangle$, on which the
flipping machine is defined, will lie on one great circle of the
Bloch sphere. This is clear from the fact that there exists exact
flipping machine for the set of states taken from one great circle.

Thus, if the exact flipping machine does exist, and is applied
locally on one subsystem of the initial pure bipartite state, then
an impossible transformation is shown to occur. Obviously this
impossibility comes through our assumption on the existence of
universal exact flipping machine. It is interesting to observe that
the arbitrary phase factor of the flipping operation we have
considered, does not make a difference in the result obtained.

\section{Possible physical reason behind this connection}

In this chapter, we have considered first two examples by which we are
able to show, how an impossible local operation is connected with
the restrictions imposed on state transformations by LOCC. One of
the root of this connection is the anti-unitary nature of the
exact universal spin-flipping operation. In the case of state
transformation criterion, the allowed local operations on each
parties are such that as a whole it can be
implemented by an unitary evolution and if we restrict further
individual local operations as unitary then bipartite entanglement
cannot be changed under such local unitary operations. We, however
considered here a local operation which is anti-unitary and it is observed
that anti-unitary operator acts in a nonphysical way.

Also, the above results show an interesting interplay between the notion of
incomparability and no-flipping principle. It indicates no-flipping
can be used to determine the interrelations between LOCC and
entanglement behavior of the quantum system. We observe that the
incomparability criterion of local state transformations is also
capable of revealing some more fundamental properties of the quantum
systems. It can detect operations which are nonphysical in nature,
such as, here it is anti-unitary. Naturally one could conjecture
that the two impossibilities are equivalent, as they both require
anti-unitary operators. Our results support this conjecture. It also
exhibit the impossibility of extending LOCC operations to
incorporate anti-unitary operators which can create an increase of
information content of the system, as anti-parallel spin states
contain more information than that of the parallel ones. In the next
chapter, we will find incomparability as a new detector of impossible
operations.

\chapter{General Impossible Operations}\footnote{Some portions of this chapter is
published in Quantum Information and Computation, {\bf 7}(4), 392
(2007).}

\section{Physical Operations and LOCC}

Quantum systems allow physical operations to perform some tasks that
seems to be impossible in classical domain. However, varying with the
nature of the operations performed, there are several restrictions
imposed on correctness or exact behavior of the operations to act
for the whole class of states of the quantum system. Possibilities
or impossibilities of various kind of such operations acting on some
specified system is thus one of the basic tasks of quantum
information processing \cite{46b}. In case of cloning and deleting,
the input states must be orthogonal to each other for the exactness
of the operations performed. Rather, if the operation considered is
spin-flipping or Hadamard type then the allowable set of input
states enhanced to the infinite set of qubits defined by any great
circle of the Bloch sphere. It indicates that any angle preserving
operation has some restrictions imposed on the set of allowable
input states. The unitary nature of all physical evolution raised
the question that whether the non-physical nature of the
anti-unitary operations is a natural constraint over the system or
not. In other words, it is nice to show how an impossible operation
like anti-unitary, evolve with the physical systems concerned.

First part of this chapter concerned with a connection between
general anti-unitary operations and evolution of a joint system
through local operations together with classical communications, in
short, LOCC. Some constraint over the system are always imposed by the
condition that the system is evolved under LOCC. Such as, performing
any kind of LOCC, the amount of entanglement between some spatially
separated subsystems can not be increased. If we further assume that
the concerned system is pure bipartite, then by Nielsen's criteria
it is possible to determine whether a pure bipartite state can be
transformed to another pure bipartite state with certainty by LOCC
or not. Consequently, we find earlier that there are pairs of pure
bipartite states, known as incomparable states which are not
interconvertible by LOCC with certainty. The existence of such class
of states prove that the amount of entanglement content does not
always determine the possibility of exact transformation of a joint
system by applying LOCC. Now, we first pose the problem of detecting
the possibility of existence of some operations. In this chapter, we
will able to detect some single qubit impossible operations.

Suppose, $\rho_{ABCD\ldots}$ be a state shared between distinct
parties situated at distant locations. They are allowed to do local
operations on their subsystems and also they may communicate any
amount of classical information among themselves. But they do not
know whether their local operations are valid physical operations or
not. By valid physical operation, we mean a completely positive map
(may be trace-preserving or not) acting on the physical system.
Sometimes an operation is confusing in the sense that it works as a
valid physical operation for a certain class of states but not as a
whole. Therefore, they want to judge their local operations using
quantum formalisms or with other physical principles, may be along with
quantum formalism or may not be. No-signalling, non-increase of
entanglement by LOCC are some of the good detectors to detect
nonphysical operations. In this chapter, we want to establish another
good detector for a large number of nonphysical operations. The
existence of incomparable states enables us to find this good
detector. Suppose, $L_A\otimes L_B \otimes L_C\otimes L_D\otimes
\cdots$ be an operation acting on the physical system described by
the state $\rho_{ABCD\ldots}$ and $\rho^{\prime}_{ABCD\ldots}$ be the
transformed state. Now, if it is known that the states
$\rho_{ABCD\ldots}$ and $\rho^{\prime}_{ABCD\ldots}$ are
incomparable by the action of any deterministic LOCC, then we can
certainly say that at least one of the operations $L_A, L_B, L_C,
L_D, \cdots$ are nonphysical. Therefore, if somehow we find two
states that are incomparable and an operation acting on the local
system of any party (or a number of parties) of one state, can
transform it to another state, then we certainly claim that the
operation is a nonphysical one.

\subsection{Separable Superoperator, LOCC and impossible operations}
In this section, we recall the notion of a physical operation
in the sense of Kraus \cite{Kraus}. Suppose a physical system is
described by a state $\rho$. By a physical operation on $\rho$, we
mean a completely positive map $\mathcal{E}$ acting on the system
and described by
\begin{equation}
\mathcal{E}(\rho )= \sum_{k} A_{k} \rho A^{\dagger}_{k}
\end{equation}
where each $A_k$ is positive linear operator that satisfies the
relation $ \sum_{k}  A^{\dagger}_{k} A_{k} \leq I$. If $ \sum_{k}
A^{\dagger}_{k} A_{k} = I$, then the operation is known trace
preserving. When the state is shared between a number of parties,
say, A, B, C, D,. .... and each $A_k$ has the form $A_k = L^A_k
\otimes  L^B_k \otimes L^C_k \otimes L^D_k \otimes \cdots$ with all
the $L^A_k, L^B_k, L^C_k, L^D_k, \cdots$ are linear positive
operators, the operator is then called as a separable superoperator.
In this context, we want to mention an interesting result concerned
with LOCC. Every LOCC is a separable superoperator but it is not
known whether the converse is also true or not. It is affirmed that
there are separable superoperators which cannot be expressed by
finite LOCC \cite{nwe}. Now, if a physical system evolved under LOCC
(may be deterministic or stochastic) then quantum mechanics does not
allow the system to behave arbitrarily. More precisely, under the
action of any LOCC one fundamental constraint arises for any
entangled system. The content of entanglement will not increase
under LOCC. This is usually known as principle of non-increase of
entanglement under LOCC. Further, for any closed system as unitarity
is the only possible evolution, the constraint is then: the
entanglement content will not change under LOCC. So, if we find some
violation of these principles under the action of any local
operation, then we certainly claim that the operation is not a
physical one. No-flipping \cite{flip}, no-cloning, no-deleting, all
those theorems are already established with these principles,
basically with the principles of non-increase of entanglement. These
kind of proof for those important no-go theorems will always give us
a more powerful physically intuitive approaches for quantum
information processing, apart from the mathematical proofs that the
dynamics should be linear as well as unitary. Linearity and
unitarity are the building blocks of every physical operation. But
within the quantum formalism we always search for more and more new
physical situations that are much useful and intuitive for quantum
information processing. Existence of incomparable states in pure
bipartite entangled states allow us to use it as a new detector. In
particular, we have proved three impossibilities, viz., exact
cloning, deleting \cite{12a} and flipping operations
\cite{incomflip} by the existence of incomparable states under LOCC.

\subsection{Two Examples of Impossible operations}

\emph{Spin flipping:} Exact flipping operation $\Omega$ acting on a
arbitrary qubit, $|\psi \rangle=\alpha |0 \rangle+\beta |1
\rangle~; ~~~|\alpha|^2+ |\beta|^2=1$ is defined by,

$$|\phi \rangle~=~\Omega|\psi \rangle~=~\overline{|\psi \rangle}$$
where, $\langle \phi |\psi \rangle =~0$, i.e., $|\phi
\rangle~=~\beta^* |0 \rangle-\alpha^* |1 \rangle$.

This operation is not in general a physical operation. The impossibility of the
operation occurs due to arbitrariness of the input states. i.e., it works exactly
for a class of states, but not for the whole class.

\emph{Hadamard gate:} Universal Hadamard gate $\Lambda$ acting on
the arbitrary qubit $|\psi \rangle$ defined by,
\begin{equation}
|\phi \rangle=\Lambda |\psi \rangle= \frac{1}{\sqrt{2}}(|\psi
\rangle+i\overline{|\psi \rangle})
\end{equation}

This is also an impossible operation for single-qubit
system \cite{46b}.

\section{Anti-Unitary operators and Incomparability}
{\bf General class of anti-unitary operations:} A general class of
anti-unitary operations can be defined in the form, $\Gamma= CU;$
where $C$ is the conjugation operation and $U$ be the most general
type of unitary operation on a qubit, in the form
$$U=\left(%
\begin{array}{cc}
  \cos\theta & e^{i\alpha}\sin\theta \\
  -e^{i\beta}\sin\theta & e^{i(\alpha+\beta)}\cos\theta \\
\end{array}%
\right)$$

Let us consider three qubit states with the spin-directions along
$x,y,z$ as, $$|0_x
\rangle=\frac{|0\rangle+|1\rangle}{\sqrt{2}}~,~|0_y
\rangle=\frac{|0\rangle+i|1\rangle}{\sqrt{2}}~,
|0_z\rangle=|0\rangle.$$

The action of the operator $\Gamma$ on these three states can be
described as,
\begin{equation}
\begin{array}{lcl}
\Gamma|0_x\rangle&=& (\frac{\cos \theta+ e^{-i \alpha}\sin
\theta}{\sqrt{2}})|0\rangle
+e^{-i \beta}(\frac{e^{-i \alpha} \cos\theta-\sin\theta}{\sqrt{2}})|1\rangle , \\
\Gamma|0_y\rangle&=& (\frac{\cos \theta-i e^{-i \alpha}\sin
\theta}{\sqrt{2}})|0\rangle -e^{-i \beta}(\frac{ie^{-i \alpha}
\cos\theta+\sin\theta}{\sqrt{2}})|1\rangle,\\
\Gamma |0_z \rangle&=& \cos \theta |0\rangle -e^{-i \beta} \sin
\theta |1\rangle
\end{array}
\end{equation}

\subsection{Relation with incomparability}

To prove that the operation $\Gamma$ is nonphysical and its
existence leads to an impossibility, we choose particular pure
bipartite state $|\chi^i \rangle_{AB}$ shared between two spatially
separated parties Alice and Bob in the form,
\begin{equation}
\begin{array}{lcl}|\chi^i\rangle_{AB} & = &\frac{1}{\sqrt{3}} \{|0\rangle_A|0_z
\rangle_B | 0_z \rangle_B + |1 \rangle_A | 0_x \rangle_B | 0_y
\rangle_B\\ & &~ ~+ |2 \rangle_A |0_y \rangle_B |0_x \rangle_B\}
\end{array}
\end{equation}

The impossibility we want to show here is that by the action of
$\Gamma$ locally we are able to convert a pair of incomparable
states deterministically. Now, to show incomparability between a pair
of pure bipartite states, the minimum Schmidt rank we require is
three. So, the joint state we consider above is a $3\times 4$ state,
where Alice has a qutrit and Bob has two qubits. The initial reduced
density matrix of Alice's side is then,
\begin{equation}
\begin{array}{lcl}
\rho_A^i &=&\frac{1}{3}~ \{P[|0\rangle]+P[|1\rangle]+P[|2\rangle]+
\frac{1}{2}(|0\rangle\langle1|+|1\rangle\langle0|\\
& & ~ ~ +|0\rangle\langle2| +|2\rangle\langle0|+|1\rangle\langle2|
+|2\rangle\langle1|) \}
\end{array}
\end{equation}

The Schmidt vector corresponding to the initial state $|\chi^i
\rangle_{AB}$ is $(\frac{2}{3}, \frac{1}{6}, \frac{1}{6} )$.
Assuming that Bob operates $\Gamma$ on one of his two qubits, say,
on the last qubit, the joint state shared between Alice
and Bob will transform to:
\begin{equation}
\begin{array}{lcl}
|\chi^f \rangle_{AB}&=& \frac{1}{\sqrt{3}} \{|0\rangle_A |0_z
\rangle_B \Gamma(| 0_z \rangle_B) + |1 \rangle_A | 0_x \rangle_B
\Gamma(| 0_y \rangle_B)\\ & & ~ ~+ |2 \rangle_A |0_y \rangle_B
\Gamma(|0_x \rangle_B)\}
\end{array}
\end{equation}

Tracing out Bob's subsystem we again consider the reduced density
matrix of Alice's subsystem. The final reduced density matrix is
\begin{equation}
\begin{array}{lcl}
\rho_A^f &=&\frac{1}{3}~ \{P[|0\rangle]+P[|1\rangle]+P[|2\rangle]+
\frac{1}{2}(|0\rangle\langle1|+|1\rangle\langle0|\\
& & ~ ~+|0\rangle\langle2| +|2\rangle\langle0|-i|1\rangle\langle2|
+i|2\rangle\langle1|) \}
\end{array}
\end{equation}

The Schmidt vector corresponding to the final state $|\chi^f
\rangle_{AB}$ is $(\frac{1}{3}+\frac{1}{2\sqrt{3}}~, \frac{1}{3}~,$
$ \frac{1}{3}-\frac{1}{2\sqrt{3}})$. Interestingly, the Schmidt
vector of the final state does not contain the arbitrary parameters
of anti-unitary operator $\Gamma$. It is now easy to check that the
final and initial Schmidt vectors are incomparable as,
$\frac{2}{3}~> \frac{1}{3}+\frac{1}{2\sqrt{3}}~> \frac{1}{3}~>
\frac{1}{6}~> \frac{1}{3}-\frac{1}{2\sqrt{3}}$. Thus we have,
$|\chi^i \rangle \not \leftrightarrow |\chi^f \rangle$ so that the
transformation of the pure bipartite state $|\chi^i \rangle$ to
$|\chi^f \rangle$ by LOCC with certainty is not possible following
Nielsen's criteria. Though by applying the anti-unitary operator
$\Gamma$ on Bob's local system the transformation $|\chi^i \rangle
\rightarrow |\chi^f \rangle$ is performed exactly. This
impossibility emerges out of the impossible operation $\Gamma$ which
we have assumed to be exist and apply it to generate the impossible
transformation.

If, instead of operating $\Gamma= CU$, we will operate only $U,$
\emph{i.e.}, the general unitary operator, the initial and final
density matrices of one side will be seen to be identical, implying
that there is not even a violation of No-Signalling principle. This
is true as we only operate the unitary operator on any qubit not
restricting on any particular choices, such as they will act
isotropically for all the qubits, etc. Thus it can not
even used to send a signal here.\\
As a particular case, we have verified the non-existence of exact
universal flipper by our above method, where we have chosen,
$\theta = \pi/2, \alpha = 0, \beta =0$.

\subsubsection{If input states lying
in one great circle:} Now, if the three input states are chosen from any
great circle of the Bloch sphere, then the anti-unitary operator
$\Gamma$ defined on them will exists uniquely, which is very
natural, as it can be generated by a 180 degree rotation of the
Bloch sphere about the axis perpendicular to that great circle.

Thus, we are now able to detect non-physical nature of anti-unitary
operators through the existence of incomparable pairs of pure entangled
states. In the next section, we will consider another class of operations,
viz., inner product preserving operations.

\section{Universal inner product preserving operations}

There is an intrinsic relation between incomparability with the
impossibility of some inner product preserving operations defined
only on the minimum number of qubits $|0_x \rangle, |0_y \rangle,
|0_z \rangle$. Here, we consider the existence of the operation
defined on these three qubits in the following manner,
\begin{equation}
\begin{array}{lcl}
|0_z \rangle &\longrightarrow & (\alpha
|0_z\rangle + \beta |1_z\rangle),\\
|0_x\rangle &\longrightarrow & (\alpha
|0_x\rangle + \beta |1_x\rangle), \\
|0_y\rangle &\longrightarrow & (\alpha |0_y\rangle + \beta
|1_y\rangle),\label{angle}%
\end{array}
\end{equation}
where $|\alpha|^2 + |\beta|^2 =1.$

This operation exactly transforms the input qubit into an arbitrary
superposition of the input qubit with its orthogonal one. To verify
the possibility or impossibility of existence of this operation as
physical one, we consider a pure bipartite state shared between
Alice and Bob as follows:
\begin{equation}
\begin{array}{lcl}|\Pi^i\rangle_{AB}& = & \frac{1}{\sqrt{3}} \{|0\rangle_A (|0_z \rangle
| 0_z \rangle)_B + |1 \rangle_A (| 0_x \rangle | 0_x \rangle)_B +|2
\rangle_A (|0_y \rangle |0_y \rangle)_B\}
\end{array}
\end{equation}

Then, the reduced density matrix of Alice's side will be of the form,
\begin{equation}
\begin{array}{lcl}
\rho_A^i &=&\frac{1}{3}~ \{P[|0\rangle]+P[|1\rangle]+P[|2\rangle]+
\frac{1}{2}(|0\rangle\langle1|+|1\rangle\langle0|\\
& & ~ ~+|0\rangle\langle2| +|2\rangle\langle0|-i|1\rangle\langle2|
+i|2\rangle\langle1|) \}
\end{array}
\end{equation}
Therefore, the Schmidt vector corresponding to the initial joint state $|\Pi^i
\rangle_{AB}$ is, $(\frac{1}{3}+\frac{1}{2\sqrt{3}}~, \frac{1}{3}~,
\frac{1}{3}-\frac{1}{2\sqrt{3}})$.

Suppose, Bob has now a machine that can operate on the three input qubits $|0_x
\rangle, |0_y \rangle, |0_z \rangle$ as defined in Eq. (\ref{angle}).
He operates that machine on a part of his local subsystem (say, on
the last qubit). Then, the joint state between Alice and Bob will
be of the form:
\begin{equation}
\begin{array}{lcl}
|\Pi^f \rangle_{AB}&=& \frac{1}{\sqrt{3}} \{|0\rangle_A |0_z
\rangle_B (\alpha |0_z\rangle + \beta |1_z\rangle)_B + |1 \rangle_A
| 0_x \rangle_B
\\& & (\alpha | 0_x \rangle + \beta |1_x\rangle)_B + |2
\rangle_A |0_y \rangle_B (\alpha|0_y \rangle+ \beta |1_y\rangle)_B\}
\end{array}
\end{equation}
And thus, the final reduced density matrix of Alice's side will be
of the form:
\begin{equation}
\begin{array}{lcl}
\rho_A^f &=&\frac{1}{3}~ \{P[|0\rangle]+P[|1\rangle]+P[|2\rangle]+
p(|0\rangle\langle1|+|1\rangle\langle0|)\\
& & ~ ~+q|0\rangle\langle2|
+\overline{q}|2\rangle\langle0|+r|1\rangle\langle2|
+\overline{r}|2\rangle\langle1|) \}
\end{array}
\end{equation}where $p=\frac{1}{2}~\{|\alpha|^2-|\beta|^2
+\alpha~\overline{\beta}+\beta~\overline{\alpha}\}$,
~$q=\frac{1}{2}~\{|\alpha|^2+~i|\beta|^2
+\alpha~\overline{\beta}-i\beta~\overline{\alpha}\}$ and
$r=\frac{1}{2}~\{\alpha~\overline{\beta}+\beta~\overline{\alpha}-i\}$.

To compare the initial and final state, we have to check whether the
initial and final eigenvalues will satisfy either of the relations
of Eq.(\ref{incom}). We write, the final eigenvalue equation as:
\begin{equation}
x^3-3Ax+B=0
\end{equation}
where, $A=\frac{1}{3}(p\overline{p} + q\overline{q} + r\overline{r})\geq0$
and $B=pr\overline{q}+\overline{pr}q$. The eigenvalues $\{\lambda_1,
~\lambda_2,~\lambda_3\}$ can then be expressed (by, Carden's method)
as:
\begin{equation}
\begin{array}{lcl}
\{\lambda_1\equiv\frac{1}{3}[1-2\sqrt{A}\cos(\frac{2\pi}{3}+\theta)],
~\lambda_2\equiv\frac{1}{3}[1-2\sqrt{A}\cos\theta],~
\lambda_3\equiv\frac{1}{3}[1-2\sqrt{A}\cos(\frac{2\pi}{3}-\theta)]\}
\end{array}
\end{equation}
where, $\cos (3\theta )=~\frac{-B}{2\sqrt{A^3}}$. We discuss the matter
case by case.

\textbf{Case-1} : For $\emph{B}<0$, we observe an incomparability
between the initial and final joint states if $A= \frac{1}{4}$. In
case of $A< \frac{1}{4}$, we observe that either there is an
incomparability between the initial and final states or the
entanglement content of the final state is larger than that of the
initial states. Lastly, if $A> \frac{1}{4}$, we also see a case of
incomparability if the condition
$2\sqrt{A}\cos(\frac{2\pi}{3}+\theta)~>~-\frac{\sqrt{3}}{2}$ holds.
Numerical searches also support that for real values of $(\alpha,~\beta)$
incomparability is observed almost everywhere in this region. Details
are as follows:

Now, $\emph{B}<0$  implies $3\theta \in [0,~\frac{\pi}{2}) \bigcup
(\frac{3\pi}{2},~2\pi].$ We analyze this in two parts.

If, $3\theta \in [0,\frac{\pi}{2})$ we have, $\frac{\sqrt{3}}{2}~<~
\cos\theta~
\leq~1~\Rightarrow~\lambda_2~\in~[\frac{1}{3}(1-2\sqrt{A}),\frac{1}{3}(1-\sqrt{3A}))$.
Again, $0~\leq~\theta~<~\frac{\pi}{6}~
\Rightarrow~-\frac{\sqrt{3}}{2} < \cos(\frac{2\pi}{3}+\theta) \leq
-\frac{1}{2} ~\Rightarrow
\lambda_1~\in~[\frac{1}{3}(1+\sqrt{A}),\frac{1}{3}(1+\sqrt{3A}))$.
Finally, $0~\leq~\theta~<~\frac{\pi}{6}~\Rightarrow~
\cos(\frac{2\pi}{3}-\theta)\in [-\frac{1}{2},0) ~\Rightarrow
\lambda_3~\in~(\frac{1}{3},\frac{1}{3}(1+\sqrt{A})]$.

Otherwise, $~3\theta \in (\frac{3\pi}{2},2\pi],$ i.e., $\theta \in
(\frac{\pi}{2},\frac{2\pi}{3}]$, we have,
$\lambda_3~\in~[\frac{1}{3}(1-2\sqrt{A}),\frac{1}{3}(1-\sqrt{3A}))$,
$\lambda_2~\in~[\frac{1}{3}(1+\sqrt{A}),\frac{1}{3}(1+\sqrt{3A}))$
and $\lambda_1~\in~(\frac{1}{3},\frac{1}{3}(1+\sqrt{A})]$.

So, in both the cases,
$\lambda^f_{MAX}~\in~[\frac{1}{3}(1+\sqrt{A}),
\frac{1}{3}(1+\sqrt{3A}))$
and \\$\lambda^f_{MIN} ~\in~ [\frac{1}{3}(1-2\sqrt{A}), \frac{1}{3}(1-\sqrt{3A}))$.\\

Thus, for $A=\frac{1}{4}$, we observe that
${\lambda^f}_{MIN}\in[0,\frac{1}{3}(1-\frac{\sqrt{3}}{2}))<{\lambda^i}_{MIN}$
and
${\lambda^f}_{MAX}\in[\frac{1}{2},\frac{1}{3}(1+\frac{\sqrt{3}}{2}))<{\lambda^i}_{MAX},$
which implies $|\Pi^i\rangle_{AB}~,|\Pi^f\rangle_{AB}$ are
incomparable.

If, $A~<~\frac{1}{4}$ then, ${\lambda^f}_{MAX}\leq {\lambda^i}_{MAX}$.
So, in case of ${\lambda^f}_{MIN}\leq {\lambda^i}_{MIN}$, the states
$|\Pi^i\rangle_{AB}~,~|\Pi^f\rangle_{AB}$ are incomparable,
otherwise, we have ${\lambda^f}_{MIN}\geq {\lambda^i}_{MIN}$, and then
$E(|\Pi^i\rangle_{AB})~<~E(|\Pi^f\rangle_{AB})$. For real values of
$\alpha,~ \beta$, we can express $A$,$B$ as:
\begin{equation}
\begin{array}{lcl}
~ ~\emph{A}= \frac{1}{4}+ \frac{1}{6}[ 2{\alpha}^2{\beta}^2+
3\alpha\beta({\alpha}^2-{\beta}^2)] \\
\emph{B}= \frac{\beta}{4} ({\alpha}^2-{\beta}^2+ 2\alpha\beta)
[\alpha(2{\alpha}^2+1)+ \beta({\alpha}^2-{\beta}^2)] \label{AB}%
\end{array}
\end{equation}
Numerical evidences support that for real $\alpha$, $\beta$ most of
the cases show incomparability between
$|\Pi^i\rangle_{AB}~,~|\Pi^f\rangle_{AB}$.

Lastly if $A~\geq~\frac{1}{4}$, then
${\lambda^i}_{MAX}~<~{\lambda^f}_{MAX}$. Thus incomparability
between $|\Pi^i\rangle_{AB}~,~|\Pi^f\rangle_{AB}$ will hold if
${\lambda^i}_{MIN}~<~{\lambda^f}_{MIN}$. For this, we get the
condition that $2\sqrt{A}\cos\phi~<~\frac{\sqrt{3}}{2}$, where
$\phi=\min\{\theta,~\frac{2\pi}{3}-\theta\}\in(\frac{\pi}{6},\frac{\pi}{3})$.
For real values of $\alpha,\beta$ from Eq.(\ref{AB}), we see that
$A>\frac{1}{4}$ implies
$\alpha\beta({\alpha}^2-{\beta}^2)>0~\Rightarrow~B>0$.
Thus, for real $\alpha,\beta$, this subcase does not arise.

\textbf{Case-2} : For $\emph{B}=0$, we found that there does not arise
a case of incomparability. It is also seen that there is always an
increase of entanglement by LOCC if $A< \frac{1}{4}$, which is the
only possibility for real values of $\alpha,\beta$.

Here, the final eigenvalues are $\{\frac{1}{3}(1+\sqrt{3A}),
~\frac{1}{3},~\frac{1}{3}(1-\sqrt{3A})\}.$ Thus,
$E(|\Pi^i\rangle_{AB})~\geq~E(|\Pi^f\rangle_{AB})$  for $A~\geq
\frac{1}{4}$. Incomparability between the initial and final joint
states $|\Pi^i\rangle_{AB}~,|\Pi^f\rangle_{AB}$ will not occur in
this case.

Hence, for all values of $\alpha,\beta$ for which $A<\frac{1}{4}$,
there is an increase of entanglement by applying the local operation
defined in Eq.(\ref{angle}) in Bob's system. This impossibility indicate
the impossibility of the operation defined in Eq.(\ref{angle}) for those
values of $\alpha,\beta$ which satisfy $A~<~\frac{1}{4}$. And for real
values of $\alpha, \beta$, in all possibilities for $\emph{B}=0$
we have $\emph{A}~<~\frac{1}{4}$.
This case also always shows an increase of entanglement.

\textbf{Case-3 : $\emph{B}>0$.} In this case, we also find similar
results like Case-1, only the condition for incomparability in case
$A> \frac{1}{4}$ if changed to the form $2\sqrt{A}\cos\varphi <
\frac{\sqrt{3}}{2}$, where
$\varphi=\min\{\theta,~(\frac{2\pi}{3}-\theta)\}\in(\frac{\pi}{6},~\frac{\pi}{3}).$
It must be noted here that for real values of $\alpha,\beta$ this subcase
do not arise at all. Now, $3\theta \in
(\frac{\pi}{2}~,\frac{3\pi}{2}) \Rightarrow \theta \in
(\frac{\pi}{6}~,\frac{\pi}{2}) ~\Rightarrow \cos \theta \in
(\frac{\sqrt{3}}{2}~ ,0)\Rightarrow
\lambda_2\in(\frac{1}{3}(1-\sqrt{3A}),\frac{1}{3})$. Again, $\theta
\in (\frac{\pi}{6}~,\frac{\pi}{2})\Rightarrow
\cos(\frac{2\pi}{3}+\theta)\in (-1,-\frac{\sqrt{3}}{2})
\Rightarrow\lambda_1\in(\frac{1}{3}(1+\sqrt{3A})~,\frac{1}{3}(1+2\sqrt{A}))$.
Lastly, $\theta \in (\frac{\pi}{6},\frac{\pi}{2}) ~\Rightarrow
\cos(\frac{2\pi}{3}-\theta)\in(0,\frac{\sqrt{3}}{2})~\Rightarrow
\lambda_3\in (\frac{1}{3}(1-\sqrt{3A}),\frac{1}{3})$.

Hence, in this case,
$\lambda^f_{MAX}~\in~(\frac{1}{3}(1+\sqrt{3A}),\frac{1}{3}(1+2\sqrt{A}))$
and $\lambda^f_{MIN} ~\in
(\frac{1}{3}(1-\sqrt{3A}),\frac{1}{3})$.
Thus, for $A~=~\frac{1}{4}$ we have,
${\lambda^i}_{MAX}~<~{\lambda^f}_{MAX}$ and
${\lambda^i}_{MIN}~<~{\lambda^f}_{MIN}$. These two relations
together implies that $|\Pi^i\rangle_{AB}~,~|\Pi^f\rangle_{AB}$ are
incomparable.

Also, for $A~\leq~\frac{1}{4}$ we see,
${\lambda^f}_{MIN}~>~{\lambda^i}_{MIN}$. Thus, if
${\lambda^f}_{MAX}~>~{\lambda^i}_{MAX}$, then the states
$|\Pi^i\rangle_{AB}~,~|\Pi^f\rangle_{AB}$ are incomparable or, if
${\lambda^f}_{MAX}~<~{\lambda^i}_{MAX}$, then
$E(|\Pi^i\rangle_{AB})~<~E(|\Pi^f\rangle_{AB})$.

Lastly, if $A~\geq~\frac{1}{4}$, then
${\lambda^i}_{MAX}~<~{\lambda^f}_{MAX}$. Incomparability between the
initial and final joint states
$|\Pi^i\rangle_{AB}~,~|\Pi^f\rangle_{AB}$ will hold if
${\lambda^i}_{MIN}~<~{\lambda^f}_{MIN}$. For this we get the
condition that
$2\sqrt{A}\cos(\frac{2\pi}{3}+\theta)~>~-\frac{\sqrt{3}}{2}$. From
Eq.(\ref{AB}) we find, for real values of $\alpha$ and
$\beta$, numerical results also support that in most of the cases there
is an incomparability between $|\Pi^i\rangle_{AB}~,~|\Pi^f\rangle_{AB}$.

In particular, if we check the values of $\alpha, \beta$ be such that
they represent the operations flipping (i.e., $\alpha = 0$) and
Hadamard (i.e., $\alpha = \beta = \frac{1}{\sqrt{2}}$) respectively,
then we find from the above that in both the cases the initial and final
states are incomparable.

In almost all the cases above, we find some kind of violation of
physical laws, which implies that in general, most of the inner product
preserving operations defined only on three states is non-physical in
nature and we observe for a large class of such inner-product-preserving
operation incomparability senses efficiently as a good detector.

\section{Incomparability as a detector of Impossible operations}

Thus, in this chapter together with the previous chapter, we find how
local state transformation criteria and correspondingly the existence
of incomparable states provide a new direction of search for the
possibility and impossibility of various local operations. This is
sometimes more preferable than the constraint on local evolution of
entangled state with the amount of entanglement content. Sometimes the
non-increase of entanglement under LOCC, may become less powerful
in showing such violations. Many known and unknown impossibilities may
be easily detected by our above method \cite{12a}.

\chapter{Bound entanglement-Strange but strong resource}{\footnote{Some portions of this chapter is
published in Physical Review A {\bf 71}, 062317 (2005).}

\section{Bound entanglement exists in nature}

Bound entanglement is quite a similar notion with that of bound
energy in classical physics. This is a peculiar kind of
characteristic, observed in bi-partite as well as multi-partite systems.
Pure state entanglement in bipartite case, is reversible in nature.
For any pure bipartite entangled state $|\phi\rangle_{AB}$, it is
physically possible to extract out all the entanglement required to
prepare the state, i.e.,
$E_F(|\phi\rangle_{AB})= E_D(|\phi\rangle_{AB})= E(|\phi\rangle_{AB})$.
This feature is remarkably absent in mixed state level. There are strong
physical restrictions on distillation process to recover all the
entanglement for some mixtures. If a state is entangled, i.e.,
the entanglement cost of preparing such states is non-zero, but
it is not possible to distill out any positive amount of
entanglement from the state as long as the parties sharing the state
will remain specially separated and are allowed to perform only
LOCC, then the state is said to be \emph{Bound entangled}. The
notion of bound entanglement was first shown by Horodecki
\emph{et.al.} in system of $3\times 3$ states \cite{horb}.

We first discuss the example provided by P. Horodecki \cite{31}
that turns out to be a bound entangled state in $3\times 3$ dimension.
The state has the form,
\begin{equation}
\begin{array}{lcl}
\varrho_a = \frac{8a}{8a+1} \varrho_{ins} +
\frac{1}{8a+1}|\phi_a\rangle\langle\phi_a|
\end{array}
\end{equation}where $ \varrho_{ins} $ is an inseparable state (as it
has negative partial transpose) described by,
\begin{equation}
\begin{array}{lcl}
\varrho_{ins} =\frac{1}{8}\{I + P[|00\rangle+|11\rangle+|22\rangle]
-P[|00\rangle]-P[|11\rangle]-P[|22\rangle]-P[|20\rangle]\}
\end{array}
\end{equation}and $|\phi_a\rangle$ be a pure product state defined
as,
\begin{equation}
\begin{array}{lcl}
|\phi_a\rangle=|2\rangle \otimes (\sqrt{\frac{1+a}{2}}|0\rangle +
\sqrt{\frac{1-a}{2}}|2\rangle),~~0\leq a \leq 1
\end{array}
\end{equation}
Now, the state $\varrho_a$ is found to be PPT, but entangled. This
also shows that PPT criteria is only sufficient but not a necessary
condition for separability. Again, it is proved that every PPT state
is undistillable \cite{horb}. Thus, existence of PPT entangled state
immediately provides the example of states, that are enriched with
some resources of entanglement, but it is truly impossible to
extract out any non-negative amount of entanglement from those
states. The states of this kind are certainly bound entangled. Next,
we will show a constructive way to find bound entangled states,
provided by Bennett \emph{et. al.} \cite{upbBennett}. Before going to
discuss this matter, we first describe the aspect of local
distinguishability of states which have a close connection with the
bound entangled states.

\subsection{Distinguishing entangled states by LOCC}
In the first chapter, we have described some notions of non-orthogonal
states and the aspect of discriminating set of states of a single
quantum system. It is also discussed that the concept of
non-orthogonality, exact distinguishability and exact cloning are
physically equivalent. If we consider any composite system (in other
words, a non-local system empowered with the source of quantum
communication allowed between its subsystems) as a whole (i.e.,
globally), the situation appears to be similar with a single system.
The situation is more complex in case of composite systems shared
between different parties situated at distant locations where it
becomes difficult to create a global set up to
distinguish a set of states. Rather, it is much more desirable to
discriminate a set of mutually orthogonal quantum states belonging
to some composite systems, by allowing only local quantum operations
on the subsystems together with the classical communications between
different locations (i.e., by LOCC). Though performing a task via
LOCC sometimes seen to be unsatisfactory. The physical reasons
behind this notion are the constraints imposed on possible evolution
of a physical system under local operations. For example,
amount of entanglement can not be increased under LOCC. In other words,
if we allow only some restricted set of operations, then it
is reasonable to accomplish less satisfactory results for some
specific tasks \cite{53,52}. Naturally we expect that
such differences in achievement by using global or local operations
would only be reflected in some entangled system, but not in any
separable set up. The first result we will discuss in this respect is quite
satisfactory indeed \cite{53,52}. It is proved that there are
orthogonal composite quantum states for which the task of
discrimination can be achieved locally, are as good as with global
discrimination processes.\\

\textbf{Local distinguishability of Pure states:}
Walgate \emph{et.al.} \cite{53} proved that any two pure orthogonal
states $|\psi\rangle,~|\phi\rangle$ bipartite or multipartite,
whether entangled or separable, can always be distinguished with
certainty by LOCC.

For simplicity, we only describe the  bipartite case. Suppose, Alice
and Bob are two persons situated at distant laboratories. Each of
them share a part of some quantum system in one of the two pure
quantum states $|\psi\rangle,~|\phi\rangle$, which are orthogonal to
each other. The task is to determine which state they share, by LOCC
with certainty, if single copy of the states are given.
Walgate \emph{et.al.} \cite{53} showed that any two pure orthogonal
bipartite states can always be represented by (ignoring the
normalization factors),
\begin{equation}
\begin{array}{lcl}
|\psi\rangle &=& |1\rangle_A |\eta_1 \rangle_B + |2\rangle_A |\eta_2
\rangle_B + \cdots +|n\rangle_A |\eta_n \rangle_B\\
|\phi\rangle &=& |1\rangle_A |\upsilon_1 \rangle_B + |2\rangle_A
|\upsilon_2 \rangle_B + \cdots +|n\rangle_A |\upsilon_n \rangle_B
\end{array}
\end{equation}
where $\{|1\rangle_A , |2\rangle_A , \cdots , |n\rangle_A \}$ form
an orthonormal basis for Alice's local system, chosen very
specifically so that $\langle\eta_i |\upsilon_i\rangle=0, ~\forall
~~i=1,2,\cdots,n$. Either of the two sets $\{|\eta_i \rangle_B  ;~
i=1, 2, \cdots,n \}$, $\{|\upsilon_i \rangle_B  ;~ i=1, 2, \cdots,n \}$
may not necessarily constitute set of orthogonal vectors for Bob's
subsystem. Thus, the form of the two states $|\psi \rangle, |\phi \rangle$,
immediately proves the local distinguishability of the states
with certainty.

In multipartite case, the same notion could also be generalized for
local distinguishability of any two pure orthogonal states. The
situation drastically changes if we consider more than only two pure
orthogonal states. In the simplest possible case of $2 \times 2$
system of states, there is an example of four Bell states
$|\Phi^{\pm}\rangle =\frac{1}{\sqrt{2}}(|00\rangle \pm |11 \rangle,
~ ~ |\Psi^{\pm}\rangle = \frac{1}{\sqrt{2}}(|01\rangle \pm |10
\rangle),$ that are not locally distinguishable, if single copy of
the states are given \cite{indisting}. Actually, any three of them
are locally indistinguishable. For pure $2\times 2$ system of
states, the analysis is almost complete \cite{dist2,52}. However,
for higher dimensions, there is no local discriminating procedure by
which we can certainly show that a set of orthogonal pure states of
quantum system are locally distinguishable. For mixed states, the
aspect of discrimination is much more difficult to deal with.
Chefles provided a criteria for perfect discrimination of mixed
states in finite case, incorporating some ideas of entanglement
witness \cite{chefles3}. Thus, observing such kind of peculiarity in
composite quantum systems, we may understood a form of non-locality
plays in quantum systems \cite{nwe,Watrous}. Now, we elaborate some
cases of local indistinguishability further for higher dimensions.

Let us consider the Bennett set of pure orthogonal maximally entangled
states in $3\times 3$ system represented by,
\begin{equation}
\begin{array}{lcl}
|\Psi_1\rangle &=& \frac{1}{\sqrt{3}}\{|0\rangle_A |0 \rangle_B +
|1\rangle_A |1\rangle_B +|2\rangle_A |2 \rangle_B\}\\
|\Psi_2\rangle &=& \frac{1}{\sqrt{3}}\{|0\rangle_A |1 \rangle_B +
|1\rangle_A |2\rangle_B +|2\rangle_A |0 \rangle_B\}\\
|\Psi_3\rangle &=& \frac{1}{\sqrt{3}}\{|0\rangle_A |2 \rangle_B +
|1\rangle_A|0\rangle_B +|2\rangle_A |1 \rangle_B\}\\
|\Psi_4\rangle &=& \frac{1}{\sqrt{3}}\{|0\rangle_A |0 \rangle_B +
\omega|1\rangle_A |1\rangle_B +\omega^2 |2\rangle_A |2 \rangle_B\}\\
|\Psi_5\rangle &=& \frac{1}{\sqrt{3}}\{|0\rangle_A |1 \rangle_B +
\omega|1\rangle_A |2\rangle_B +\omega^2 |2\rangle_A |0 \rangle_B\}\\
|\Psi_6\rangle &=& \frac{1}{\sqrt{3}}\{|0\rangle_A |2 \rangle_B +
\omega|1\rangle_A|0\rangle_B +\omega^2 |2\rangle_A |1 \rangle_B\}\\
|\Psi_7\rangle &=& \frac{1}{\sqrt{3}}\{|0\rangle_A |0 \rangle_B +
\omega^2 |1\rangle_A |1\rangle_B + \omega|2\rangle_A |2 \rangle_B\}\\
|\Psi_8\rangle &=& \frac{1}{\sqrt{3}}\{|0\rangle_A |1 \rangle_B +
\omega^2 |1\rangle_A |2\rangle_B +\omega|2\rangle_A |0 \rangle_B\}\\
|\Psi_9\rangle &=& \frac{1}{\sqrt{3}}\{|0\rangle_A |2 \rangle_B +
\omega^2 |1\rangle_A|0\rangle_B +\omega|2\rangle_A |1 \rangle_B\}
\end{array}
\end{equation}
where $\omega,~\omega^2$ are the distinct cube roots of unity.
In a similar manner, one can also describe the Bennett set of pure
orthogonal maximally entangled states in $d \times d$ system. It is
proved that any set of $d+1$ number of states chosen arbitrarily
from Bennett set of $d^2$ states in $d \times d, ~ d \geq 2$ system,
can not be distiguishable locally if single copy of the states are
given \cite{dist1}. Nathanson later extended this result using the
idea of mutually unbiased bases for bipartite systems
\cite{nathanson}. Fan showed that it is not possible to distinguish
locally $l$ number of pure maximally entangled states if $l(l-1)
\leq 2d$ \cite{fan}. However, it is interesting to note that if two
copies of the set of maximally entangled states are given, then they
are distinguishable by LOCC with certainty \cite{dist3,dist1}.
Related with the local or global distinguishability of set of
states, there is also an important issue of conclusive
discrimination. Conclusive distinguishability relates the complete
discrimination of the states with some probability. Inconclusive
discrimination of a set of states can be described as partly
discriminated set of states where recovery of only a part of the
information about the states is possible. Unlike this case of
distinguishability of states, discrimination of unitary operators is
somehow a less constrained phenomena \cite{acinUnitary,chefles4}.
Now, we will discuss a constructive approach to find bound entangled
states through unextendible product bases, where local
indistinguishability is inherent.

\subsection{Unextendible product bases and bound entanglement}
Naturally, the general notion of quantum non-locality was directly
associated with the existence of entangled states. Surprisingly,
with the discovery of Bennett group \cite{nwe}, this assumption was
found to be wrong. They provide a complete orthonormal bipartite
product basis in $3 \times 3$ system, (i.e., a set of orthogonal
quantum states) that can not be perfectly distinguished by LOCC.
This peculiar phenomena of local indistinguishability is termed as
non-locality of a set of quantum states without entanglement.
Another, very interesting class of local indistinguishable states is
unextendible product basis that also generates bound entangled
states \cite{upbBennett,upb}.

\textbf{Unextendible product basis:} An incomplete product basis of
a multipartite quantum system described in the Hilbert space
$H=~\bigotimes_{i=1}^m H_i$, where the dimension of the $i^{th}$
Hilbert space is $d_i~;~ i=1,2,\cdots,m$, is a set of pure
orthogonal product states spanning a proper subspace $H_s$ of $H$.
Unextendible Product Basis (in short, UPB) is a product basis whose
complementary subspace $H-H_s$ contains no product state
\cite{upbBennett}. In other words, a set of product orthogonal
vectors in $H$ is such that: (a) it has fewer elements than the
dimension of the space, and (b) there does not exist any product
vector orthogonal to all of them is called an unextendible product
basis (UPB).\\

\emph{\textbf{Example}:} Consider the five states of $3 \times 3$
system defined as,
\begin{equation}
\begin{array}{lcl}
|\psi_0 \rangle~=~ \frac{1}{\sqrt{2}}|0 \rangle(|0 \rangle-|1
\rangle)\\
|\psi_1 \rangle~=~ \frac{1}{\sqrt{2}}(|0 \rangle-|1
\rangle)|2\rangle\\
|\psi_2 \rangle~=~ \frac{1}{\sqrt{2}}|2 \rangle(|1 \rangle-|2 \rangle)\\
|\psi_3 \rangle~=~ \frac{1}{\sqrt{2}}(|1 \rangle-|2 \rangle)|0
\rangle\\
|\psi_4 \rangle~=~ \frac{1}{3}(|0 \rangle+|1 \rangle+|2
\rangle)(|0 \rangle+|1 \rangle+|2
\rangle) \label{upb}%
\end{array}
\end{equation}

This set of states forms an unextendible product basis of $3 \times
3$ system. The states are locally indistinguishable
\cite{upbBennett}. There arises a natural question that whether a
complete product basis is always distinguishable by LOCC or not. The
answer is in negative and the criteria determining this result is
given below as a theorem.

\emph{Theorem:} Any complete orthogonal product basis of a bipartite
system $H= H_A \otimes H_B$ is distinguishable by LOCC if and only
if it can not be decomposed into two disjoint sets of states
spanning the subspaces $H_A \otimes H'_B$ and $H_A \otimes
{H'_B}^\bot$ (or, the two subspaces $H'_A \otimes H_B$ and
${H'_A}^\bot \otimes H_B$), in any manner \cite{rinaldis}.

The following result is a constructive way of finding bound
entangled states corresponding to any unextendible product basis .

\emph{Theorem:} \cite{upbBennett} Let $S=\{|\psi_j \rangle \equiv
\bigotimes_{i=1}^m |\varphi_j \rangle ;~ j=1,2,\cdots,n \}$ be an
unextendible product basis of the m-partite system $H= H_1\otimes
H_2 \otimes \cdots \otimes H_m$, such that $\dim{(H)} = d$. Then the
state corresponding to the uniform mixture on the space
complementary to the orthogonal UPB $S$ is a bound entangled state,
described by,
\begin{equation}
\overline{\rho}~=~\frac{1}{d-n}~(1-\sum_{j=1}^n |\psi_j \rangle
\langle \psi_j |)
\end{equation}

\subsection{Notion of bound entanglement in general}
Entanglement is generally defined as a resource of the system and
studied through various protocols of computational tasks performed
on entangled states. Protocols are usually given on Maximally
entangled or other well known states. On other states they usually
act as, first distill out some entanglement in terms of known states
and implement the tasks on such known states. Bound entangled states
are inactive in this sense. This notion is also supported by a work
of Linden \emph{et.al.}, where it is shown that for teleportation
some PPT bound entangled states can not be used to obtain a better
result than classical case \cite{popescu}. It is also demonstrated
that presence of an infinite amount of bound entanglement can not
enhance the amount of distillable entanglement of a state
\cite{Vedral3}. Quite opposite to all those results, it is also
investigated that sometimes bound entanglement could be used even to
perform otherwise impossible tasks \cite{Ishizaka}. For some special
kind of bound entangled states, however, it is possible to perform
some computational tasks by means of some global operations allowed
by quantum mechanics. Such processes of performing tasks on the
states that are seemed to be inactive resource are designated as
different activation processes.

Now, for bipartite systems, bound entanglement is defined clearly.
But, in a multipartite setting, due to several distinct spatially
separated configurations the definition of bound entanglement is
not unique. A multipartite quantum state is said to be bound entangled
if there is no distillable entanglement between any subset of parties as
long as all the parties remain spatially separated from each other
\cite{19}. However, one may allow some of the parties to group together
and perform local operations collectively. It leads us to two qualitatively
different classes of bound entangled states. In 2001, Smolin found
\cite{Smolin} a bound entangled state defined as \emph{Unlockable Bound
Entangled State}, that actually provides us the scheme for classifying
bound entangled states into two different way. They are specified as
activable and inactivable bound entangled states.

\emph{Activable Bound Entanglement:} The states that are not
distillable when every party is separated from each other, but
becomes distillable if certain parties decide to group together.
i.e., there is at least one bipartite cut where the state is
negative under partial transposition (NPT).

\emph{Non-activable Bound Entanglement:} Non-activable bound
entangled states are not distillable under any modified
configuration as long as there are at least two spatially separated
groups. In other words, such states are always positive under partial
transposition across any bipartite partition.

\section{A class of activable bound entangled states}

Recently, we discovered \cite{bcbe} that there is a large class of bound
entangled states in multi-qubit systems. This is a generalization of
the state given by Smolin, but besides of simple generalization we
find some new results about the structure of the concerned
Hilbert spaces. We will now analyze the matter step by step starting
from the cases with lower number of qubits. The states are constructed
by mixing up the four Bell states,
\begin{equation}
\begin{array}{lcl}
|\Phi^{\pm} \rangle =\frac{|00\rangle \pm |11\rangle}{\sqrt{2}}~,~~
|\Psi^{\pm} \rangle =\frac{|01\rangle \pm
|10\rangle}{\sqrt{2}}\label{bell}
\end{array}
\end{equation}which are pure maximally entangled
states of two-qubit system.

\subsection{Bound entangled states in Hilbert space of four qubits}

Consider the equiprobable Bell mixture,
\begin{equation}
\begin{array}{lcl}
\rho_4^+ &=&\frac{1}{4}~ \{P[|\Phi^+\rangle]\otimes
P[|\Phi^+\rangle]+P[|\Phi^-\rangle]\otimes P[|\Phi^-\rangle]+ P[|\Psi^+\rangle]
\otimes P[|\Psi^+\rangle]\\&
&+P[|\Psi^-\rangle]\otimes P[|\Psi^-\rangle] \}\label{smolin}
\end{array}
\end{equation}
which we recognize as the unlockable bound entangled state, presented
by Smolin, where for simplicity, we have used the notation $P[\cdot]$
as projector corresponding to a pure state.

Applying Pauli operators $\sigma_z, \sigma_x, i\sigma_y$ on any one
qubit of $\rho_4^+$, say, on the last qubit, three other four qubit
states of this type will be obtained as follows,
\begin{equation}
\begin{array}{lcl}
\rho_4^- &=&\frac{1}{4}~ \{P[|\Phi^+\rangle]\otimes
P[|\Phi^-\rangle]+P[|\Phi^-\rangle]\otimes P[|\Phi^+\rangle]+
P[|\Psi^+\rangle]\otimes
P[|\Psi^-\rangle]\\&  &+P[|\Psi^-\rangle]\otimes P[|\Psi^+\rangle] \}\\

\sigma_4^+ &=&\frac{1}{4}~ \{P[|\Phi^+\rangle] \otimes P[|\Psi^ +\rangle]+P[|\Phi^-\rangle]
\otimes P[|\Psi^ -\rangle]+ P[|\Psi^+\rangle]\otimes P[|\Phi^
+\rangle]\\&  &+P[|\Psi^-\rangle] \otimes P[|\Phi^ -\rangle] \}\\

\sigma_4^- &=&\frac{1}{4}~ \{P[|\Phi^+\rangle] \otimes P[|\Psi^ -\rangle]+P[|\Phi^-\rangle]
\otimes P[|\Psi^ +\rangle]+ P[|\Psi^+\rangle]\otimes P[|\Phi^ -\rangle]\\&  &+
P[|\Psi^-\rangle]\otimes P[|\Phi^ +\rangle] \}\label{bcbe4bell}
\end{array}
\end{equation}

As the four Bell states are mutually orthogonal then so also the
four states $\rho_4^{\pm}, \sigma_4^{\pm}$. The states can also be
rewritten in the symmetric form,
\begin{equation}
\begin{array}{lcl}
\rho_4^{\pm}  ~  &=&\frac{1}{8}(P[|0000\pm 1111\rangle]+
P[|0011 \pm 1100\rangle]~ +P[|0101\pm 1010\rangle]
+P[| 0110 \pm 1001\rangle]  )\\
\sigma_4^{\pm}  ~ & =&\frac{1}{8}(P[|0001 \pm 1110\rangle]+
P[|0010 \pm 1101\rangle] ~ +P[|0100 \pm 1011\rangle]
+P[|0111 \pm 1000\rangle] )\label{bcbe4sym}
\end{array}
\end{equation}

Following Smolin's argument, it is easy to show that the states are
activable bound entangled. Assume that any one of the four states
$\rho_4^{\pm}, \sigma_4^{\pm}$ are shared among four distant parties
Alice, Bob, Charlie and Daniel. We allow Charlie and Daniel to come
to one lab. Then they will make a projective Bell measurement on their
two qubit subsystem and communicates their results to Alice and Bob
by any classical means. After this measurement Alice and Bob will
able share one Bell state among them. Thus the state $\rho_4^{\pm},
\sigma_4^{\pm}$ are entangled. Now, to express our results in a more
convenient way, we will also use the notations for the four states
$\rho_4^{\pm},\sigma_4^{\pm}$ as $\rho_{ABCD}^{\pm},\sigma_{ABCD}^{\pm}$
and with these notations we again, express the four states as,
\begin{equation}
\begin{array}{lcl}
\rho_{ABCD}^+ &=&\frac{1}{4}~ \{P[|\Phi^+\rangle]_{AB} \otimes
P[|\Phi^+\rangle]_{CD}+P[|\Phi^-\rangle]_{AB} \otimes P[|\Phi^-\rangle]_{CD}\\& &+
P[|\Psi^+\rangle]_{AB}\otimes P[|\Psi^+\rangle]_{CD}+P[|\Psi^-\rangle]_{AB}
\otimes P[|\Psi^-\rangle]_{CD} \}\\

\rho_{ABCD}^- &=&\frac{1}{4}~ \{P[|\Phi^+\rangle]_{AB} \otimes
P[|\Phi^-\rangle]_{CD}+P[|\Phi^-\rangle]_{AB} \otimes P[|\Phi^+\rangle]_{CD}\\& &+
P[|\Psi^+\rangle]_{AB} \otimes P[|\Psi^-\rangle]_{CD}+P[|\Psi^-\rangle]_{AB}
\otimes P[|\Psi^+\rangle]_{CD} \}\\

\sigma_{ABCD}^+ &=&\frac{1}{4}~ \{P[|\Phi^+\rangle]_{AB} \otimes
P[|\Psi^+\rangle]_{CD}+P[|\Phi^-\rangle]_{AB}  \otimes P[|\Psi^ -\rangle]_{CD}\\& & +
P[|\Psi^+\rangle]_{AB}\otimes P[|\Phi^+\rangle]_{CD} +P[|\Psi^-\rangle]_{AB}
\otimes P[|\Phi^ -\rangle]_{CD}\}\\

\sigma_{ABCD}^- &=&\frac{1}{4}~ \{P[|\Phi^+\rangle]_{AB}  \otimes
P[|\Psi^-\rangle]_{CD}+ P[|\Phi^-\rangle]_{AB}  \otimes P[|\Psi^+\rangle]\\& &+
P[|\Psi^+\rangle]_{AB} \otimes P[|\Phi^-\rangle]_{CD} +P[|\Psi^-\rangle]_{AB} \otimes
P[|\Phi^+\rangle]_{CD}\}
\end{array}
\end{equation}
The above expressions immediately tell us, that the four states are
separable in the bipartite cut $AB: CD$. Again, from the permutation
symmetry of the states, the form of the states are identical if we
express them in any combination of the parties like ABDC, ACBD,
BCAD, etc. For example, if we exchange the positions of the parties
$B$ and $C$, then the state $\rho_{4}^+$ will also be represented as,
\begin{equation}
\begin{array}{lcl}
\rho_{ABCD}^+ = \rho_{ACBD}^+ &=&\frac{1}{4}~ \{P[|\Phi^+\rangle]_{AC} \otimes
P[|\Phi^+\rangle]_{BD}+P[|\Phi^-\rangle]_{AC} \otimes P[|\Phi^-\rangle]_{BD}\\& &+
P[|\Psi^+\rangle]_{AC} \otimes P[|\Psi^+\rangle]_{BD}+P[|\Psi^-\rangle]_{AC} \otimes
P[|\Psi^-\rangle]_{BD} \}
\end{array}
\end{equation}
This will imply that the state $\rho_{4}^+$ is also separable in the
bipartite cut $AC:BD$. In a similar manner, we can show that each of
the four states are separable in every possible $2:2$ bipartite cut. Now,
if it could be possible to distill out any non-zero amount of
entanglement from any of the four states by LOCC when the parties are
far apart, then it would certainly violate the result that the states
are separable across any $2:2$ bipartite cut. This shows that the states
$\rho_4^{\pm}, \sigma_4^{\pm}$ are bound entangled, i.e., they are
activable bound entangled.

Now, if we denote the support of the density matrices
corresponding to the four states as the sets $S_1, S_2, S_3, S_4$,
then,
\begin{equation}
\begin{array}{lcl}
S_1  &=&\{|0000+ 1111\rangle,  |0011+ 1100\rangle,  |0101+ 1010\rangle  ,  |0110+ 1001\rangle\}\\
S_2  &=&\{|0000- 1111\rangle,  |0011- 1100\rangle,  |0101- 1010\rangle  ,  |0110- 1001\rangle\}\\
S_3  &=&\{|0001+ 1110\rangle,  |0001+1110\rangle,  |0100+ 1011\rangle  ,  |0111+ 1000\rangle\}\\
S_4  &=&\{|0001- 1110\rangle,  |0001- 1110\rangle,  |0100-1011\rangle  ,  |0111- 1000\rangle\}\\
\end{array}
\end{equation}

The above form shows that the four sets $S_1, S_2, S_3, S_4$
together span the total Hilbert space of four qubits. Also, the sets
are mutually exclusive and equal in size. Thus, the Hilbert space of
the four qubit system has been divided into four disjoint sections, i.e.,
four subspaces mutually orthogonal to each other.

Having the result of four qubit system in our hand, we now gradually
proceed to higher number of parties each holding a qubit system in
a similar geometric way. For this, we try to split the total
Hilbert space by dividing the usual basis in four parts.
Unfortunately, such splitting that may construct bound entangled
states will only seen in even number of qubit systems. For
odd number of qubit systems such Hilbert space symmetry is absent.

\subsection{Bound entangled states in Hilbert space of six qubits}

Following the symmetric structure, it is now easy to construct similar
states of six qubits system. Firstly, the state similar in structure to the
Smolin state can be constructed as,
\begin{equation}
\begin{array}{lcl}
\rho_6^+ &=&\frac{1}{32}~ \{P[|000000+111111\rangle]+P[|000011+111100\rangle]+
P[|001100+110011\rangle] \\ & & +P[|000101+111010\rangle]+
P[|001010+110101\rangle]+P[|000110+111001\rangle]\\ & & + P[|001001+110110\rangle]
+ P[|010001+101110\rangle]+P[|010010+101101\rangle]\\ & & +
P[|010100+101011\rangle]+ P[|011000+100111\rangle]+
P[|001111+110000\rangle]\\ & & +P[|010111+101000\rangle]+
P[|011011+100100\rangle]\\ & &+P[|011101+100010\rangle]+P[|011110+100001\rangle] \}
\end{array}
\end{equation}

If we express the last two qubits of this mixed state in terms of
Bell states then this state can be expressed as Bell mixture of the
four states of four qubit system, i.e.,
\begin{equation}
\begin{array}{lcl}
\rho_{6} ^{+} &=& \frac{1}{4}~ \{ \rho_{4}^{+} \otimes
P[|\Phi^{+}\rangle]+\rho_{4}^{-} \otimes P[|\Phi^{-}\rangle]+
\sigma_{4}^{+} \otimes P[|\Psi^{+}\rangle]\\&  &+\sigma_{4}^{-}
\otimes P[|\Psi^{-}\rangle] \}
\end{array}
\end{equation}

One can now generate three other mutually orthogonal activable bound
entangled states of six qubit by operating Pauli operators on any
one qubit. Thus, the four states of six qubit system are expressed as,

\begin{equation}
\begin{array}{lcl}
\rho_{6} ^{\pm} &=& \frac{1}{4}~ \{ \rho_{4}^{+} \otimes
P[|\Phi^{\pm}\rangle]+\rho_{4}^{-} \otimes P[|\Phi^{\mp}\rangle]+
\sigma_{4}^{+} \otimes P[|\Psi^{\pm}\rangle]\\&  &+\sigma_{4}^{-}
\otimes P[|\Psi^{\mp}\rangle] \}\\
\sigma_{6} ^{\pm} &=& \frac{1}{4}~ \{ \rho_{4}^{+} \otimes
P[|\Psi^{\pm}\rangle]+\rho_{4}^{-} \otimes P[|\Psi^{\mp}\rangle]+ \sigma_{4}^{+}
\otimes P[|\Phi^{\pm}\rangle]\\&  &+\sigma_{4}^{-} \otimes P[|\Phi^{\mp}\rangle] \}
\end{array}
\end{equation}

\subsection{General class of bound entangled states in higher dimension}
By mathematical induction formula, we find in any even number of qubit
system starting from four, there are only four activable bound entangled
states belonging to this class. We will show this in a constructive way.

\textbf{Constructive approach:} Consider the usual basis of the
Hilbert space of $2N+2$ qubits. First, we divide the basis set in two
section according as the first qubit is in the state $|0\rangle$ or
$|1\rangle$. Then we subdivide the two subsets according as the
number of $0$'s are even or odd. so, each of the four subsets have
exactly $\frac{2^{2N+2}}{4}=2^{2N}$ number of elements. The explicit
form of the sets are given below,

\begin{equation}
\begin{array}{lcl}
\Gamma_{2N+2}^1~=~\{|p_{2N+2}^i\rangle:~|p_{2N+2}^i\rangle=|a_0^i\rangle|a_1^i\rangle
\cdots |a_{2N+1}^i\rangle \}
\end{array}
\end{equation}
where $a_0^i=0,~\forall i= 1,2, \cdots 2^{2N}$ and $\sum_{j=0}^{{2N+1}}a_j^i=0 ~(mod 2)$,
i.e., $|p_{2N+2}^i\rangle$ has a string of 0's and 1's of length $2N+2$ such
that the first element of the string is $0$ and there is an even
number of $0$'s in the string.

Now, we consider the subset with elements
$|\overline{p_{2N+2}^i}\rangle$ orthogonal to $|p_{2N+2}^i\rangle$
as,
\begin{equation}
\begin{array}{lcl}
\Gamma_{2N+2}^2~=~\{|\overline{p_{2N+2}^i}\rangle:
|\overline{p_{2N+2}^i}\rangle= |\overline{a_0^i}\rangle|\overline{a_1^i}\rangle
\cdots |\overline{a_{2N+1}^i}\rangle\}
\end{array}
\end{equation}
Thus, the form of elements of the second subset will be determined by
$$\langle a_j^i |\overline{a_j^i}\rangle~=~0~~\forall ~i,j$$
i.e., $|\overline{p_{2N+2}^i}\rangle$ has a string of 0's and 1's of length
$2N+2$ such that the $j$-th element of the $i$-th string of
$|\overline{p_{2N+2}^i}\rangle$ is $0$ or $1$, according as the $j$-th
element of the $i$-th string of $|p_{2N+2}^i\rangle$, will
be $1$ or $0$, for all $i,j$. By construction, there is also an even number
of $0$'s in the strings of this set. Next, we consider other two class of
sets.

\begin{equation}
\begin{array}{lcl}
\Gamma_{2N+2}^3~=~\{|q_{2N+2}^i\rangle:~
|q_{2N+2}^i\rangle=|b_0^i\rangle|b_1^i\rangle \cdots
|b_{2N+1}^i\rangle\}
\end{array}
\end{equation}
such that $b_0^i=0,~\forall i = 1,2, \cdots 2^{2N}$ and $\sum_{j=0}^{2N+1}b_j^i=1 (mod 2)$,
i.e., $|q_{2N+2}^i\rangle$ has a string of 0's and 1's of length $2N+2$
such that the first element of the string is $0$ and there is an odd
number of $0$'s in the string.

Lastly, we consider the set with elements orthogonal to the elements
of the third set as,
\begin{equation}
\begin{array}{lcl}
\Gamma_{2N+2}^4~=~\{|\overline{q_{2N+2}^i}\rangle:
~|\overline{q_{2N+2}^i}\rangle=
|\overline{b_0^i}\rangle|\overline{b_1^i}\rangle \cdots
|\overline{b_{2N+1}^i}\rangle\}
\end{array}
\end{equation} such that $\langle b_j^i  |\overline{b_j^i}\rangle ~=~0,~~\forall i,j$,
i.e., $|\overline{q_{2N+2}^i}\rangle$ has a string of 0's and 1's of length
$2N+2$ such that the $j$-th element of the $i$-th
string of $|\overline{q_{2N+2}^i}\rangle$ is $0$ or $1$, according as the $j$-th
element of the $i$-th string of $|q_{2N+2}^i\rangle$, will be
$1$ or $0$, for all $i,j$. There will be an odd number of $0$'s in
each string of this set.

Now, we reconstruct four new sets with the elements of the above four sets as
follows:
\begin{equation}
\begin{array}{lcl}

S_{2N+2}^+ &=&\{|s_{2N+2}^i\rangle~:~|s_{2N+2}^i\rangle
=\frac{1}{\sqrt{2}}(|p_{2N+2}^i\rangle~+~|\overline{p_{2N+2}^i}\rangle),
~\forall i = 1,2, \cdots 2^{2N}\}\\

S_{2N+2}^-&=&\{|\overline{s_{2N+2}^i}\rangle~:~|\overline{s_{2N+2}^i}\rangle
=\frac{1}{\sqrt{2}}(|p_{2N+2}^i\rangle~-~|\overline{p_{2N+2}^i}\rangle),
~\forall i = 1,2, \cdots 2^{2N}\}\\

R_{2N+2}^+ &=&\{|r_{2N+2}^i\rangle~:~|r_{2N+2}^i\rangle
=\frac{1}{\sqrt{2}}(|q_{2N+2}^i\rangle~+~|\overline{q_{2N+2}^i}\rangle),
~\forall i = 1,2, \cdots 2^{2N}\}\\

R_{2N+2}^-
&=&\{|\overline{r_{2N+2}^i}\rangle~:~|\overline{r_{2N+2}^i}\rangle
=\frac{1}{\sqrt{2}}(|q_{2N+2}^i\rangle~-~|\overline{q_{2N+2}^i}\rangle),
~\forall i = 1,2, \cdots 2^{2N}\}
\end{array}
\end{equation}

Then, we prepare four mixed states by taking an equal mixture of all
the states from each of the four sets. The four mixed states will be
of the following form,
\begin{equation}
\begin{array}{lcl}

\rho_{2N+2}^+&=&\frac{1}{2^{2N}} ~\sum_{i=1}^{2^{2N}} P[|s_{2N+2}^i \rangle]\\

\rho_{2N+2}^- &=&\frac{1}{2^{2N}} ~\sum_{i=1}^{2^{2N}} P[|\overline{s_{2N+2}^i}
\rangle]\\

\sigma_{2N+2}^+&=&\frac{1}{2^{2N}} ~\sum_{i=1}^{2^{2N}} P[|r_{2N+2}^i \rangle]\\

\sigma_{2N+2}^- &=&\frac{1}{2^{2N}} ~\sum_{i=1}^{2^{2N}} P[|\overline{r_{2N+2}^i}
\rangle]
\end{array}
\end{equation}

So far we have described the process of constructing the four mixed states in
any system consists of an even number of local qubit subsystem. We
will now discuss the inter-relations of this class of states. To
investigate the properties and uses of this class of states, we need
a simple rule of connection between two successive systems. We obtain
a nice correlation between two successive systems. This relation plays
the most fundamental step of generalization of this class. We represent the
result in the following theorem.\\

\textbf{Theorem:} \emph{The four states $\rho_{2N+2} ^{\pm}$,
$\sigma_{2N+2} ^{\pm}$ of  $2N+2$ qubit system can be expressed as
equal Bell-mixture of the four states of the previous states
$\rho_{2N} ^{\pm}$, $\sigma_{2N} ^{\pm}$ of $2N$ qubit system.}\\

\textbf{Proof:} The states of the $2N+2$ qubit system $|s_{2N+2}^i \rangle$,
can be expressed as,
\begin{equation}
\begin{array}{lcl}
|s_{2N+2}^i \rangle
&=&\frac{1}{\sqrt{2}}(|p_{2N+2}^i\rangle~+~|\overline{p_{2N+2}^i}\rangle)
\end{array}
\end{equation}

Now, the string of $|p_{2N+2}^i\rangle$ has its first element as $0$ and an even
number of $0$'s in total. Thus, it must be either in the form $|00\rangle
|\mu_{2N}^1\rangle$ or $|00\rangle |\mu_{2N}^2\rangle$ or
$|01\rangle |\nu_{2N}^1\rangle $ or $|01\rangle |\nu_{2N}^2\rangle$,
where $|\mu_{2N}^1\rangle$ has a string of $0$'s and $1$'s of
length $2N$ with the first element to be 0 and an even number of
$0$'s in it. Similarly,  $|\mu_{2N}^2\rangle$ has a string of $0$'s and $1$'s
of length $2N$ with the first element to be 1 and an even number of
$0$'s in it, $|\nu_{2N}^1\rangle $ has a string of $0$'s and $1$'s of length
$2N$ with the first element to be 0 and an odd number of $0$'s in it
and lastly, $|\nu_{2N}^2\rangle$ has a string of $0$'s and $1$'s of length
$2N$ with the first element to be 1 and an odd number of $0$'s in
it. It is obvious that $|\mu_{2N}^1\rangle \in \Gamma_{2N}^1$, $|\mu_{2N}^2\rangle
\in \Gamma_{2N}^2$, $|\nu_{2N}^1\rangle \in \Gamma_{2N}^3$ and $|\nu_{2N}^2\rangle
\in \Gamma_{2N}^4$. In other words, the set $S_{2N+2}^+$ can be
divided into four disjoint parts as, $S_{2N+2}^+~=~ \Upsilon_{2N+2}^1
\bigcup \Upsilon_{2N+2}^2 \bigcup \Upsilon_{2N+2}^3 \bigcup
\Upsilon_{2N+2}^4$ where,

\begin{equation}
\begin{array}{lcl}

\Upsilon_{2N+2}^1&=& \{|\psi^i\rangle:
|\psi^i\rangle=\frac{1}{\sqrt{2}}(|00\rangle |p_{2N}^i\rangle +
|11\rangle |\overline{p_{2N}^i}\rangle), ~\forall~ |p_{2N}^i\rangle
\in \Gamma_{2N}^1\},\\

\Upsilon_{2N+2}^2&=& \{|\phi^i\rangle:
|\phi^i\rangle=\frac{1}{\sqrt{2}}(|00\rangle |
\overline{p_{2N}^i}\rangle + |11\rangle | p_{2N}^i\rangle),
~\forall~ |p_{2N}^i\rangle \in \Gamma_{2N}^1\},\\

\Upsilon_{2N+2}^3&=& \{|\chi^i\rangle:
|\chi^i\rangle=\frac{1}{\sqrt{2}}(|01\rangle | q_{2N}^i\rangle +
|10\rangle | \overline{q_{2N}^i}\rangle), ~\forall~ |q_{2N}^i\rangle
\in \Gamma_{2N}^3\},\\

\Upsilon_{2N+2}^4&=& \{|\varphi^i\rangle:
|\varphi^i\rangle=\frac{1}{\sqrt{2}}(|01\rangle |
\overline{q_{2N}^i}\rangle + |10\rangle | q_{2N}^i\rangle),
~\forall~
|q_{2N}^i\rangle \in \Gamma_{2N}^3\}
\end{array}
\end{equation}

Number of elements of each of the four sets above will be
$2^{2N-2}$. As the state $\rho_{2N+2}^+$ is formed as equal mixture
of all possible states $|s_{2N+2}^i \rangle~\in ~S_{2N+2}^+$, therefore,
we can represent it as,
\begin{equation}
\begin{array}{lcl}
\rho_{2N+2}^+ &=& \frac{1}{2^{2N}}\sum_{i=1}^{2^{2N-2}}\{
|\psi^i\rangle \langle \psi^i| + |\phi^i\rangle \langle \phi^i|
+|\chi^i\rangle \langle \chi^i| +|\varphi^i\rangle \langle
\varphi^i|\}\label{gen}
\end{array}
\end{equation}Now, using the form of Bell basis in terms of the usual
product basis, we can write,
\begin{equation}
\begin{array}{lcl}
|00\rangle &=& \frac{1}{\sqrt{2}}(|\Phi^{+} \rangle + |\Phi^{-}
\rangle)\\
|11\rangle &=& \frac{1}{\sqrt{2}}(|\Phi^{+} \rangle - |\Phi^{-}
\rangle)\\
|01\rangle &=& \frac{1}{\sqrt{2}}(|\Psi^{+} \rangle + |\Psi^{-}
\rangle)\\
|10\rangle &=& \frac{1}{\sqrt{2}}(|\Psi^{+} \rangle - |\Psi^{-}
\rangle)
\end{array}
\end{equation}
Thus, we may express,
\begin{equation}
\begin{array}{lcl}
|\psi^i\rangle &=& \frac{1}{\sqrt{2}}\{|00\rangle | p_{2N}^i\rangle
+ |11\rangle | \overline{p_{2N}^i}\rangle\}\\ &=&
\frac{1}{2}\{(|\Phi^{+}\rangle + |\Phi^{-}\rangle)|p_{2N}^i\rangle +
(|\Phi^{+}\rangle - |\Phi^{-}\rangle)
|\overline{p_{2N}^i}\rangle\}\\ &=& \frac{1}{2}\{|\Phi^{+}\rangle
(|p_{2N}^i\rangle+|\overline{p_{2N}^i}\rangle) + |\Phi^{-} \rangle
(|p_{2N}^i\rangle-|\overline{p_{2N}^i}\rangle)\}
\end{array}
\end{equation}Similarly,
\begin{equation}
\begin{array}{lcl}
|\phi^i\rangle &=& \frac{1}{2}\{|\Phi^{+} \rangle
(|p_{2N}^i\rangle+|\overline{p_{2N}^i}\rangle) - |\Phi^{-} \rangle
(|p_{2N}^i\rangle-|\overline{p_{2N}^i}\rangle)\},\\
|\chi^i\rangle &=& \frac{1}{2}\{|\Psi^{+} \rangle
(|q_{2N}^i\rangle+|\overline{q_{2N}^i}\rangle) + |\Psi^{-} \rangle
(|q_{2N}^i\rangle-|\overline{q_{2N}^i}\rangle)\},\\
|\varphi^i\rangle &=& \frac{1}{2}\{|\Psi^{+} \rangle
(|q_{2N}^i\rangle +|\overline{q_{2N}^i}\rangle) - |\Psi^{-} \rangle
(|q_{2N}^i\rangle-|\overline{q_{2N}^i}\rangle)\}
\end{array}
\end{equation}

Inserting all these relations in Eq.(\ref{gen}), we obtain,
\begin{equation}
\begin{array}{lcl}
\rho_{2N+2}^+ &=& \frac{1}{2^{2N}}
\frac{1}{2^{2}}\sum_{i=1}^{2^{2N-2}}\{ P[|\Phi^{+} \rangle
(|p_{2N}^i\rangle+|\overline{p_{2N}^i}\rangle) + |\Phi^{-} \rangle
(|p_{2N}^i\rangle-|\overline{p_{2N}^i}\rangle)]\\ &+& P[|\Phi^{+}
\rangle (|p_{2N}^i\rangle+|\overline{p_{2N}^i}\rangle) - |\Phi^{-}
\rangle (|p_{2N}^i\rangle-|\overline{p_{2N}^i}\rangle)]\\ &+&
P[|\Psi^{+} \rangle (|q_{2N}^i\rangle+|\overline{q_{2N}^i}\rangle) +
|\Psi^{-} \rangle (|q_{2N}^i\rangle-|\overline{q_{2N}^i}\rangle)]\\
&+& P[|\Psi^{+} \rangle
(|q_{2N}^i\rangle+|\overline{q_{2N}^i}\rangle) - |\Psi^{-} \rangle
(|q_{2N}^i\rangle-|\overline{q_{2N}^i}\rangle)]\}\\
&=&
\frac{1}{2^{2N+2}}\sum_{i=1}^{2^{2N-2}}2\{P[|\Phi^{+}\rangle]P[|p_{2N}^i\rangle+|\overline{p_{2N}^i}\rangle]
+P[|\Phi^{-} \rangle]P[|p_{2N}^i\rangle-|\overline{p_{2N}^i}\rangle]\\
&+& P[|\Psi^{+}\rangle]P[|q_{2N}^i\rangle+|\overline{q_{2N}^i}\rangle]
+P[|\Psi^{-} \rangle]P[|q_{2N}^i\rangle-|\overline{q_{2N}^i}\rangle]\}\\
&=&\frac{1}{2^{2N+1}}\sum_{i=1}^{2^{2N-2}}2\{
P[|\Phi^{+}\rangle]~P[|s_{2N}^i\rangle] +
P[|\Phi^{-} \rangle]~P[|\overline{s_{2N}^i}\rangle]\\
&+& P[|\Psi^{+} \rangle]~P[|r_{2N}^i\rangle] +P[|\Psi^{-}
\rangle]~P[|\overline{r_{2N}^i}\rangle]\}\\
&=&\frac{1}{4} \{P[|\Phi^{+}\rangle] \otimes\rho_{2N}^+  + P[|\Phi^{-}
\rangle]\otimes\rho_{2N}^- + P[|\Psi^{+} \rangle]\otimes\sigma_{2N}^+
+P[|\Psi^{-} \rangle]\otimes\sigma_{2N}^-\}
\end{array}
\end{equation}

Thus, we obtain a nice Bell-correlation between the states of two
successive systems for the state $ \rho_{2N+2}^+ $. In a similar way,
we can obtain the Bell-correlated form of other three states
$\rho_{2N+2} ^{-}$, $\sigma_{2N+2}^{\pm}$. If, instead of considering
the first two qubits in the above analysis, we proceed with the
last two qubits, then the Bell-correlated form of the four states of
$2N+2$ qubit system will be given by,
\begin{equation}
\begin{array}{lcl}
\rho_{2N+2} ^{\pm} &=& \frac{1}{4}~ \{ \rho_{2N}^{+} \otimes
P[|\Phi^{\pm}\rangle]+\rho_{2N}^{-} \otimes P[|\Phi^{\mp}\rangle]+
\sigma_{2N}^{+} \otimes P[|\Psi^{\pm}\rangle]\\&  &+\sigma_{2N}^{-}
\otimes P[|\Psi^{\mp}\rangle] \}\\
\sigma_{2N+2} ^{\pm} &=& \frac{1}{4}~ \{ \rho_{2N}^{+} \otimes
P[|\Psi^{\pm}\rangle]+\rho_{2N}^{-} \otimes P[|\Psi^{\mp}\rangle]+ \sigma_{2N}^{+}
\otimes P[|\Phi^{\pm}\rangle]\\&  &+\sigma_{2N}^{-} \otimes P[|\Phi^{\mp}\rangle] \}
\label{bcbe}
\end{array}
\end{equation}

Now, by the above correlated formula, it is possible to prove that the four
states of $2N+2$ qubit system, for any $N\geq 2$, are mutually
orthogonal activable bound entangled states. This correlation
also enables one to generate the whole class of states from the four
qubit states mentioned earlier by a recursive process. Next, we will
explain the various properties of this large class of states.\\

\textbf{Property-1:} The whole class of states are symmetric over
all the parties concerned, i.e., the states remain invariant under
the interchange of any two parties. The four states
$\rho_{2N+2}^{\pm}$, $\sigma_{2N+2} ^{\pm}$ can be expressed as,
$$\rho_{2N+2} ^{\pm} = \frac{1}{2^{2N}}~ \{~ \sum_{i=1}^{2^{2N}}
P[|p^i_{2N+2}\rangle \pm
\overline{|p^i_{2N+2}\rangle}]~\}$$
$$\sigma_{2N+2} ^{\pm} = \frac{1}{2^{2N}}~ \{ ~\sum_{i=1}^{2^{2N}}
P[|q^i_{2N+2}\rangle\pm \overline{|q^i_{2N+2}\rangle}]~\}$$ where,
$|p^k_{2N+2}\rangle$, $|q^j_{2N+2}\rangle$, for $k,j= 1, 2,\cdots 2^{2N}$,
are given by,
\begin{equation}
\begin{array}{lcl}
|p^k_{2N+2}\rangle &=& |a^k_0\rangle \otimes |a^k_1\rangle
\otimes \cdots \otimes|a^k_{2N+1}\rangle
\end{array}
\end{equation}with
$a^k_i \in \{0,1\},~ ~\forall ~i=0,1,\cdots,2N+1$ and $a^k_0=0 $,
\begin{equation}
\begin{array}{lcl}
|q^j_{2N+2}\rangle &=& |b^j_0\rangle \otimes |b^j_1\rangle
\otimes \cdots \otimes|b^j_{2N+1}\rangle
\end{array}
\end{equation} with
$b^j_i \in \{0,1\},~ ~\forall ~i=0,1,\cdots,2N+1$ and $b^j_0=0 $,
such that $$\sum_{i=0}^{2N+1} a^k_i=~0~(mod~2)~,~\sum_{i=0}^{2N+1}
b^j_i =~1~(mod~2)$$ The states
$\overline{|p^k_{2N+2}\rangle}$ and
$\overline{|q^j_{2N+2}\rangle}$ are orthogonal to the states
$|p^k_{2N+2}\rangle$ and $|q^j_{2N+2}\rangle$ respectively,
for all possible values of $k$ and $j$. The set
$\{|p^k_{2N+2}\rangle: \forall k \}$ contains all possible
permutations of strings of $0$'s and
$1$'s with an even number of zeros. Thus, any permutation of the
positions of different parties will interchange all the
$|p^k_{2N+2}\rangle$, in between themselves and their orthogonals
$\overline{|p^k_{2N+2}\rangle}$. This proves the permutation symmetry of
all the four states $\rho_{2N+2}^{\pm}$, $\sigma_{2N+2} ^{\pm}$.

Explicitly, from the Eq.(\ref{bcbe4sym}) of the four qubit systems, i.e.,
$$\rho_4^{\pm}  ~  =~\frac{1}{8}(P[|0000\pm 1111\rangle]+
P[|0011 \pm 1100\rangle]~ +P[|0101\pm 1010\rangle]
+P[| 0110 \pm 1001\rangle]  )$$

$$\sigma_4^{\pm}  ~ =~\frac{1}{8}(P[|0001 \pm 1110\rangle]+
P[|0010 \pm 1101\rangle] ~ +P[|0100 \pm 1011\rangle]
+P[|0111 \pm 1000\rangle] )$$ we find clearly the permutation symmetry
of the states over all the parties concerned.\\

\textbf{Property-2:}  From Eq.(\ref{bcbe}), it is clear that the four
states of $2N+2$ qubit system are orthogonal to each other if the
$2N$ qubit states are so. Also, from Eq.(\ref{smolin}) and
Eq.(\ref{bcbe4bell}), we observe that the four states
$\rho_4^\pm~,~\sigma_4^\pm$ of four qubit system are mutually
orthogonal. Thus, in a recursive way it provides us the mutual
orthogonality of the four activable bound entangled states of any
even qubit system, starting from four. This is also evident from
the orthogonality of the support sets $S_{2N+2}^+,
~S_{2N+2}^{-},~R_{2N+2}^+,~R_{2N+2}^-$ corresponding to those four
states of $2N+2$ qubit system.\\

\textbf{Property-3:} All the states above are
mixed entangled states. Suppose, the four states of
$2N+2$ qubit system are shared between the parties denoted by $A_1,
A_2, \cdots A_{2N},~B,~C$ as,
\begin{equation}
\begin{array}{lcl}
\rho_{2N+2} ^{\pm} &=& \frac{1}{4}~ \{ (\rho_{2N}^{+})_{A_1 A_2
\cdots A_{2N}} \otimes (P[|\Phi^{\pm}\rangle])_{BC}+(\rho_{2N}^{-})_{A_1 A_2
\cdots A_{2N}} \otimes (P[|\Phi^{\mp}\rangle] )_{BC} \\& &+
(\sigma_{2N}^{+})_{A_1 A_2 \cdots A_{2N}} \otimes
(P[|\Psi^{\pm}\rangle])_{BC}+
(\sigma_{2N}^{-})_{A_1 A_2 \cdots A_{2N}} \otimes (P[|\Psi^{\mp}\rangle])_{BC} \}\\
\sigma_{2N+2}^{\pm} &=& \frac{1}{4}~ \{ (\rho_{2N}^{+})_{A_1 A_2
\cdots A_{2N}}  \otimes (P[|\Psi^{\pm}\rangle])_{BC}+(\rho_{2N}^{-} )_{A_1
A_2 \cdots A_{2N}}   \otimes (P[| \Psi^{\mp}\rangle])_{BC}\\& &+
(\sigma_{2N}^{+})_{A_1 A_2 \cdots A_{2N}} \otimes
(P[|\Phi^{\pm}\rangle])_{BC}+(\sigma_{2N}^{-} )_{A_1 A_2 \cdots A_{2N}}
\otimes (P[|\Phi^{\mp}\rangle])_{BC} \} \label{bcbeent}
\end{array}
\end{equation}
Now, the states $\rho_{2N} ^{\pm}$, $\sigma_{2N} ^{\pm}$, are
orthogonal to each other (follows from the above property-2). Then, the
first $2N$ parties  $A_1 A_2 \cdots A_{2N}$ can join together and perform a
projective measurement on their system of $2N$ qubit in those four
mixed orthogonal states $\rho_{2N}^{\pm}$, $\sigma_{2N}^{\pm}$. The
result of their measurement is then communicated to $B$ and $C$ and
consequently, the remaining state shared between $B$ and $C$ will
be the corresponding Bell state determined by the above expression.
For example, when the parties share the state $\rho_{2N+2}^{+}$,
and if the result of the orthogonal measurement
of the $2N$ parties is $\sigma_{2N}^{+}$, then the state shared
between the other two paries  $B$ and $C$, who are still remain
specially separated is $|\Psi^{+}\rangle$. As all the four Bell
states are maximally entangled state in $2\times 2$, it is always
possible to extract non-zero amount of entanglement from the state
$\rho_{2N+2}^{+}$ from the above procedure. Thus, the mixed state
$\rho_{2N+2}^{+}$ is entangled and so also the others. Therefore,
by induction method, the whole class of states are entangled.\\

\textbf{Property-4:} States are activable bound entangled. From the
Eq.(\ref{bcbe}), we have obtained that for each $N\geq 2$, the
$2N+2$ qubit states $\rho_{2N+2} ^{\pm}$, $\sigma_{2N+2} ^{\pm}$,
can be expressed in Bell-correlated form with the $2N$ qubit states
$\rho_{2N}^{\pm}, \sigma_{2N} ^{\pm}$. This form also tells us that
they are separable across a $2N:2$ cut. Now,
proceeding in the same manner, we could further divide the $2N$ qubit
states in terms of the $2N-2$ qubit states and so on. Thus the
states of $2N+2$ qubit system could be represented as states separable
across any $2K:2N+2-2K~ (~1\leq K \leq N)$ cut. Again, as the states
are symmetric over all possible permutations of the concerned
parties, therefore, it is clear that the states of this class are
separable in any bipartite cut, if the number of parties in each of
the two groups are even. Thus, in any $2K:2N+2-2K$ cut, i.e., when we
join all the concerned parties in any two separate groups so that
the number of parties in each group is even, the states are separable.
So, it is not possible to distill out any entanglement from the states
by LOCC, when all the $2N+2$ parties remain spatially separated.
It shows the bound entangled character of all the four states of the
$2N+2$ qubit systems, for all $N\geq 2$.

Also, if any of the $2N$ parties of $\rho_{2N+2} ^{\pm}$,
$\sigma_{2N+2} ^{\pm}$, join together in a lab and make a projective
measurement on the four orthogonal states $\rho_{2N} ^{\pm}$,
$\sigma_{2N} ^{\pm}$, then this operation with classical communications
results in sharing a Bell state between the other two separated parties.
Hence, the states could be activated by collective operations on a subset
of the concerned parties. So, the states are Activable, or in other
words, Unlockable bound entangled states.\\

\textbf{Property-5:} The class of states are Bell-correlated. It is
proved earlier before mentioning property-1. From Eq.(\ref{bcbe}),
we find each of the four states $\rho_{2N+2} ^{\pm}$, $\sigma_{2N+2}
^{\pm}$ for $N \geq 2$, is expressed as equiprobable mixture of
$\rho_{2N} ^{\pm}$, $\sigma_{2N} ^{\pm}$ with the four Bell states
$|\Phi^{\pm}\rangle,~|\Psi^{\pm}\rangle$, taken in a specific order.
Explicitly, the four qubit states $\rho_{4} ^{\pm}$, $\sigma_{4}
^{\pm}$ are themselves written in terms of Bell states. In fact, the
whole class of states can be written explicitly as equiprobable
mixture of products of Bell states taken in some specific order.\\

\textbf{Property-6:} The four states $\rho_{2N+2} ^{\pm}$,
$\sigma_{2N+2} ^{\pm}$, for $N \geq 1$, are locally Pauli-connected
in one party, i.e., starting from any one of the four states and
operating the Pauli matrices on the local subsystem of any one party
of the $2N+2$ qubit system, one would able to get the other three
states. Thus, denoting,
$\rho^1_{2N+2}=\rho_{2N+2}^{+},~\rho^2_{2N+2}=\rho_{2N+2}^{-},
~\rho^3_{2N+2}=\sigma_{2N+2}^{+},~\rho^4_{2N+2}=\sigma_{2N+2}^{-}$
and the Pauli matrices as, $\sigma^1=I, \sigma^2=\sigma_z,
\sigma^3=\sigma_x, \sigma^4=i\sigma_y$, for each $i=1,2,3,4$, we can
express, $\rho^i_{2N+2}={\Gamma^i}_k\rho^1_{2N+2}{\Gamma^i}_k$, for
any $k=1,2,\cdots ,2N+2$, where, ${\Gamma^i}_k=I_1 \otimes I_2
\otimes \cdots \otimes I_{k-1} \otimes \sigma^i \otimes
I_{k+1}\otimes \cdots \otimes I_{2N+2}$, is the operator whose
action can be described
as the operation by the Pauli operator $\sigma^i$ on k-th qubit.\\

\textbf{Property-7:} The support of the density matrices
corresponding to the four states $\rho_{2N+2} ^{\pm}$,
$\sigma_{2N+2} ^{\pm}$, for $N \geq 1$, are the four sets
$S_{2N+2}^+, ~S_{2N+2}^{-},~R_{2N+2}^+,~R_{2N+2}^-$ respectively,
described earlier. This four sets will together span the total
Hilbert space of $2N+2$ qubit system. Also, the four sets are
mutually disjoint and equal in size. Thus, any two states of the
$2N+2$ qubit Hilbert space, belonging to the support of two different
sets. Geometrically, the Hilbert spaces of any even qubit systems are
divided into four equal and mutually orthogonal parts.\\

\textbf{Property-8:} The entanglement cost of preparing the $2N$
qubit states is $N$ e-bits {\cite{cost}}, shown by possing a lower
bound on entanglement cost and then achieving the bound by an
explicit protocol.\\

Now, we will establish an important property of the class of four
activable bound entangled states in any $2N+2$ qubit system for
$N\geq 1$. The four states are local indistinguishable.
We will represent it as a theorem.\\

\textbf{Theorem:} For all $N\geq2$, the probability for
distinguishing the four states $\rho_{2N}^{\pm}$,
$\varrho_{2N}^{\pm}$ exactly, by LOCC is zero.\\

To prove it, let us assume that for some value of
$N\geq 2$, the four states $\rho_{2N}^{\pm}$, $\sigma_{2N} ^{\pm}$
are locally distinguishable. Now, consider the state,
$$\rho_{2N+2}^{+} = \frac{1}{4}~ \{ \rho_{2N}^{+} \otimes
P[|\Phi^{+}\rangle]+\rho_{2N}^{-}
\otimes P[|\Phi^{-}\rangle]+ \sigma_{2N}^{+} \otimes P[|\Psi^{+}\rangle]\\
+\sigma_{2N}^{-} \otimes P[|\Psi^{-}\rangle] \}$$ where the first $2N$
parties are $A_1, A_2, \ldots, A_{2N-1}, B_1$ and the last two parties
are $A_{2N}, B_2$, i.e., the state is separable by construction in
$A_1 A_2 \ldots A_{2N-1} B_1 : A_{2N} B_2$ cut. Again, the state is
symmetric with respect to the interchange of any two parties, i.e.,
$\rho_{2N+2}^{+}$ has the same form if the first $2N$ parties are
$A_1, A_2, \ldots, A_{2N}$  and the last two parties are $B_1, B_2$. If,
the four states $\rho_{2N}^{\pm}$, $\sigma_{2N} ^{\pm}$ are locally
distinguishable, then by LOCC only, $A_1, A_2, \ldots, A_{2N}$ are
able to share a Bell state among $B_1$ and $B_2$, which is
impossible as initially there is no entanglement in between $B_1$
and $B_2$. So, all the four states $\rho_{2N}^{\pm}$,
$\sigma_{2N}^{\pm}$ are locally indistinguishable for any $N\geq 2$.
Our protocol also suggest that the states are even probabilistically
indistinguishable for any $N\geq 2$, as it is impossible to share
any entanglement by LOCC between $B_1$ and $B_2$. Let us assume that
the four states are locally indistinguishable with probability
$p>0$, then having the shared state $\rho_{2N+2}^{+}$ among the
$2N+2$ parties, any set of $2N$ parties may able to distinguish
their joint local system with probability $1>p>0$ and
correspondingly share one Bell state among the other two parties. In
this way, it is possible to extract on average non-zero amount of
entanglement by performing LOCC only. This contradicts with
the bound entangled nature of $\rho_{2N+2}^{+}$. Thus, the four
states of $2N$ qubit system are even probabilistically locally
indistinguishable. So, the probability for distinguishing the
four states $\rho_{2N}^{\pm}$, $\sigma_{2N}^{\pm}$ by LOCC is
zero, for any $N\geq 2$.

\subsection{Conclusion}
In this chapter, we find interesting nature of the general class
of Activable Bound Entangled States. Besides of being activable
bound entangled, which itself is a peculiar feature, they have some
other special properties, never attainable by any other known state.
We focussed our attention here centrally on the property of local
indistinguishability. This is because it is a very special kind of
non-locality observed in composite quantum systems. For multi-partite
systems, it is rare to observe and this non-locality could be used
for practical purposes. Previously, Smolin state is being used
successfully for performing many computational tasks \cite{41,Shor2}.
One may search whether those tasks can also be performed by employing
our multipartite activable bound entangled states. Such generalizations
are always important for practical implementation of any computational
schemes. Our next aim is to build some hiding protocol with the class of
states. In the next chapter, we will elaborate this concept in detail.

\chapter{An Application- Quantum Data Hiding}\footnote{Some portions of this chapter is
published in Physics Letter A, {\bf 365}, 273 (2007).}

\section{Introduction}
Information theory concerns largely with encoding, securing and
manipulating information in terms of physical states. Quantum
information theory is a newly developed subject with a vast field of
applicability known basically after the discovery of quantum
teleportation, dense coding, etc., \cite{crypto, teleport,
densecoding,bitcommitment}. Quantum non-locality or, more
specifically the quantum entanglement is the key invention of
quantum information theory that has been used for lots of
information processing schemes, which could not be possible by any
classical protocol. Protocols are now supported by practical
laboratory experiments or on the verge of implementations.
Information theory cares largely with the objects that maintains the
secrecy of a data. Unlike teleportation, where the objective is to
send information from one place to another so that the secrecy of
the data is not explored by any eavesdropper, the aim in this area
is to share some information among a number of different parties
situated at distant laboratories, so that any one or some subset of
the concerned parties cannot break the security or change the secret
data \cite{quantumsecret}. Thus, protocols are proposed against the
cheating attempts of some parties sharing the states. This
completely changes the old scenario. Obviously, the investigations
become more successful by using the nonlocal character of the
quantum states. Entanglement is now used for posing different
protocols on quantum secret sharing or on data hiding. Secret
sharing is comparatively more known to us, whereas data hiding is a
new branch of information theory. It is mainly concerned with the
sharing of information to some parties situated at distant places,
so that each of the associated parties have access of some part of
the data. Though, there are strong restrictions on recovering the
secret information. Now, the difficulty in data hiding by using
entangled states is the very important requirement, i.e., local
indistinguishability of the states. We have observed in the previous
chapter that our activable bound entangled states have the property
of local indistinguishability. In this chapter, we will explore the
possibility and limitations of that class of activable bound
entangled states to build a data hiding protocol. Before that, we
first describe some basic ideas of different hiding schemes.

\section{Sharing information secretly among distant parties}

Early ideas came with the hiding classical information in
terms of classical bits (in short, cbits). The process is known as classical
secret sharing protocol \cite{classicalsecret}. Later, we find the
concept of keeping quantum information secret in various purposes.
The protection of secret information can be defined either against
eavesdropper attack or, against the cheating of the concerned parties.
In the first case, it is sufficient to detect the presence of an
eavesdropper having some (obviously, unwanted and secret) access of the
communication channels between the associated parties {\cite{cryp}}.
The classical one-time-pad schemes are proposed for using a classical
state as the secret channel, where the information is processed only
once and then discarded by the choice of another classical state.
This scheme is generalized to quantum case by sharing a private quantum
channel between some parties so that the hidden information remains
secured against any eavesdroppers attempt to explore the secret.
A more complicated situation may arise when the data is kept secret
from the parties themselves. The security of the hidden data is built
on the premises that the data cannot be perfectly revealed by any kind
of cheating attack (local or global) of the associated parties. Altogether,
we have two different branches of research in this area, known as, Secret
Sharing and Data Hiding.

Now, in a secret sharing scheme the basic task is to share some information
classical or quantum, among a number of parties situated at different labs
so that each of the parties have some access of a part of the information,
though the information is not known to them. The information is hidden
from the parties and thus the protection of it is defined against various
kinds of cheating attempts maid by the associated parties. This is
done by encoding the information in some classical or quantum state
and sharing the state among all the parties. The associated parties
(may be only some of them) are not allowed to know the secret.
Thus, to reveal the hidden information they have to retrieve the secret
state shared by them altogether. Schemes are proposed in
this direction must have a properly defined set of allowable
operations that the parties can perform on their shared part and a
properly chosen security parameter specifying the level of
perfection of security of the secret information. Thus any computational
scheme proposed for this purpose is characterized by the following
properties.

(1) The amount of information (in terms of cbits or qubits or qudits, etc.)
that can be hidden in the system.

(2) The maximum number of parties among which the system is
distributed.

(3) Maintaining the security criteria, what is the allowable set of
operation that the associated parties can perform to reveal the
data.

(4) The maximum number of unfaithful parties among all the
associated parties. Here, it is to be noticed that the number of
parties is important; i.e., a symmetric structure of the states
is required.

(5) What is the amount of (zero for perfect security and otherwise
an asymptotically small quantity) of secret information that the unfaithful
parties can at most retrieve from the system by their joint attack.

Usually, a $(k,n)$ threshold scheme is proposed for sharing secret, where
$n$ is the total number of parties sharing the state and $k$ of them can be
used to reconstruct the state, but $k-1$ or fewer have absolutely no
information about the state \cite{quantumsecret,Hilery}.

\section{Hiding classical data}
A further restricted and secured scenario can be found in the schemes of
data hiding where a very strong constraint is always imposed over the
system. Here, classical information is kept secret in terms of orthogonal
quantum states shared among some parties situated at distant locations. The
involved parties know which quantum state is used to encode which
classical bit, but do not know the actual state they are sharing.
Even considering all of the concerned parties to be unfaithful, the
security in such schemes must guarantee the requirement that the
parties can not retrieve the secret by LOCC only. This requirement
turns out to be the local indistinguishability of the hiding
states \cite{hide02,53}. This imply, in a quantum data hiding scheme
the hiding states must be locally indistinguishable. The aim of such
a hiding scheme is to build a considerably high level of security and
to minimize the number of faithful parties, required to maintain the
secrecy. We can imagine a situation where an employer is sharing some
data to a number of his/her employees, without knowing the actual data.
The aim of the scheme is, according to his/her own convenience, the
employer can reveal the data. However, the parties could not be able
to know the hidden data by their cooperations. For example, $\log_{2}k$
number of classical bits are shared among $n$ number of distant parties
with $k$ number of orthogonal quantum states, say,
$\{\rho_i ; i=1, \ldots, k \}$. The hiding states,
$\{\rho_i ; i=1, \ldots, k \}$ are chosen to be locally indistinguishable.
Also, it is desirable that the hiding states must be symmetric over all possible
interchanges of the positions of the parties. Then, the next step is to
check whether the states can be distinguished locally with some
non-zero probability or not. This probability requirement indicates the
amount of data that can be recovered at most locally. Here, one should
note that the data is known to all the concerned parties with the guessing
probability. Therefore, it is expected that in a hiding protocol, the unfaithful
parties or anyone who wants to know the hidden data, could not be able to
retrieve the hidden data more than the guessing probability.
Thus, data hiding is really a challenging area of research where
one have to define the security of the protocol, first by defining
the number of unfaithful parties who are able to communicate through
some quantum channels and then, by the class of operations that the
parties can perform. If the system of orthogonal states are chosen
cleverly, then we have to check only such sets of operations which
will maintain the symmetry of the states.

Now, the idea of hiding information in quantum states so that parties
concerned could not be able to know the secret information by LOCC, is
came after the work of Terhal \emph{et.al.} {\cite{51}}. In their
protocol, a classical bit $b=0,1$ is hidden among Alice and Bob by sharing
$n$ number of Bell states chosen at random with uniform probability, with
an odd or even number of singlets corresponding to the values $0,1$. The
protocol is asymptotically secure as the amount of the data revealed
locally by Alice and Bob, decreases with $n$.
After this, Eggeling and Werner {\cite{20}} provided a protocol for
hiding a classical bit in four qubit system. In their work, the security
of the hidden data is defined against quantum communication between
some predefined partition of the four parties and with a predefined
level of security which can be chosen to be arbitrarily high. Next, we
describe the notion of hiding quantum data.

\section{Hiding quantum data}

Quantum data hiding is a subject where the basic tasks is to hide
information in terms of quantum states, like qubits, qudits, etc.
Protocols proposed for hiding qubits are almost similar with the
hiding classical data. DiVincenzo~\emph{et.al.}{\cite{hidequantum}}
showed that in bipartite case, there is an equivalence between
hiding one qubit and two cbits hiding with a similar security
criteria. It is further shown that any quantum data hiding scheme
for hiding $2k$ cbits with a security parameter $\epsilon$, can be
converted into a $\delta$-secure $k$-qubit hiding scheme, for
$\delta= 2^{k+1}\epsilon$.

In multipartite case, data hiding is rather complex to describe. As
here quantum communications may include the joint operations on any
subset of the all $n$ number of parties. Security is defined by
allowing a fixed number of parties to communicate within themselves
quantum mechanically. Hayden et.al. {\cite {28}} provided a general
scheme for hiding qubit information in multipartite quantum state.
The hiding states belongs to a class of nearly perfect locally
indistinguishable states, rather than only perfectly indistinguishable
class. The scheme is proposed with $(k,n)$ access structure through
suitably defined encoding and decoding maps with the required level of
security. The quantum information is hidden in multipartite quantum
states by some completely positive trace-preserving maps. Corresponding
to the hiding states, the encoding map is defined as,
\begin{equation}
E(\Phi)= ~\frac{1}{r} ~\sum_i \{U_i \Phi U_i^{\dagger} \otimes \rho^i\}
\end{equation}
where, $\Phi$ is a multipartite state, $\rho^i$ are hiding states, $U_i$
are some unitary operations and $r$ is the normalizing factor. Then, the
encoded state is almost indistinguishable from maximally mixed state by
arbitrary LOCC and quantum communications among any $k-1$ number of parties.

The security of the protocol will be guaranteed by the condition,
\begin{equation}
\parallel L\{\rho^i\}-L\{\rho^j\} \parallel _1~\leq~\epsilon
\end{equation}
where $L$ is any admissible local operation.
Again, the correctness of the scheme is guaranteed by the requirement
that the data is correctly decoded,
\begin{equation}
\parallel(D^{(X)}\otimes E)(\Phi)-\Phi \parallel _1
~\leq~\delta
\end{equation}
where $D^{(X)}$ is the decoding map and in all the above two equations
$\parallel \cdot \parallel_1 $ represents trace norm.

Now, we shall discuss an important property of our activable bound
entangled states defined previously and explore the possibilities of
hiding information.

\section{Activable bound entangled states- a property}

In the previous chapter, we found that activable bound entanglement
is a very special kind of entanglement. They have a highly symmetric
structure with some other very special characteristics. There are
very few classes of mixed entangled states in multi-qubit systems,
which are used in so many computational tasks like the Smolin state
\cite{18,Shor2}. Previously, we have discussed local indistinguishability
of the four activable bound entangled states $\rho_{2N+2}^{\pm}$ and
$\sigma_{2N+2}^{\pm}$ of $2N+2$ qubit system. It provides a suitable
ground for proposing a natural multipartite data hiding scheme. For
this purpose, we investigate how much information of the global states
could be gathered by the co-operation of any proper subset of the the
concerned parties. This is reflected from the nature of the reduced
density matrices corresponding to the four states of the same system.
We first show that all the reduced density matrices of the $2N+1$ qubit
systems from $\rho_{2N+2}^{\pm}$ and $\sigma_{2N+2}^{\pm}$ are maximally mixed
state or in other words, the subsystems do not have any short of
information of the global system.

\emph{\textbf{Theorem:}} \emph{For all $N\geq1$ all the reduced
density matrices corresponding to the four states
$\rho_{2N+2}^{\pm}$, $\sigma_{2N+2}^{\pm}$ are maximally mixed
state.} Ignorance of any one party (i.e., by tracing out one qubit
system) from any of the four states $\rho_{2N+2}^{\pm}$,
$\sigma_{2N+2}^{\pm}$ would result in the state
$\frac{1}{2^{2N+1}}I^{2N+1}$.

To establish this result, let us first consider the four states of
four qubit system. The Smolin state is,
\begin{equation}
\begin{array}{lcl}
\rho_4^{+}&=&\frac{1}{4}~\{P[\Phi^+]\otimes
P[|\Phi^+\rangle]+P[|\Phi^-\rangle]\otimes P[|\Phi^-\rangle]+
P[|\Psi^+\rangle]\otimes P[|\Psi^+\rangle]\\&
&+P[|\Psi^-\rangle]\otimes P[|\Psi^-\rangle]\}\\ &=&
\frac{1}{4}~\{P[\frac{|00 \rangle+ |11\rangle}{\sqrt{2}}]\otimes
P[|\Phi^+\rangle]+P[\frac{|00 \rangle- |11\rangle}{\sqrt{2}}]\otimes
P[|\Phi^-\rangle]\\& &+ P[\frac{|01\rangle+ |10\rangle}{\sqrt{2}}]
\otimes P[|\Psi^+\rangle]+P[\frac{|01 \rangle- |10\rangle}{\sqrt{2}}]\otimes P[|\Psi^-\rangle]\}\\
&=& \frac{1}{8}~\{(P[|00 \rangle]+ P[|11\rangle])\otimes
(P[|\Phi^+\rangle]+P[|\Phi^-\rangle])+ (P[|01 \rangle]\\& &+
P[|10\rangle])\otimes
(P[|\Psi^+\rangle]+P[|\Psi^-\rangle])+(|00\rangle \langle11|+ |11
\rangle \langle 00|)\otimes(P[|\Phi^+\rangle]\\&
&-P[|\Phi^-\rangle])+ (|01\rangle\langle10|+|10\rangle\langle01|)
\otimes(P[|\Psi^+\rangle]-P[|\Psi^-\rangle])\}
\end{array}
\end{equation}
Thus, tracing out the first party of the Smolin state
we have,

\begin{equation}
\begin{array}{lcl}
\rho'_3&=&\frac{1}{8}~\{(P[|0 \rangle]+ P[|1\rangle])\otimes
(P[\Phi^+]+P[\Phi^-])\\& &~~~~+ (P[|0\rangle]+ P[|1\rangle])\otimes
(P[\Psi^+]+P[\Psi^-])\}\\ &=&\frac{1}{8}~(P[|0 \rangle]+
P[|1\rangle])\otimes(P[\Phi^+]+P[\Phi^-]+P[\Psi^+]+P[\Psi^-])\\
&=&\frac{1}{2^{3}}I\otimes I^{2}\\
&=&\frac{1}{2^{3}}I^{3}
\end{array}
\end{equation}

As the state is invariant under all possible permutations of the
concerned parties, therefore, by tracing out any one qubit system
of the Smolin state, we would obtain a maximally mixed state. The
other three activable bound entangled states of the four qubit system
are Pauli-connected with the Smolin state in any one qubit. For
example, if we consider the state $\sigma_4^{-}$, then we can express
it in the form,
\begin{equation}
\begin{array}{lcl}
\sigma_4^{-}&=&(I\otimes I\otimes i\sigma_y \otimes I) ~\rho_4^{+}~
(I \otimes I \otimes i\sigma_y \otimes I)
\end{array}
\end{equation}

This form reflects clearly that tracing out the third party, we would
obtain just the same as that of $\rho_4^{+}$, i.e.,
$\frac{1}{2^{3}}I^{3}$. Following this argument, we conclude that all
the four states $\rho_4^{\pm},~\sigma_4^{\pm}$ have this property.
The next step is to prove this property for the whole class of states.
It would follow from the mathematical induction process prescribed in
Eq.(\ref{bcbe}). The process certainly ensures that if the statement
of the property is true for the $2N$ qubit states then so also for the
$2N+2$ qubit states and thus proceeding from the four qubit states to
the six qubit activable bound entangled states, then from six to eight
and so on. To prove this, let us assume that for some integer $N$,
the four states $\rho_{2N}^{\pm}$, $\sigma_{2N} ^{\pm}$ have this
property. Thus, tracing out the first qubit system of
$\rho_{2N} ^{\pm}$, $\sigma_{2N} ^{\pm}$, we have,
$\frac{1}{2^{2N-1}}I^{2N-1}$. Then applying the relation Eq.(\ref{bcbe}),
we find the state $\frac{1}{2^{2N+1}}I^{2N+1}$ by tracing out the first
qubit system of the state $\rho_{2N+2} ^{\pm}$. As, tracing out the first
qubit system of $\rho_{2N+2} ^{\pm}$ we have,
\begin{equation}
\begin{array}{lcl}
\rho'_{2N+1}&=&~\frac{1}{4}\cdot\frac{1}{2^{2N-1}}I^{2N-1} ~\otimes
(P[\Phi^+]+P[\Phi^-]+P[\Psi^+]+P[\Psi^-])\\ &=&
\frac{1}{2^{2N+1}}~I^{2N-1}\otimes I^{2}\\ &=&
\frac{1}{2^{2N+1}}~I^{2N+1}
\end{array}
\end{equation}
Similarly, for the states $\sigma_{2N+2} ^{\pm}$. Therefore, by the
mathematical induction formula we have the result. Thus, through a
recursive method we obtain that the property is true for the whole
class of activable bound entangled states. As the states are
symmetric over permutations of all the parties concerned, thus tracing
out anyone party we find the same result. It would also imply that the
individual density matrices of each party is a maximally mixed
state, i.e., $\frac{1}{2}I$.\\

This property of the states of generalized class (Eq. (\ref{bcbe})) is
very important for implementing the data hiding protocol using these
states as the hiding states.

\section{Hiding of two classical bits}
The properties of the above class of states enable us to construct a
protocol \cite{hideclass} to hide two cbits of information
among any even number of spatially separated parties starting from
four. Here, we assume that two cbits of information is hidden in the
shared state between $2N+2$ number of parties separated by distance,
i.e., corresponding to the classical information $b=0,1,2,3,$ the
four states $\rho_{2N+2} ^{\pm}$, $\sigma_{2N+2} ^{\pm}$, for $N
\geq 1$ are shared among themseves. The hidden data is secured
against every possible LOCC among all the parties and against any
sort of quantum communication among $2N + 1$ parties as the hidden
data cannot retrieved perfectly, until and unless all the parties
remain separated or all of the $2N + 2$ parties are dishonest.

\textbf{Secured against LOCC:} The protocol appears to be secured
against any LOCC attack as the four states of same system are
locally indistinguishable, so that even probabilistically no
information can be extracted by all possible LOCC among all
concerned parties and no security bound is required against LOCC.
However, this conclusion is not perfectly correct as the states
are inconclusively discriminated by LOCC. We discuss this limitation
later on.

\textbf{Security against global operation:} The data remains secure
under the action of any $2N + 1$ number of dishonest distant parties,
who are allowed to make global operations, by joining in some
laboratories and make collective operations on their joint system.
It follows directly from the maximal ignorance property of the
activable bound entangled states, as ignorance of the system
of the honest party (there should be at least one such or, otherwise
the states are obviously globally distinguishable as being orthogonal
to each other) gives the reduced density matrix of the others to be
the maximally mixed state. Thus, the quantum communication is allowable
among a maximum number of parties, i.e., $2N + 1$.
It is interesting to note that in the above protocol we need only one
honest party, not allowed to communicate with the others through some
quantum channel. The hider may not be a part of the system. It is also
not necessary that the hider herself encrypt the bit in the quantum state
and thus knows the hidden data.

\subsection{Limitations regarding inconclusive distinguishability}
Though the above situation appears to be quite nice to maintain the
secrecy of hidden data in very stronger manner, but it has some
limitations. So far we have only considered perfect distinguishability
of the states. Precisely, it implies to discriminate which state is
given from the whole set (here the set of four states from $2N+2$
qubit system). Though it may possible to determine whether the given
state belongs to some particular subset of the whole set of states.
Such as here, though it is impossible to distinguish perfectly the
four states $\rho_{2N+2} ^{\pm}$, $\sigma_{2N+2} ^{\pm}$, for $N
\geq 1$ even with an arbitrarily small probability by LOCC, but
it is possible to determine by LOCC, either the given state belongs
to the subset $\rho_{2N+2} ^{\pm}$ or, from the subset $\sigma_{2N+2}
^{\pm}$. It is possible simply by measuring on $\sigma_z$ basis in each
party and checking only the parity (even or odd number of
zeros or ones). The basic fact of this set discrimination, taken two
together, is that the four states are locally Pauli connected.

\section{Conclusive remarks}
In conclusion our scheme is proposed to hide two bits of classical
information or, equivalently one quantum bit among $2N+2$ number of
parties, for any $N \geq 1$. The advantage of our protocol is that
the number of parties can be extended in pairs up to any desired
level keeping the individual systems only with dimension two. The
hidden information cannot be exactly revealed by any classical
attack of the corresponding parties and also against every quantum
attack, as long as one party remains honest. However, the hiding
scheme has some limitations from the viewpoint of set
distinguishability. The states are nice for practical preparation
by sharing Bell mixtures among distant parties. This class of
locally Pauli connected but locally indistinguishable states with
the power of activable boundness has created a new direction to
investigate the relation between non-locality and local distinguishability.

\cleardoublepage
\chapter*{Conclusion}

\thispagestyle{plain}
\addcontentsline{toc}{chapter}{Conclusion}

In conclusion, this work highlights on some peculiar features of some special
classes of entangled systems. For bipartite systems, we explore
the nature of incomparability regarding the state transformation
by LOCC with certainty, proposed by M. Nielsen. Our basic object
was to search for the origin of the existence of such pair. We
have provided methods to resolve this non-transferability of
incomparable pairs by LOCC. This feature is further
connected with various impossible operations defined on single
systems. We have considered here the connection of incomparability
with universal anti-unitary operations and universal angle-preserving
operations. As a particular case, both include the universal
spin-flipping operation. The existence of incomparable states also
provides us to consider some kind of irreversibility that plays in pure
state level also. We have succeeded in posing incomparability as
a good detector of impossible operations. We have also observed that the
action of a local operation on multipartite systems is a very peculiar
phenomena. It is not always predictable the behaviour of an operation
in a composite system from its action on a single system. In mixed state level we
have found a general class of Activable Bound Entangled states in
multi-qubit systems. Apart from the nice Bell-correlation and other
important properties, these classes of bound entangled states have
the remarkable feature of local indistinguishability in quantum systems.
The four states of $2N$ qubit system are even probabilistically locally
indistinguishable for each $N\geq 2$. We explore this feature to prescribe a
hiding protocol with these classes. Our protocol, though has its own
limitations regarding inconclusive distinguishability, applicable to any
even number of multi-partite systems. We expect in future those queers will gloom
into some fundamental properties of entangled systems.

\cleardoublepage

\bibliographystyle{unsrt}
\end{document}